\documentclass{article}
\usepackage{amsfonts}

\title{Affine symmetry in mechanics of collective and
internal modes. Part II. Quantum models}

\author{ J. J. S\l awianowski, V. Kovalchuk, A. S\l awianowska,\\
B. Go\l ubowska, A. Martens, E. E. Ro\.zko, Z. J. Zawistowski\\
Institute of Fundamental Technological Research,\\
Polish Academy of Sciences,\\
21 \'{S}wi\c{e}tokrzyska str., 00-049 Warsaw, Poland\\
e-mails: jslawian@ippt.gov.pl, vkoval@ippt.gov.pl,\\
aslawian@ippt.gov.pl, bgolub@ippt.gov.pl,\\ 
amartens@ippt.gov.pl, erozko@ippt.gov.pl,\\ 
zzawist@ippt.gov.pl}

\begin{document}

\maketitle
\begin{abstract}

Discussed is the quantized version of the classical description of
collective and internal affine modes as developed in Part I. We
perform the Schr\"odinger quantization and reduce effectively the
quantized problem from $n^{2}$ to $n$ degrees of freedom. Some
possible applications in nuclear physics and other quantum
many-body problems are suggested. Discussed is also the
possibility of half-integer angular momentum in composed systems
of spin-less particles.

\end{abstract}


\section*{Introduction}

A fascinating feature of our models of affine collective dynamics
is their extremely wide range of applications. It covers the
nuclear and molecular dynamics, micromechanics of structured
continua, perhaps nanostructure and defects phenomena, macroscopic
elasticity and astrophysical phenomena like vibration of stars and
clouds of cosmic dust. Obviously, microphysical applications must
be based on the quantized version of the theory. And one is
dealing then with a very curious convolution of quantum theory
with mathematical methods of continuum mechanics. It is worth to
mention that there were even attempts, mainly by Barut and
R\c{a}czka \cite{Bar-Racz77}, to describe the dynamics of strongly
interacting elementary particles (hadrons) in terms of some
peculiar, quantized continua. By the way, as French say, the
extremes teach one another; it is not excluded that the dynamics
of cosmic objects like neutron stars must be also described in
quantum terms. They are though giant nuclei, very exotic ones,
because composed exclusively of neutrons (enormous "mass numbers"
and vanishing "atomic numbers").

\section{Quantization of classical geodetic systems}

As usual, before quantizing the classical model, one has to
perform some preliminary work on the level of its classical
Hamiltonian dynamics
\cite{Mart03,JJS02_2,JJS04,JJS04_2,JJS-VK04_2}.

Let us consider a classical geodetic system in a Riemannian
manifold $(Q,\Gamma)$, where $Q$ denotes the configuration space,
and $\Gamma$ is the "metric" tensor field on $Q$ underlying the
kinetic energy form. In terms of generalized coordinates or in
Hamiltonian terms we have, respectively,
\[
T=\frac{1}{2}\Gamma_{\mu\nu}\frac{dq^{\mu}}{dt}\frac{dq^{\nu}}{dt},\qquad
\mathcal{T}=\frac{1}{2}\Gamma^{\mu\nu}p_{\mu}p_{\nu},
\]
where, obviously,
$\Gamma^{\mu\alpha}\Gamma_{\alpha\nu}=\delta^{\mu}{}_{\nu}$,
$p_{\mu}=\partial T/\partial
\dot{q}^{\mu}=\Gamma_{\mu\nu}(dq^{\nu}/dt)$.

As usual, the metric tensor $\Gamma$ gives rise to the natural
measure $\mu_{\Gamma}$ on $Q$,
\[
d\mu_{\Gamma}(q)=\sqrt{\left|\det[\Gamma_{\mu\nu}]\right|}dq^{1}\cdots
dq^{f},
\]
where $f$ denotes the number of degrees of freedom, i.e., $f=\dim
Q$. For simplicity the square-root expression will be always
denoted by $\sqrt{\left|\Gamma\right|}$. The mathematical
framework of Schr\"odinger quantization is based on
L$^{2}(Q,\mu_{\Gamma})$, i.e., the Hilbert space of complex-valued
wave functions on $Q$ square-integrable in the
$\mu_{\Gamma}$-sense. Their scalar product is given by the usual
formula:
\[
<\Psi_{1}|\Psi_{2}>=\int
\overline{\Psi}_{1}(q)\Psi_{2}(q)d\mu_{\Gamma}(q).
\]

The classical kinetic energy expression is replaced by the
operator $\mathbf{T}=-(\hbar^{2}/2)\Delta(\Gamma)$, where $\hbar$
denotes the ("crossed") Planck constant, and $\Delta(\Gamma)$ is
the Laplace-Beltrami operator corresponding to $\Gamma$, i.e.,
\[
\Delta(\Gamma)=\frac{1}{\sqrt{|\Gamma|}}\sum_{\mu,\nu}\partial_{\mu}
\sqrt{|\Gamma|}\Gamma^{\mu\nu}\partial_{\nu}=
\Gamma^{\mu\nu}\nabla_{\mu}\nabla_{\nu}.
\]
In the last expression $\nabla_{\mu}$ denotes the Levi-Civita
covariant differentiation in the $\Gamma$-sense. Therefore, the
kinetic energy operator $\mathbf{T}$ is formally obtained from the
corresponding classical expression $\mathcal{T}$ (kinetic
Hamiltonian) by the substitution $p_{\mu}\mapsto
\mathbf{p}_{\mu}=(\hbar/i)\nabla_{\mu}$.

If the problem is non-geodetic and some potential $V(q)$ is
admitted, the corresponding Hamilton (energy) operator is given by
$\mathbf{H}=\mathbf{T}+\mathbf{V}$, where the operator
$\mathbf{V}$ acts on wave functions simply multiplying them by
$V$, i.e., $\left(\mathbf{V}\Psi\right)(q)=V(q)\Psi(q)$. This is
the reason why very often one does not distinguish graphically
between $\mathbf{V}$ and $V$.

\section{Problems concerning quantization}

There are, obviously, many delicate problems concerning
quantization which cannot be discussed here and, fortunately, do
not interfere directly with the main subjects of our analysis.
Nevertheless, we mention briefly some of them. Strictly speaking,
wave functions are not scalars but complex densities of the weight
$1/2$ so that the bilinear expression $\overline{\Psi}\Psi$ is a
real scalar density of weight one, thus, a proper object for
describing probability distributions \cite{Mac63}. But in all
realistic models, and the our one is not an exception, the
configuration space is endowed with some Riemannian structure. And
this enables one to factorize scalar (and tensor) densities into
products of scalars (tensors) and some standard densities built of
the metric tensor. Therefore, the wave function may be finally
identified with the complex scalar field (multicomponent one when
there are internal degrees of freedom).

There are also some arguments for modifying $\mathbf{T}$ by some
scalar term proportional to the curvature scalar. Of course, such
a term may be always formally interpreted as some correction
potential. And besides, we usually deal with Riemannian manifolds
of the constant Riemannian curvature, and then such additional
terms result merely in the over-all shifting of energy levels.

In Riemann manifolds the Levi-Civita affine connection preserves
the scalar product; because of this, the operator $\nabla_{\mu}$
is formally anti-self-adjoint and $(\hbar/i)\nabla_{\mu}$,
$\mathbf{T}=-(\hbar^{2}/2)\Gamma^{\mu\nu}\nabla_{\mu}\nabla_{\nu}$
are formally self-adjoint. They are, however, differential
operators, thus, the difficult problem of self-adjoint extensions
appears. And besides, being differential operators, they are
unbounded in the usual sense, thus, their spectral analysis also
becomes a difficult and delicate subject. All such problems will
be neglected and considered in the zeroth-order approximation of
the mathematical rigor, just as it is usually done in practical
physical applications. This is also justified by the fact that, as
a rule, our first-order differential operators generate some
well-definite global transformation groups admitting a lucid
geometrical interpretation. It is typical that in such situation
all subtle problems on the level of functional analysis, like the
common domains, etc., may be successfully solved.

Therefore, from now on we will proceed in a "physical" way and all
terms like "self-adjoint", "Hermitian", etc. will be used in a
rough way characteristic for physical papers and applied
mathematics.

We shall deal almost exclusively with stationary problems when the
Hamilton operator $\mathbf{H}$ is time-independent, thus, the
Schr\"odinger equation
\[
i\hbar\frac{\partial \psi}{\partial t}=\mathbf{H}\psi
\]
will be replaced by its stationary form, i.e., by the
eigenequation $\mathbf{H}\Psi=E\Psi$, where, obviously,
\[
\psi=\exp\left(-\frac{i}{\hbar}Et\right)\Psi
\]
and $\Psi$ is a time-independent wave function on the
configuration space.

\section{Multi-valuedness of wave functions}

There is another delicate point concerning fundamental aspects of
quantization which, however, may be of some importance and will be
analyzed later on. Namely, it is claimed in all textbooks in
quantum mechanics that wave functions solving reasonable
Schr\"odinger equations must satisfy strong regularity conditions,
and first of all they must be well-defined one-valued functions
all over the configuration space, in addition, continuous together
with their derivatives. This demand is mathematically essential in
the theory of Sturm-Liouville equations and besides it has to do
with quantization or, more precisely, discrete spectra of certain
physical quantities. By the way, these two things are not
independent.

There are, however, certain arguments that some physical systems
may admit multi-valued wave functions. It is so when the
configuration space is not simply connected and its fundamental
group is finite. Physically it is only the squared modulus
$\overline{\Psi}\Psi$ that is to be one-valued because, according
to the Born statistical interpretation, it represents the
probability distribution of detecting a system in various regions
of the configuration space. But for the wave function $\Psi$
itself it is sufficient to be "locally" one-valued and
sufficiently smooth, i.e., to be defined on the universal covering
manifold $\overline{Q}$ of the configuration space $Q$. This may
lead to a consistent quantum mechanics, perhaps with some kind of
superselection rules. It is so in quantum mechanics of rigid body,
which is sometimes expected to be a good model of the elementary
particles spin \cite{ABB95,ABMB95,BBM92}. The configuration space
of the rigid body without translational motion may be identified
with the proper rotation group SO$(3,\mathbb{R})$
(SO$(n,\mathbb{R})$ in $n$ dimensions), obviously, when some
reference orientation and Cartesian coordinates are fixed. But it
is well-known that SO$(3,\mathbb{R})$ is doubly-connected (and so
is SO$(n,\mathbb{R})$ for any $n\geq 3$). Its covering group is
SU$(2)$ (Spin$(n)$ for any $n\geq 3$). Therefore, it is really an
instructive exercise, and perhaps also a promising physical
hypothesis, to develop the rigid top theory with SU$(2)$ as
configuration space \cite{ABB95,ABMB95,BBM92}. In affinely-rigid
body mechanics we are dealing with a similar situation, namely,
GL$(3,\mathbb{R})$ and SL$(3,\mathbb{R})$ (more generally,
GL$(n,\mathbb{R})$ and SL$(n,\mathbb{R})$ for $n>3$) are
doubly-connected. This topological property is simply inherited
from the corresponding one for SO$(3,\mathbb{R})$
(SO$(n,\mathbb{R})$) on the basis of the polar decomposition
\cite{Bar-Racz77,Zhel78,Zhel83}. Therefore, the standard
quantization procedure in a manifold should be modified by using
wave amplitudes defined on the covering manifolds $\overline{{\rm
GL}(n,\mathbb{R})}$, $\overline{{\rm SL}(n,\mathbb{R})}$. By the
way, some difficulty and mathematical curiosity appears then
because these covering groups are non-linear (do not admit
faithful realizations in terms of finite-dimensional matrices).
This fact, known long ago to E. Cartan, was not known to
physicists; a rather long time and enormous work has been lost
because of this.

\section{Classical background for quantization}

Before going into such details we must go back to certain
classical structures underlying quantization procedure. They were
touched earlier in sections 2 and 3 of Part I \cite{part1} but in
a rather superficial way, and besides, we concentrated there on
the collective modes ruled by the linear and affine groups. This
is really the main objective of our study, nevertheless, not
exceptional one; it is also clear that, injecting the subject into
a wider context, one attains a deeper understanding, free of
accidental details.

In section 2 of Part I \cite{part1} Lie-algebraic objects
$\Omega,\hat{\Omega}\in G^{\prime}$ were introduced. It is an
important fact from the Lie group theory that they give rise to
some vector fields $X$, $Y$ on $G$ invariant, respectively, under
right and left translations on $G$. Namely, for any fixed
$\Omega,\hat{\Omega}\in G^{\prime}$, they are given by
$X_{g}[\Omega]:=\Omega g$, $Y_{g}[\hat{\Omega}]:=g\hat{\Omega}$.

Affine velocities introduced in section 3 of Part I \cite{part1}
are just the special case of Lie-algebraic objects. In the same
section the dual objects $\Sigma$, $\hat{\Sigma}$, i.e., affine
spin in two representations, were introduced. These dual
quantities exist also in the general case when $G$ is an arbitrary
Lie group. They are then elements of the dual space, i.e., Lie
co-algebra, $\Sigma,\hat{\Sigma}\in G^{\prime\ast}$. Their
relationship with canonical momenta $p$ and configurations $g$ is
given by the following formula involving evaluations of co-vectors
on vectors: $\langle
p,\dot{g}\rangle=\langle\Sigma,\Omega\rangle=\langle\hat{\Sigma},\hat{\Omega}\rangle$,
where $\dot{g}\in T_{g}G$, $p\in T^{\ast}_{g}G$, and $g$,
$\dot{g}$ are arbitrary. Denoting the adjoint transformation of
Ad$_{g}$ by the usual symbol Ad$^{\ast}_{g}$, we have that
$\Sigma={\rm Ad}^{\ast -1}_{g}\hat{\Sigma}$, the obvious
generalization of the corresponding relationship between
laboratory and co-moving representation of affine (or usual
metrical) spin. And just as in this special case, the quantities
$\Sigma$, $\hat{\Sigma}$ are Hamiltonian generators of the groups
of left and right regular translations $L_{G}$, $R_{G}$ on $G$.

In applications we are usually dealing with some special Lie
groups for which many important formulas and relationships may be
written in a technically simple form avoiding the general abstract
terms.

As mentioned, throughout this series of articles we are dealing
almost exclusively with linear groups $G\in$ GL$(W)\subset$
L$(W)$, where $W$ is a linear space, e.g., some $\mathbb{R}^{n}$
or $\mathbb{C}^{n}$.

All the mentioned simplifications follow from the obvious
canonical isomorphism between L$(W)$ and its dual L$(W)^{\ast}$,
based on the pairing $\langle C,D\rangle={\rm Tr}\left(CD\right)$.
The Lie algebra $G^{\prime}$ is a linear subspace of L$(W)$,
therefore, its dual space $G^{\prime\ast}$ may be canonically
identified with the quotient space L$(W)^{\ast}/{\rm
An}G^{\prime}$, where An$G^{\prime}$ denotes the subspace of
linear functions vanishing on $G^{\prime}$. But, according to the
above identification between L$(W)^{\ast}$ and L$(W)$ itself,
An$G^{\prime}$ may be identified with some linear subspace of
L$(W)$; we shall denote it by $G^{\prime\bot}$. Therefore, the Lie
co-algebra $G^{\prime\ast}$ is canonically isomorphic with the
corresponding quotient, i.e., $G^{\prime\ast}\simeq {\rm
L}(W)/G^{\prime\bot}$. This is the general fact for linear groups
and their Lie algebras. However, in some special cases, just ones
of physical relevance, this quotient space admits a natural
canonical isomorphism onto some distinguished linear subspace of
L$(W)$ consisting of natural representants of cosets, e.g., in the
most practical cases $G^{\prime\ast}$ is canonically isomorphic
with $G^{\prime}$ itself. For example, it is so for
SO$(n,\mathbb{R})$, SL$(n,\mathbb{R})$, where the Lie algebras
SO$(n,\mathbb{R})^{\prime}$, SL$(n,\mathbb{R})^{\prime}$ may be
identified with the duals SO$(n,\mathbb{R})^{\prime\ast}$,
SL$(n,\mathbb{R})^{\prime\ast}$. By the way, for certain reasons
it is more convenient to use the pairing $\langle
A,B\rangle=-(1/2){\rm Tr}(AB)$ for the orthogonal group
SO$(n,\mathbb{R})$.

Just as in the special case of affine objects, transformation
rules for $\Sigma$, $\hat{\Sigma}$ are analogous to those for
$\Omega$, $\hat{\Omega}$; we mean transformations under regular
translations:
\begin{eqnarray}
L_{k}&:&\quad \Sigma\mapsto {\rm Ad}_{k}^{\ast -1}\Sigma,\qquad
\hat{\Sigma}\mapsto \hat{\Sigma},
\nonumber\\
R_{k}&:&\quad \Sigma\mapsto \Sigma,\qquad\quad \hat{\Sigma}\mapsto
{\rm Ad}_{k}^{\ast}\hat{\Sigma}.\nonumber
\end{eqnarray}
Using the identifications mentioned above (assuming that they
work), we can write these rules in a form analogous to that for
non-holonomic velocities,
\begin{eqnarray}
L_{k}&:&\quad \Sigma\mapsto k\Sigma k^{-1},\qquad
\hat{\Sigma}\mapsto \hat{\Sigma},\nonumber
\\
R_{k}&:&\quad \Sigma\mapsto \Sigma,\qquad \hat{\Sigma}\mapsto
k^{-1}\hat{\Sigma}k,\nonumber
\end{eqnarray}
i.e., just as it is for the affine spin.

Geometrical meaning of $\Sigma$ and $\hat{\Sigma}$ is that of the
momentum mappings induced, respectively, by the group of left and
right regular translations. And the relationship between two
versions of $\Sigma$-objects is as follows:
$\Sigma=g\hat{\Sigma}g^{-1}$. The objects $\Sigma$ and
$\hat{\Sigma}$ may be also interpreted in terms of right- and
left-invariant differential forms (co-vector fields), i.e.,
Maurer-Cartan forms $A$, $B$ on the group $G$. Assuming the
afore-mentioned identification, we can express $A$, $B$ for any
fixed $\Sigma$, $\hat{\Sigma}$ in the following forms:
$A_{g}[\Sigma]=g^{-1}\Sigma$,
$B_{g}[\hat{\Sigma}]=\hat{\Sigma}^{-1}g$.

Just as in the special case of affine systems, Poisson bracket
relations of $\Sigma$- and $\hat{\Sigma}$-components are given by
structure constants of $G$. Those for $\hat{\Sigma}$ have opposite
signs to those for $\Sigma$, and the mutual ones vanish (left
regular translations commute with the right ones).

\section{Hamiltonian systems on Lie group spaces}

Geodetic Hamiltonian systems on Lie group spaces were studied by
various research groups; let us mention, e.g., the prominent
mathematicians like Hermann, Arnold, Mi\-shchenko, Fomenko, and
others. Obviously, the special stress was laid on models with
kinetic energies (Riemann structures on $G$) invariant under left
or right regular translations. As expected, models invariant
simultaneously under left and right translations have some special
properties and due to their high symmetries are computationally
simplest.

From now on we assume that our configuration space $Q$ is a Lie
group $G$ or, more precisely, its homogeneous space with trivial
isotropy groups. Also in a more general situation when isotropy
groups are nontrivial (even continuous) a large amount of analysis
performed on group spaces remains useful.

Obviously, just as in the special case of affinely-rigid bodies,
left- and right-invariant kinetic energies $T$ are, respectively,
quadratic forms of $\hat{\Omega}$ and $\Omega$ with constant
coefficients. Their underlying Riemannian structures on $G$ are
locally flat if and only if $G$ is Abelian.

In both theoretical and practical problems the Hamilton language
based on Poisson brackets is much more lucid and efficient than
that based on Lagrange equations. If besides of geodetic inertia
the system is influenced only by potential forces derivable from
some potential energy term $V(q)$, then, obviously, the classical
Hamiltonian is given by the following expression:
\[
H=\mathcal{T}+V(q)=\frac{1}{2}\Gamma^{\mu\nu}(q)p_{\mu}p_{\nu}+V(q).
\]
It is very convenient to express the Hamiltonian and all other
essential quantities in terms of non-holonomic velocities and
their conjugate non-holonomic (Poisson-non-commuting) momenta.

Let $\{E_{\mu}\}$ be some basis in the Lie algebra $G^{\prime}$
and $q^{\mu}$ be the corresponding canonical coordinates of the
first kind on $G$, i.e., $g(q)=\exp\left(q^{\mu}E_{\mu}\right)$.
Lie-algebraic objects $\Omega,\hat{\Omega}\in G^{\prime}$ will be,
respectively, expanded as follows: $\Omega=\Omega^{\mu}E_{\mu}$,
$\hat{\Omega}=\hat{\Omega}^{\mu}E_{\mu}$. Using the expansion
coefficients $\Omega^{\mu}$, $\hat{\Omega}^{\mu}$ one obtains the
following simple expressions for the left- and right-invariant
kinetic energies:
\[
T_{\rm
left}=\frac{1}{2}\mathcal{L}_{\mu\nu}\hat{\Omega}^{\mu}\hat{\Omega}^{\nu},
\qquad T_{\rm
right}=\frac{1}{2}\mathcal{R}_{\mu\nu}\Omega^{\mu}\Omega^{\nu},
\]
where the matrices $\mathcal{L}$, $\mathcal{R}$ are constant,
symmetric, and non-singular. The positive definiteness problem is
a more delicate matter, and there are some hyperbolic-signature
structures of some relevance both for physics and pure geometry.

For potential systems Legendre transformation may be easily
described with the use of non-holonomic objects, respectively,
\[
\hat{\Sigma}_{\mu}=\frac{\partial T_{\rm left}}{\partial
\hat{\Omega}^{\mu}}=\mathcal{L}_{\mu\nu}\hat{\Omega}^{\nu},\qquad
\Sigma_{\mu}=\frac{\partial T_{\rm right}}{\partial
\Omega^{\mu}}=\mathcal{R}_{\mu\nu}\Omega^{\nu},
\]
where, obviously, $\hat{\Sigma}_{\mu}$, $\Sigma_{\mu}$ are
expansion coefficients of $\hat{\Sigma}$, $\Sigma$ with respect to
the dual basis $\{E^{\mu}\}$ of the Lie co-algebra, i.e.,
$\hat{\Sigma}=\hat{\Sigma}_{\mu}E^{\mu}$,
$\Sigma=\Sigma_{\mu}E^{\mu}$. The resulting Hamiltonians have,
respectively, the following forms:
\[
H=\mathcal{T}_{\rm
left}+V(q)=\frac{1}{2}\mathcal{L}^{\mu\nu}\hat{\Sigma}_{\mu}\hat{\Sigma}_{\nu}
+V(q),
\]
\[
H=\mathcal{T}_{\rm right
}+V(q)=\frac{1}{2}\mathcal{R}^{\mu\nu}\Sigma_{\mu}\Sigma_{\nu}
+V(q),
\]
where, obviously, the matrices $[\mathcal{L}^{\mu\nu}]$,
$[\mathcal{R}^{\mu\nu}]$ are reciprocal to
$[\mathcal{L}_{\mu\nu}]$, $[\mathcal{R}_{\mu\nu}]$.

If structure constants of $G^{\prime}$ with respect to the basis
$\{E_{\mu}\}$ are defined according to the convention
$[E_{\mu},E_{\nu}]=E_{\lambda}C^{\lambda}{}_{\mu\nu}$, then the
Poisson brackets of $\Sigma$-objects are given as follows:
\[
\{\Sigma_{\mu},\Sigma_{\nu}\}=\Sigma_{\lambda}C^{\lambda}{}_{\mu\nu},\qquad
\{\hat{\Sigma}_{\mu},\hat{\Sigma}_{\nu}\}=-\hat{\Sigma}_{\lambda}C^{\lambda}{}_{\mu\nu},
\qquad \{\Sigma_{\mu},\hat{\Sigma}_{\nu}\}=0.
\]

\section{Basic differential operators}

Let us define basic differential operators generating left and
right regular translations on $G$. We denote them respectively by
${\bf L}_{\mu}$ and $\mathbf{R}_{\mu}$. Their action on complex-
or vector-valued functions $F$ on $G$ is defined as follows:
\begin{equation}\label{tr15}
\left({\bf L}_{\mu}F\right)(g):=\left.\frac{\partial}{\partial
q^{\mu}}F\left(k(q)g\right)\right|_{q=0},\qquad
\left(\mathbf{R}_{\mu}F\right)(g):=\left.\frac{\partial}{\partial
q^{\mu}}F\left(gk(q)\right)\right|_{q=0}.
\end{equation}
Their Lie-bracket (commutator) relations differ from the above
Poisson rules for $\Sigma$-quantities by signs:
\[
[{\bf L}_{\mu},{\bf L}_{\nu}]=-{\bf
L}_{\lambda}C^{\lambda}{}_{\mu\nu},\qquad
[\mathbf{R}_{\mu},\mathbf{R}_{\nu}]=\mathbf{R}_{\lambda}C^{\lambda}{}_{\mu\nu},
\qquad [{\bf L}_{\mu},\mathbf{R}_{\nu}]=0.
\]
Poisson brackets between $\Sigma$-objects and functions $F$
depending only on coordinates $q$ (pull-backs of functions defined
on the configuration space $Q=G$) are given by
\[
\{\Sigma_{\mu},F\}=-{\bf L}_{\mu}F,\qquad
\{\hat{\Sigma}_{\mu},F\}=-\mathbf{R}_{\mu}F.
\]
The system of Poisson brackets quoted above is sufficient for
calculating any other Poisson bracket with the help of well-known
properties of this operation. Thus, e.g., for any pair of
functions $A$, $B$ depending in general on all phase-space
variables we have the following expression:
\[
\{A,B\}=\Sigma_{\lambda}C^{\lambda}{}_{\mu\nu}\frac{\partial
A}{\partial \Sigma_{\mu}}\frac{\partial B}{\partial \Sigma_{\nu}}-
\frac{\partial A}{\partial \Sigma_{\mu}}{\bf
L}_{\mu}B+\frac{\partial B}{\partial \Sigma_{\mu}}{\bf L}_{\mu}A,
\]
and, when the phase space is parameterized in terms of quantities
$q^{\mu}$, $\hat{\Sigma}_{\mu}$, we have the similar expression:
\[
\{A,B\}=-\hat{\Sigma}_{\lambda}C^{\lambda}{}_{\mu\nu}\frac{\partial
A}{\partial \hat{\Sigma}_{\mu}}\frac{\partial B}{\partial
\hat{\Sigma}_{\nu}}- \frac{\partial A}{\partial
\hat{\Sigma}_{\mu}}\mathbf{R}_{\mu}B+\frac{\partial B}{\partial
\hat{\Sigma}_{\mu}}\mathbf{R}_{\mu}A.
\]
Obviously, the finite regular translations may be expressed in
terms of the following exponential formulas:
\begin{equation}\label{O16}
F\left(k(q)g\right)=\exp\left(q^{\mu}{\bf L}_{\mu}\right)F,\qquad
F\left(gk(q)\right)=\exp(q^{\mu}\mathbf{R}_{\mu})F,
\end{equation}
with all known provisos concerning exponentiation of differential
operators.

Non-holonomic velocities $\Omega$, $\hat{\Omega}$ depend linearly
on generalized velocities $\dot{q}$, i.e.,
$\Omega^{\mu}=\Omega^{\mu}{}_{\nu}(q)\dot{q}^{\nu}$,
$\hat{\Omega}^{\mu}=\hat{\Omega}^{\mu}{}_{\nu}(q)\dot{q}^{\nu}$.
Similarly, $\Sigma$ and $\hat{\Sigma}$ depend contragradiently on
the conjugate momenta $p$, i.e.,
$\Sigma_{\mu}=p_{\alpha}\Sigma^{\alpha}{}_{\mu}(q)$,
$\hat{\Sigma}_{\mu}=p_{\alpha}\hat{\Sigma}^{\alpha}{}_{\mu}(q)$,
where, obviously,
$\Sigma^{\alpha}{}_{\mu}\Omega^{\mu}{}_{\beta}=\delta^{\alpha}{}_{\beta}$,
$\hat{\Sigma}^{\alpha}{}_{\mu}\hat{\Omega}^{\mu}{}_{\beta}=\delta^{\alpha}{}_{\beta}$.
This leads to the following expressions for generators:
\[
{\bf L}_{\mu}=\Sigma^{\alpha}{}_{\mu}\frac{\partial}{\partial
q^{\alpha}},\qquad
\mathbf{R}_{\mu}=\hat{\Sigma}^{\alpha}{}_{\mu}\frac{\partial}{\partial
q^{\alpha}}.
\]

Many of the above statements remain true for the general
non-holonomic velocities and their conjugate momenta without
group-theoretical background \cite{God03}. Nevertheless, there are
also important facts depending on the group structure and on the
properties of $\Sigma_{\mu}$, $\hat{\Sigma}_{\mu}$ respectively as
the basic right- and left-invariant co-vector fields
(Maurer-Cartan forms). This concerns mainly invariant volumes,
scalar products, Hermiticity of basic operators, and structure of
the Laplace-Beltrami operator.

In group manifolds we are usually interested in left- or
right-invariant kinetic energies. Even in the special case of the
double invariance the definition-based direct calculation of the
corresponding Laplace-Beltrami operator and the volume element may
be rather complicated. However, if the corresponding kinetic
metrics is left- or right-invariant, then so is the resulting
volume element. Therefore, the L$^{2}$-structure on $G$ may be
directly based on the integration with respect to the Haar
measure. As known from the theory of locally compact groups, this
measure is unique up to the constant normalization factor. In the
special case of compact groups this normalization may be fixed by
the natural demand that the total (finite in this case) volume
equals to unity. In any case, the normalization is non-essential.
In applications one deals usually with so-called unimodular
groups, where the left and right measures coincide
\cite{Loom53,Maur68}. Obviously, for the left- or right-invariant
kinetic energies the measures $\mu_{\Gamma}$ built of the
underlying metrics $\Gamma$ are also left- or right-invariant.
Therefore, they coincide with the Haar measure. This enables one
to use the Haar measure from the very beginning as the integration
prescription underlying the scalar product definition. This is
very convenient for two reasons. First of all, for typical Lie
groups appearing in physical applications the Haar measures are
explicitly known. Another nice and reasonable feature of such a
procedure is that once fixing the normalization we are given the
standard integration procedure, whereas the use of
$d\mu_{\Gamma}=\sqrt{|\Gamma|}dq^{1}\cdots dq^{f}$ changes the
scalar product normalization for various models of $T$ (of
$\Gamma$). This constant factor change is not very essential, but
its dependence on various inertial parameters like the above $I$,
$A$, $B$ obscures the comparison of various models.

\section{Unitary transformations}

It follows from the very nature of the Haar measure $\mu$ that on
the level of wave functions the left and right regular
translations are realized by unitary transformations on
L$^{2}(G,\mu)$. More precisely, let us define for any $k\in G$ the
operators ${\bf L}(k)$, $\mathbf{R}(k)$ given by $\left({\bf
L}(k)\Psi\right)(g):=\Psi(kg)$,
$\left(\mathbf{R}(k)\Psi\right)(g):=\Psi(gk)$ for any $g\in G$. It
is clear that ${\bf L}(k)$, $\mathbf{R}(k)$ preserve the space
L$^{2}(G,\mu)$, moreover, they are unitary transformations,
\[
\left<{\bf L}(k)\Psi_{1}|{\bf
L}(k)\Psi_{2}\right>=\left<\mathbf{R}(k)\Psi_{1}|\mathbf{R}(k)\Psi_{2}\right>
=\left<\Psi_{1}|\Psi_{2}\right>.
\]
The assignments $G\ni k\mapsto {\bf L}(k),\mathbf{R}(k)$ are,
respectively, a unitary anti-represen\-tation and representation
of $G$ in L$^{2}(G,\mu)$, i.e., 
\[
{\bf L}(k_{1}k_{2})={\bf
L}(k_{2}){\bf L}(k_{1}),\quad
\mathbf{R}(k_{1}k_{2})=\mathbf{R}(k_{1})\mathbf{R}(k_{2}). 
\]
To
convert ${\bf L}$ into representation it is sufficient to replace
$\Psi(kg)$ by $\Psi(k^{-1}g)$. Obviously, the difference is rather
cosmetical and related to the conventions concerning the
definition of the superposition of mappings. Nevertheless, any
neglect may lead to the accumulation of sign errors and finally to
numerically wrong results.

The operators ${\bf L}_{\mu}$, $\mathbf{R}_{\mu}$ generate the
above representations, thus, we have
\[
{\bf L}\left(\exp(q^{\mu}E_{\mu})\right)=\exp(q^{\mu}{\bf
L}_{\mu}),\qquad
\mathbf{R}\left(\exp(q^{\mu}E_{\mu})\right)=\exp(q^{\mu}\mathbf{R}_{\mu}),
\]
with all known provisos concerning domains and exponents of
evidently unbounded differential operators. It is important to
remember that the left-hand sides are always well-defined bounded
unitary operators acting on the whole L$^{2}(G,\mu)$. Unlike this,
${\bf L}_{\mu}$, $\mathbf{R}_{\mu}$ act only on differentiable
functions, they are unbounded, and the problems of domain and
convergence appear on the right-hand sides of the above equations.

Unitarity of ${\bf L}$, $\mathbf{R}$ implies that their generators
${\bf L}_{\mu}$, $\mathbf{R}_{\mu}$ are formally anti-self-adjoint
(physicists tell roughly: anti-Hermitian), i.e., 
\[
\left<{\bf
L}_{\mu}\Psi_{1}|\Psi_{2}\right>=-\left<\Psi_{1}|{\bf
L}_{\mu}\Psi_{2}\right>,\quad
\left<\mathbf{R}_{\mu}\Psi_{1}|\Psi_{2}\right>=
-\left<\Psi_{1}|\mathbf{R}_{\mu}\Psi_{2}\right>, 
\]
assuming that
the left- and right-hand sides are well-defined (this is the case,
e.g., for differentiable compactly supported functions on $G$).

Now, let us introduce the following operators:
\begin{equation}\label{!20}
{\bf \Sigma}_{\mu}:=\frac{\hbar}{i}{\bf L}_{\mu},\qquad {\bf
\hat{\Sigma}}_{\mu}:=\frac{\hbar}{i}\mathbf{R}_{\mu}.
\end{equation}
They are formally self-adjoint, i.e., "Hermitian" in the rough
language of quantum physicists: 
\[
\langle{\bf
\Sigma}_{\mu}\Psi_{1}|\Psi_{2}\rangle=\langle\Psi_{1}|{\bf
\Sigma}_{\mu}\Psi_{2}\rangle,\quad \langle{\bf
\hat{\Sigma}}_{\mu}\Psi_{1}|\Psi_{2}\rangle=\langle\Psi_{1}|{\bf
\hat{\Sigma}}_{\mu}\Psi_{2}\rangle, 
\]
with the same as previously
provisos concerning the functions $\Psi_{1}$, $\Psi_{2}$.
Obviously, $\hbar$ denotes the ("crossed") Planck constant.

The operators ${\bf \Sigma}_{\mu}$, ${\bf \hat{\Sigma}}_{\mu}$ are
quantized counterparts of classical physical quantities
$\Sigma_{\mu}$, $\hat{\Sigma}_{\mu}$. They may be expressed as
follows:
\[
{\bf
\Sigma}_{\mu}=\frac{\hbar}{i}\Sigma^{\alpha}{}_{\mu}(q)\frac{\partial}{\partial
q^{\alpha}},\qquad {\bf
\hat{\Sigma}}_{\mu}=\frac{\hbar}{i}\hat{\Sigma}^{\alpha}{}_{\mu}(q)\frac{\partial}{\partial
q^{\alpha}}.
\]

There is no problem of ordering of $q$-variables and differential
operators $\partial/\partial q^{\alpha}$. This ordering is exactly
as above, just due to the interpretation of ${\bf L}_{\mu}$ and
$\mathbf{R}_{\mu}$ as infinitesimal generators of one-parameter
subgroups.

\section{Quantum Poisson bracket}

In virtue of the above group-theoretical arguments the quantum
Poisson-bracket rules are analogous to the classical ones,
\[
{}_{Q}\{{\bf \Sigma}_{\mu},{\bf \Sigma}_{\nu}\}={\bf
\Sigma}_{\lambda}C^{\lambda}{}_{\mu\nu},\qquad {}_{Q}\{{\bf
\hat{\Sigma}}_{\mu},{\bf \hat{\Sigma}}_{\nu}\}=-{\bf
\hat{\Sigma}}_{\lambda}C^{\lambda}{}_{\mu\nu},\qquad {}_{Q}\{{\bf
\Sigma}_{\mu},{\bf \hat{\Sigma}}_{\nu}\}=0.
\]
Let us remind that the quantum Poisson bracket of operators is
defined as
\[
{}_{Q}\{{\bf A},{\bf B}\}:=\frac{1}{i\hbar}[{\bf A},{\bf
B}]=\frac{1}{i\hbar}({\bf A}{\bf B}-{\bf B}{\bf A}).
\]

One can show (see, e.g., \cite{God03}) that the kinetic energy
operators for the left- and right-invariant models are given
simply by the formerly quoted formulas with the classical
generators $\Sigma_{\mu}$, $\hat{\Sigma}_{\mu}$ replaced by the
corresponding operators ${\bf \Sigma}_{\mu}$, ${\bf
\hat{\Sigma}}_{\mu}$, i.e.,
\[
{\bf \mathcal{T}}_{\rm left}=\frac{1}{2}\mathcal{R}^{\mu\nu}{\bf
\hat{\Sigma}}_{\mu}{\bf \hat{\Sigma}}_{\nu}=-\frac{\hbar^{2}}{2}
\mathcal{R}^{\mu\nu}\mathbf{R}_{\mu}\mathbf{R}_{\nu},
\]
\[
{\bf
\mathcal{T}}_{\rm right}=\frac{1}{2}\mathcal{L}^{\mu\nu}{\bf
\Sigma}_{\mu}{\bf \Sigma}_{\nu}=-\frac{\hbar^{2}}{2}
\mathcal{L}^{\mu\nu}{\bf L}_{\mu}{\bf L}_{\nu}.
\]
As mentioned, the literal calculation of the Laplace-Beltrami
operator in terms of local coordinates $q^{\mu}$ is usually very
complicated and the resulting formula is, as a rule, quite
obscure, non-readable, and because of this practically non-useful.
Unlike this, the above block expression in terms of generators is
geometrically lucid and well apt for solving procedure of the
Schr\"odinger equation. In various problems it is sufficient to
operate algebraically with quantum Poisson brackets. To complete
the above system of brackets let us quote expressions involving
generators and position-type variables. The latter ones are
operators which multiply wave functions by other functions on the
configuration space, i.e., $\left({\bf
F}\Psi\right)(q):=F(q)\Psi(q)$. If there is no danger of
misunderstanding, we will not distinguish graphically between
${\bf F}$ and $F$. Just as on the classical level we have
\[
{}_{Q}\{{\bf \Sigma}_{\mu},{\bf F}\}=-{\bf L}_{\mu}F,\qquad
{}_{Q}\{{\bf \hat{\Sigma}}_{\mu},{\bf F}\}=-\mathbf{R}_{\mu}F.
\]
Obviously, two position-type operators mutually commute.

\noindent {\bf Remark:} Obviously, only for generators and
position quantities the quantum and classical Poisson rules are
identical. For other quantities it is no longer the case,
moreover, there are problems with the very definition of quantum
counterparts of other classical quantities. The very existence of
the above distinguished family of physical quantities is due to
the group-theoretical background of degrees of freedom.

\section{Corresponding Haar measures}

Let us now return to the main subject of our analysis, i.e., to
the quantization of affine systems. For technical purposes we
again fix some Cartesian coordinates $x^{i}$, $a^{K}$ in $M$, $N$
and identify analytically the configuration space $Q={\rm
LI}(U,V)\times M$ with the affine group GAf$(n,\mathbb{R})\simeq
{\rm GL}(n,\mathbb{R})\times_{s}\mathbb{R}^{n}$. Similarly, the
internal configuration space $Q_{\rm int}={\rm LI}(U,V)$ is
identified with GL$(n,\mathbb{R})$. The corresponding Haar
measures will be denoted respectively by $\alpha$, $\lambda$,
i.e., $d\alpha(\varphi,x)=(\det\varphi)^{-n-1}dx^{1}\cdots
dx^{n}d\varphi^{1}{}_{1}\cdots
d\varphi^{n}{}_{n}=(\det\varphi)^{-1}d\lambda(\varphi)dx^{1}\cdots
dx^{n}$,
$d\lambda(\varphi)=(\det\varphi)^{-n}d\varphi^{1}{}_{1}\cdots
d\varphi^{n}{}_{n}$. In terms of the binary decomposition we have
the following expression:
\[
d\lambda(\varphi)=d\lambda(l;q;r)=\prod_{i\neq j}\left|{\rm
sh}(q^{i}-q^{j})\right|d\mu(l)d\mu(r)dq^{1}\cdots dq^{n},
\]
where $\mu$ denotes the Haar measure on SO$(n,\mathbb{R})$. Due to
the compactness of SO$(n,\mathbb{R})$ we can, but of course need
not, normalize $\mu$ to unity, $\mu\left({\rm
SO}(n,\mathbb{R})\right)=1$.

The Haar measure on SL$(n,\mathbb{R})$ used in quantum mechanics
of incompressible objects may be symbolically written with the use
of Dirac distribution as follows:
\[
d\lambda_{SL}(\varphi)=\prod_{i\neq j}\left|{\rm
sh}(q^{i}-q^{j})\right|d\mu(l)d\mu(r)\delta(q^{1}+\cdots+q^{n})dq^{1}\cdots
dq^{n}.
\]

\section{Kinetic energy operators for affine models}

Affine spin and its co-moving representation are, respectively,
given by the following formally self-adjoint operators:
\[
{\bf \Sigma}^{a}{}_{b}:=\frac{\hbar}{i}{\bf L}^{a}{}_{b}=
\frac{\hbar}{i}\varphi^{a}{}_{K}\frac{\partial}{\partial
\varphi^{b}{}_{K}},\qquad {\bf
\hat{\Sigma}}^{A}{}_{B}:=\frac{\hbar}{i}\mathbf{R}^{A}{}_{B}=
\frac{\hbar}{i}\varphi^{m}{}_{B}\frac{\partial}{\partial
\varphi^{m}{}_{A}}.
\]
The usual spin and vorticity operators are respectively given by
\begin{equation}\label{7.1}
{\bf S}^{a}{}_{b}:={\bf \Sigma}^{a}{}_{b}-g^{ac}g_{bd}{\bf
\Sigma}^{d}{}_{c},\qquad {\bf V}^{A}{}_{B}:={\bf
\hat{\Sigma}}^{A}{}_{B}-\eta^{AC}\eta_{BD}{\bf
\hat{\Sigma}}^{D}{}_{C}.
\end{equation}
Kinetic energy operators corresponding to the formerly described
classical models of internal kinetic energies are simply obtained
by replacing the classical quantities $\Sigma^{a}{}_{b}$,
$\hat{\Sigma}^{A}{}_{B}$ by the above operators ${\bf
\Sigma}^{a}{}_{b}$, ${\bf \hat{\Sigma}}^{A}{}_{B}$ without any
attention to be paid to the ordering problem (just because of the
group-theoretic interpretation of these quantities).

Thus, for the affine-affine model (affine both in space and in the
material) we have
\[
\mathbf{T}^{\rm aff-aff}_{\rm int}=\frac{1}{2A}{\bf
\Sigma}^{i}{}_{j}{\bf \Sigma}^{j}{}_{i}-\frac{B}{2A(A+nB)}{\bf
\Sigma}^{i}{}_{i}{\bf \Sigma}^{j}{}_{j}
\]
\[
=\frac{1}{2A}{\bf
\hat{\Sigma}}^{A}{}_{B}{\bf
\hat{\Sigma}}^{B}{}_{A}-\frac{B}{2A(A+nB)}{\bf
\hat{\Sigma}}^{A}{}_{A}{\bf \hat{\Sigma}}^{B}{}_{B}.
\]
Similarly, for models with the mixed metrical-affine and
affine-metrical invariance we have, respectively,
\begin{eqnarray}
\mathbf{T}^{\rm met-aff}_{\rm
int}&=&\frac{1}{2\widetilde{I}}g_{ik}g^{jl}{\bf
\Sigma}^{i}{}_{j}{\bf
\Sigma}^{k}{}_{l}+\frac{1}{2\widetilde{A}}{\bf
\Sigma}^{i}{}_{j}{\bf
\Sigma}^{j}{}_{i}+\frac{1}{2\widetilde{B}}{\bf
\Sigma}^{i}{}_{i}{\bf
\Sigma}^{j}{}_{j},\nonumber\\
\mathbf{T}^{\rm aff-met}_{\rm
int}&=&\frac{1}{2\widetilde{I}}\eta_{AB}\eta^{CD}{\bf
\hat{\Sigma}}^{A}{}_{C}{\bf
\hat{\Sigma}}^{B}{}_{D}+\frac{1}{2\widetilde{A}}{\bf
\hat{\Sigma}}^{A}{}_{B}{\bf
\hat{\Sigma}}^{B}{}_{A}+\frac{1}{2\widetilde{B}}{\bf
\hat{\Sigma}}^{A}{}_{A}{\bf \hat{\Sigma}}^{B}{}_{B},\nonumber
\end{eqnarray}
where $\widetilde{I}=\left(I^{2}-A^{2}\right)/I$,
$\widetilde{A}=\left(A^{2}-I^{2}\right)/A$,
$\widetilde{B}=-\left(I+A\right)\left(I+A+nB\right)/B$

Similarly, the corresponding expressions for ${\bf
\mathcal{T}}_{\rm tr}$ have the following forms:
\[
\mathbf{T}^{\rm met-aff}_{\rm tr}=\frac{m}{2}g^{ij}{\bf P}_{i}{\bf
P}_{j}=\frac{m}{2}\widetilde{G}^{AB}{\bf \hat{P}}_{A}{\bf
\hat{P}}_{B},
\] 
\[
\mathbf{T}^{\rm aff-met}_{\rm
tr}=\frac{m}{2}\widetilde{C}^{ij}{\bf P}_{i}{\bf
P}_{j}=\frac{m}{2}\eta^{AB}{\bf \hat{P}}_{A}{\bf \hat{P}}_{B},
\]
where ${\bf P}_{i}$, ${\bf \hat{P}}_{A}$ are linear momentum
operators respectively in laboratory and co-moving
representations,
\[
{\bf P}_{a}=\frac{\hbar}{i}\frac{\partial}{\partial x^{a}},\qquad
{\bf \hat{P}}_{K}=\varphi^{a}{}_{K}{\bf
P}_{a}=\frac{\hbar}{i}\varphi^{a}{}_{K}\frac{\partial}{\partial
x^{a}}.
\]
Just as previously, $\widetilde{C}$, $\widetilde{G}$ are
contravariant reciprocals of deformation tensors:
$\widetilde{C}^{ik}C_{kj}=\delta^{i}{}_{j}$,
$\widetilde{G}^{AC}G_{CB}=\delta^{A}{}_{B}$. As mentioned, there
are no affine-affine models of $\mathbf{T}_{\rm tr}$, and
therefore, no affine-affine models of $\mathbf{T}$. The
corresponding "metric tensors" on GAf$(n,\mathbb{R})$ would have
to be singular.

Another important physical quantity is the canonical momentum
conjugate to the dilatational coordinate $q$. On the quantum level
it is represented by the formally self-adjoint operator
\[
{\bf p}=\frac{\hbar}{i}\frac{\partial}{\partial q}.
\]

It is also convenient to use the deviatoric (shear) parts of the
affine spin,
\[
\mathbf{s}^{a}{}_{b}:={\bf \Sigma}^{a}{}_{b}-\frac{{\bf
p}}{n}\delta^{a}{}_{b},\qquad \mathbf{\hat{s}}^{A}{}_{B}:={\bf
\hat{\Sigma}}^{A}{}_{B}-\frac{{\bf p}}{n}\delta^{A}{}_{B};
\]
obviously, ${\bf p}={\bf \Sigma}^{a}{}_{a}={\bf
\hat{\Sigma}}^{A}{}_{A}$.

Due to the group-theoretical structure of the above objects as
generators, the classical splitting of $\mathbf{T}$ into
incompressible (shear-rotational) and dilatational parts remains
literally valid, namely, we have the following expressions:
\begin{eqnarray}
\mathbf{T}^{\rm aff-aff}_{\rm int}&=&\frac{1}{2A}{\bf C}_{{\rm
SL}(n)}(2)
+\frac{1}{2n(A+nB)}{\bf p}^{2},\nonumber\\
\mathbf{T}^{\rm met-aff}_{\rm int}&=&\frac{1}{2(I+A)}{\bf C}_{{\rm
SL}(n)}(2)+\frac{1}{2n(I+A+nB)}{\bf
p}^{2}+\frac{I}{2(I^{2}-A^{2})}\|{\bf S}\|^{2},\nonumber\\
\mathbf{T}^{\rm aff-met}_{\rm int}&=&\frac{1}{2(I+A)}{\bf C}_{{\rm
SL}(n)}(2)+\frac{1}{2n(I+A+nB)}{\bf
p}^{2}+\frac{I}{2(I^{2}-A^{2})}\|{\bf V}\|^{2},\nonumber
\end{eqnarray}
where, obviously, ${\bf C}_{{\rm SL}(n)}(k):={\bf s}^{a}{}_{b}{\bf
s}^{b}{}_{c}\cdots{\bf s}^{r}{}_{s}{\bf s}^{s}{}_{a}={\bf
\hat{s}}^{A}{}_{B}{\bf \hat{s}}^{B}{}_{C}\cdots{\bf
\hat{s}}^{R}{}_{S}{\bf \hat{s}}^{S}{}_{A}$, $k$ terms in these
expressions, and $\|{\bf S}\|^{2}=-(1/2){\bf S}^{a}{}_{b}{\bf
S}^{b}{}_{a}$, $\|{\bf V}\|^{2}=-(1/2){\bf V}^{A}{}_{B}{\bf
V}^{B}{}_{A}$.

As mentioned, the SL$(n,\mathbb{R})$-part of ${\bf T}$ has both
discrete and continuous spectrum and predicts the bounded
oscillatory solutions even if no extra potential on
SL$(n,\mathbb{R})$ is used (classically this is the geodetic model
with an open subset of bounded trajectories in the complete
solution). In particular, there is an open range of inertial
parameters $(A,B,C)\in \mathbb{R}^{3}$ for which the spectrum is
positive or at least bounded from below.

One can hope that on the basis of commutation relations for the
Lie algebra SL$(n,\mathbb{R})^{\prime}$ some information
concerning spectra and wave functions may be perhaps obtained
without the explicit solving of differential equations.

There are GL$(n,\mathbb{R})$-problems where the separation of the
isochoric SL$(n,\mathbb{R})$-terms is not necessary, sometimes it
is even undesirable. Then it is more convenient to use the
quantized version of (\cite{part1}4.45)\footnote{this kind of
references means that, e.g., in Part I \cite{part1} the expression
could be found in section 4 with label 45}, (\cite{part1}4.46),
(\cite{part1}4.47), i.e.,
\[
{\bf T}^{\rm aff-aff}_{\rm int}=\frac{1}{2A}{\bf
C}(2)-\frac{B}{2A(A+nB)}{\bf p}^{2},
\]
\[
{\bf T}^{\rm met-aff}_{\rm int}=\frac{1}{2\alpha}{\bf
C}(2)+\frac{1}{2\beta}{\bf p}^{2} +\frac{1}{2\mu}\|{\bf
S}\|^{2},
\] 
\[
{\bf T}^{\rm aff-met}_{\rm
int}=\frac{1}{2\alpha}{\bf C}(2)+\frac{1}{2\beta}{\bf
p}^{2}+\frac{1}{2\mu}\|{\bf V}\|^{2},
\]
where $\alpha:=I+A$, $\beta:=-(I+A)(I+A+nB)/B$,
$\mu:=(I^{2}-A^{2})/I$, and ${\bf C}(k)$ are operators of the full
GL$(n,\mathbb{R})$-Casimirs, i.e., we have 
\[
{\bf C}(k):={\bf
\Sigma}^{a}{}_{b}{\bf \Sigma}^{b}{}_{c}\cdots{\bf
\Sigma}^{r}{}_{s}{\bf \Sigma}^{s}{}_{a}={\bf
\hat{\Sigma}}^{A}{}_{B}{\bf \hat{\Sigma}}^{B}{}_{C}\cdots{\bf
\hat{\Sigma}}^{R}{}_{S}{\bf \hat{\Sigma}}^{S}{}_{A}; 
\]
the above
contracted products contain $k$ terms. In particular, 
\[
{\bf
C}(2):={\bf \Sigma}^{a}{}_{b}{\bf \Sigma}^{b}{}_{a}={\bf
\hat{\Sigma}}^{A}{}_{B}{\bf \hat{\Sigma}}^{B}{}_{A},\quad {\bf
C}(1):={\bf \Sigma}^{a}{}_{a}={\bf \hat{\Sigma}}^{A}{}_{A}. 
\]
In
particular, if the inertial constant $B$ vanishes, then the model
${\bf T}^{\rm aff-aff}_{\rm int}$ may be interpreted in terms of
one-dimensional multi-body problems in the sense of Calogero,
Moser, Sutherland \cite{AKS89,JJS-AKS93}, etc., quite
independently of our primary motivation, i.e., $n$-dimensional
affine systems.

As mentioned, on GL$(n,\mathbb{R})$, i.e., for compressible
objects with dilatations, some dilatation-stabilizing potential
$V(q)$ must be introduced if the system has to possess bound
states. For more general doubly isotropic potentials
$V(q^{1},\ldots,q^{n})$ depending only on deformation invariants,
there is no possibility of avoiding differential equations (with
the help of ladder procedures). Nevertheless, the problem is then
still remarkably simplified in comparison with the general case,
because the quantum dynamics of deformation invariants is
autonomous (in this respect the quantum problem is in a sense
simpler than the classical one). The procedure is based then on
the two-polar decomposition, which by the way is also very
convenient on the level of purely geodetic models. In certain
problems, e.g., spatially isotropic but materially anisotropic
ones, the polar decomposition is also convenient.

\section{Two-polar decomposition in quantum case}

Let us go back to classical expressions for
$\hat{\rho},\hat{\tau}\in{\rm SO}(n,\mathbb{R})^{\prime}$,
$\rho\in{\rm SO}(V,g)^{\prime}$, $\tau\in{\rm
SO}(U,\eta)^{\prime}$, $M:=-\hat{\rho}-\hat{\tau}$,
$N:=\hat{\rho}-\hat{\tau}$. On the quantum level the classical
quantities $\rho=S$, $\tau=-V$ become the operators of spin and
minus vorticity (\ref{7.1}) ${\bf S}$, $-{\bf V}$, i.e., Hermitian
generators of the unitary groups of spatial and material rotations
$\varphi\mapsto A\varphi$, $\varphi\mapsto \varphi B^{-1}$, where
$A\in {\rm SO}(V,g)$, $B\in {\rm SO}(U,\eta)$, acting
argument-wise on wave functions. Classical quantities
$\hat{\rho}$, $\hat{\tau}$ were co-moving representants of tensors
$\rho=S$, $\tau=-V$, i.e., their projections onto principal axes
of the Cauchy and Green deformation tensors. Their quantum
counterparts, i.e., operators ${\bf \hat{r}}$, ${\bf \hat{t}}$ are
also co-moving representants of $\mathbf{r}=\mathbf{S}$,
$\mathbf{t}=-\mathbf{V}$, i.e.,
\begin{equation}\label{x29}
\mathbf{\hat{r}}^{a}{}_{b}=L^{a}{}_{i}L^{j}{}_{b}{\bf
S}^{i}{}_{j},\qquad
\mathbf{\hat{t}}^{a}{}_{b}=-R^{A}{}_{b}R^{a}{}_{B}{\bf
V}^{B}{}_{A}.
\end{equation}
They are Hermitian generators of the argument-wise right-hand side
action (\cite{part1}6.63) of SO$(n,\mathbb{R})$ on the wave
functions. Just as in classical theory, it is convenient to
introduce operators
\begin{equation}\label{kw29}
{\bf
M}^{a}{}_{b}:=-\mathbf{\hat{r}}^{a}{}_{b}-\mathbf{\hat{t}}^{a}{}_{b},\qquad
{\bf
N}^{a}{}_{b}:=\mathbf{\hat{r}}^{a}{}_{b}-\mathbf{\hat{t}}^{a}{}_{b}.
\end{equation}
Commutation relations for operators ${\bf S}$, ${\bf V}$, ${\bf
\hat{r}}$, ${\bf \hat{t}}$, ${\bf M}$, ${\bf N}$ are directly
isomorphic with those for the generators of SO$(n,\mathbb{R})$ and
are expressed in a straightforward way in terms of
SO$(n,\mathbb{R})$-structure constants.

Now we are ready to write down explicitly our kinetic energy and
Hamiltonian operators in terms of the two-polar splitting. We
begin with the traditional integer spin models, and later on we
show how half-integer angular momentum of extended bodies may
appear in a natural way.

Quantum operators ${\bf S}^{i}{}_{j}$, $-{\bf V}^{A}{}_{B}$ have
the following form:
\begin{equation}\label{tr30}
{\bf
S}^{i}{}_{j}=\frac{\hbar}{i}{\mathbf{\Lambda}}^{i}{}_{j}(L),\qquad
-{\bf V}^{A}{}_{B}=\frac{\hbar}{i}{\mathbf{\Lambda}}^{A}{}_{B}(R),
\end{equation}
where, according to the formulas (\ref{tr15}), (\ref{O16}),
(\ref{!20}), ${\mathbf{\Lambda}}^{i}{}_{j}(L)$ and
${\mathbf{\Lambda}}^{A}{}_{B}(R)$ are real first-order
differential operators generating left regular translations on
SO$(n,\mathbb{R})$, or, more precisely, on the isometric factors
$L:\mathbb{R}^{n}\rightarrow V$, $R:\mathbb{R}^{n}\rightarrow U$
of the two-polar splitting, i.e.,
\begin{eqnarray}
F\left(W(\omega)L\right)=\left(\exp\left(\frac{1}{2}\omega^{j}{}_{i}
{\mathbf{\Lambda}}^{i}{}_{j}\right)F\right)(L),\nonumber\\
\label{kw30}\\
F\left(W(\omega)R\right)=\left(\exp\left(\frac{1}{2}\omega^{B}{}_{A}
{\mathbf{\Lambda}}^{A}{}_{B}\right)F\right)(R).\nonumber
\end{eqnarray}
In the formulas above, $F$ are functions on the manifolds of
isometries from $(\mathbb{R}^{n},\delta)$ to $(V,g)$ and from
$(\mathbb{R}^{n},\delta)$ to $(U,\eta)$. Analytically, in
Cartesian coordinates they are simply functions on
SO$(n,\mathbb{R})$. Matrices $[\omega^{a}{}_{b}]$,
$[\omega^{A}{}_{B}]$ are respectively $g$- and
$\eta$-antisymmetric:
$\omega^{a}{}_{b}=-g^{ac}g_{bd}\omega^{d}{}_{c}$,
$\omega^{A}{}_{B}=-\eta^{AC}\eta_{BD}\omega^{D}{}_{C}$. Their
independent components are canonical coordinates of the first kind
on SO$(V,g)$, SO$(U,\eta)$ (roughly, on SO$(n,\mathbb{R})$),
\begin{equation}\label{kw301}
W(\omega)=\exp\left(\frac{1}{2}\omega^{b}{}_{a}E^{a}{}_{b}\right),\qquad
W(\omega)=\exp\left(\frac{1}{2}\omega^{B}{}_{A}E^{A}{}_{B}\right),
\end{equation}
where $E^{a}{}_{b}\in {\rm SO}(V,g)^{\prime}$, $E^{A}{}_{B}\in
{\rm SO}(U,\eta)^{\prime}$ are basic elements corresponding to
some (arbitrary) choice of bases in $V$, $U$, i.e.,
$(E^{a}{}_{b})^{i}{}_{j}=\delta^{a}{}_{j}\delta^{i}{}_{b}-
g^{ai}g_{bj}$,
$(E^{A}{}_{B})^{C}{}_{D}=\delta^{A}{}_{D}\delta^{C}{}_{B}-
\eta^{AC}\eta_{BD}$.

One could reproach against our permanent changing between the
simplified analytical description based on $\mathbb{R}^{n}$,
GL$^{+}(n,\mathbb{R})$, SO$(n,\mathbb{R})$ and the careful
geometric distinguishing between the material and physical spaces
$U$, $V$ and the manifolds LI$(U,V)$,
O$^{+}(\mathbb{R}^{n},\delta;V,g)$,
O$^{+}(\mathbb{R}^{n},\delta;U,\eta)$; the latter two denoting the
manifolds of orientation-preserving isometries between indicated
Euclidean spaces (equivalently, manifolds of positively oriented
orthonormal frames F$^{+}(V,g)$, F$^{+}(U,\eta)$). However, this
"monkey" way of changing branches has some advantages, provided
that done carefully. There are relationships easily representable
for computational purposes in matrix terms, however, in certain
fundamental formulas this may be misleading and risky.

And now, at some final stage of our discussion there appear some
expressions where the calculus on $\mathbb{R}^{n}$ as such (not on
$\mathbb{R}^{n}$ base-identified with $U$, $V$) becomes not only
temporarily admissible but just mathematically proper one. Namely,
it is just the matrix group SO$(n,\mathbb{R})$ that acts on the
right on the objects $L\in {\rm O}^{+}(\mathbb{R}^{n},\delta;V,g)$
and $R\in {\rm O}^{+}(\mathbb{R}^{n},\delta;U,\eta)$. As said
above, on the classical level the corresponding Hamiltonian
generators, i.e., momentum mappings, are given by
$[\hat{\rho}^{a}{}_{b}]$, $[\hat{\tau}^{a}{}_{b}]$. In quantized
theory the same role is played by the formally self-adjoint
differential operators ${\bf \hat{r}}^{a}{}_{b}$, ${\bf
\hat{t}}^{a}{}_{b}$,
\begin{eqnarray}\label{7.2}
F\left(LW(\omega)\right)&=&\left(\exp\left(\frac{1}{2}
\omega^{b}{}_{a}{\mathbf{\Upsilon}}^{a}{}_{b}\right)F\right)(L)=
\left(\exp\left(\frac{i}{2\hbar}\omega^{b}{}_{a}{\bf
\hat{r}}^{a}{}_{b}\right)F\right)(L),\\
F\left(RW(\omega)\right)&=&\left(\exp\left(\frac{1}{2}
\omega^{b}{}_{a}{\mathbf{\Upsilon}}^{a}{}_{b}\right)F\right)(R)=
\left(\exp\left(\frac{i}{2\hbar}\omega^{b}{}_{a}{\bf
\hat{t}}^{a}{}_{b}\right)F\right)(R).\nonumber
\end{eqnarray}

Here the skew-symmetry of $[\omega^{a}{}_{b}]$ is meant in the
literal Kronecker-delta sense; nothing like $g$ and $\eta$ is
implicitly assumed:
$\omega^{a}{}_{b}=-\omega_{b}{}^{a}=-\delta^{ac}\delta_{bd}\omega^{d}{}_{c}$.
Just $\mathbb{R}^{n}$ as such with its numerical metric is used
here. In the physical three-dimensional case one uses the duality
between skew-symmetric tensors and axial vectors, thus, on the
quantum operator level we use the quantities ${\bf \hat{r}}_{a}$,
${\bf \hat{t}}_{a}$, ${\mathbf{\Upsilon}}(L)_{a}$,
${\mathbf{\Upsilon}}(R)_{a}$, where
\[
{\bf \hat{r}}^{a}{}_{b}=\epsilon^{a}{}_{b}{}^{c}{\bf
\hat{r}}_{c},\qquad {\bf
\hat{t}}^{a}{}_{b}=\epsilon^{a}{}_{b}{}^{c}{\bf
\hat{t}}_{c},\qquad
{\mathbf{\Upsilon}}^{a}{}_{b}=\epsilon^{a}{}_{b}{}^{c}{\mathbf{\Upsilon}}_{c},
\]
\[
{\bf \hat{r}}_{a}=\frac{1}{2} \epsilon_{ab}{}^{c}{\bf
\hat{r}}^{b}{}_{c},\qquad {\bf \hat{t}}_{a}=\frac{1}{2}
\epsilon_{ab}{}^{c}{\bf \hat{t}}^{b}{}_{c},\qquad
{\mathbf{\Upsilon}}_{a}=\frac{1}{2}
\epsilon_{ab}{}^{c}{\mathbf{\Upsilon}}^{b}{}_{c}.
\]
Obviously, the expressions ${\mathbf{\Upsilon}}^{a}{}_{b}$,
${\mathbf{\Upsilon}}_{a}$ are meant in two versions, as acting on
the $L,R$-variables, thus, puristically we should have used the
symbols ${\mathbf{\Upsilon}}^{a}{}_{b}(L)$,
${\mathbf{\Upsilon}}_{a}(L)$, ${\mathbf{\Upsilon}}^{a}{}_{b}(R)$,
${\mathbf{\Upsilon}}_{a}(R)$, however, when non-necessary, we
prefer to avoid the crowd of symbols. Commutation relations are in
both cases: $[{\mathbf{\Upsilon}}_{a},{\mathbf{\Upsilon}}_{b}]=
\epsilon_{ab}{}^{c}{\mathbf{\Upsilon}}_{c}$, i.e., in terms of
quantum Poisson brackets:
\[
\frac{1}{i\hbar}[{\bf \hat{r}}_{a},{\bf
\hat{r}}_{b}]=-\epsilon_{ab}{}^{c}{\bf \hat{r}}_{c},\qquad
\frac{1}{i\hbar}[{\bf \hat{t}}_{a},{\bf
\hat{t}}_{b}]=-\epsilon_{ab}{}^{c}{\bf \hat{t}}_{c}.
\]
It is clear that $[{\bf \hat{r}}_{a},{\bf \hat{t}}_{b}]=0$,
$[{\mathbf{\Upsilon}}_{a}(L),{\mathbf{\Upsilon}}_{b}(R)]=0$.
Obviously, the raising and lowering of indices is meant here in
the trivial Krone\-cker-delta sense, so it is written only for
cosmetic reasons, e.g.,
$\epsilon^{a}{}_{b}{}^{c}=\delta^{ak}\delta^{cl}\epsilon_{kbl}$,
etc. What concerns the $V$- and $U$-space objects like ${\bf
S}^{i}{}_{j}={\bf r}^{i}{}_{j}$, ${\bf V}^{A}{}_{B}=-{\bf
t}^{A}{}_{B}$, analogous expressions are true when one uses
orthonormal coordinates, i.e., when $g_{ij}=_{\ast}\delta_{ij}$,
$\eta_{AB}=_{\ast}\delta_{AB}$. When more general rectilinear
coordinates are used, the formulas become more complicated because
various expressions involving $\det[g_{ij}]$, $\det[\eta_{AB}]$
appear; there is, however, no practical need to use this
representation.

In orthonormal coordinates in $V$ and $U$ spaces we have again the
following expressions in terms of axial vectors: ${\bf
r}^{i}{}_{j}={\bf S}^{i}{}_{j}=\epsilon^{i}{}_{j}{}^{k}{\bf
r}_{k}=\epsilon^{i}{}_{j}{}^{k}{\bf S}_{k}$, ${\bf
t}^{A}{}_{B}=-{\bf V}^{A}{}_{B}=\epsilon^{A}{}_{B}{}^{C}{\bf
t}_{C}=-\epsilon^{A}{}_{B}{}^{C}{\bf V}_{C}$. These quantities are
expressed through differential operators
${\mathbf{\Lambda}}^{i}{}_{j}(L)$,
${\mathbf{\Lambda}}^{A}{}_{B}(R)$, cf. (\ref{tr30}), for which the
same dual representation will be used, i.e.,
\[
{\mathbf{\Lambda}}^{i}{}_{j}(L)=\epsilon^{i}{}_{j}{}^{k}{\mathbf{\Lambda}}_{k}(L),\qquad
{\mathbf{\Lambda}}_{k}(L)=\frac{1}{2}
\epsilon_{ij}{}^{k}{\mathbf{\Lambda}}^{j}{}_{k}(L),
\]
\[
{\mathbf{\Lambda}}^{A}{}_{B}(R)=\epsilon^{A}{}_{B}{}^{C}{\mathbf{\Lambda}}_{C}(R),\qquad
{\mathbf{\Lambda}}_{A}(R)=\frac{1}{2}
\epsilon_{AB}{}^{C}{\mathbf{\Lambda}}^{B}{}_{C}(R).
\]
When using the convention of "small" and "capital" indices, one
can omit the $L$- and $R$-labels at ${\mathbf{\Lambda}}$-symbols.
Obviously, we have ${\bf S}_{i}={\bf
r}_{i}=(\hbar/i){\mathbf{\Lambda}}_{i}$, ${\bf V}_{A}=-{\bf
t}_{A}=-(\hbar/i){\mathbf{\Lambda}}_{A}$. One should be careful
with some subtle sign problems in commutation relations,
\[
[{\mathbf{\Lambda}}_{i},{\mathbf{\Lambda}}_{j}]=-\epsilon_{ij}{}^{k}{\mathbf{\Lambda}}_{k},\qquad
[{\mathbf{\Lambda}}_{A},{\mathbf{\Lambda}}_{B}]=-\epsilon_{AB}{}^{C}{\mathbf{\Lambda}}_{C},\qquad
[{\mathbf{\Lambda}}_{i},{\mathbf{\Lambda}}_{A}]=0,
\]
therefore,
\[
\frac{1}{i\hbar}[{\bf S}_{i},{\bf S}_{j}]=\epsilon_{ij}{}^{k}{\bf
S}_{k},\qquad \frac{1}{i\hbar}[{\bf V}_{A},{\bf
V}_{B}]=-\epsilon_{AB}{}^{C}{\bf V}_{C},\qquad [{\bf S}_{i},{\bf
V}_{A}]=0.
\]
Let us also notice that
\[
[{\mathbf{\Lambda}}_{i},{\mathbf{\Upsilon}}_{a}(L)]=0,\qquad
[{\mathbf{\Lambda}}_{i},{\mathbf{\Lambda}}_{A}]=0,\qquad
[{\mathbf{\Lambda}}_{i},{\mathbf{\Upsilon}}_{a}(R)]=0,
\]
\[
[{\mathbf{\Lambda}}_{A},{\mathbf{\Upsilon}}_{a}(L)]=0,\qquad
[{\mathbf{\Upsilon}}_{a}(L),{\mathbf{\Upsilon}}_{a}(R)]=0,\qquad
[{\mathbf{\Lambda}}_{A},{\mathbf{\Upsilon}}_{a}(R)]=0.
\]

\section{Rotation-vector-space language}

Obviously, "coordinates" $\omega^{a}{}_{b}$ on SO$(n,\mathbb{R})$
are redundant, unless we restrict ourselves to
$\omega_{ab}=\delta_{ac}\omega^{c}{}_{b}$, $a<b$ (or conversely).
If $n=3$, one uses so-called "rotation vector" $k^{a}$, where
$\omega^{a}{}_{b}=-\epsilon^{a}{}_{bc}k^{c}$,
$k^{a}=-(1/2)\epsilon^{a}{}_{b}{}^{c}\omega^{b}{}_{c}$. It is
convenient to use the "magnitude"
$k=\sqrt{(k^{1})^{2}+(k^{2})^{2}+(k^{3})^{2}}$. In this
parameterization, SO$(3,\mathbb{R})$ is covered by the ball $k\leq
\pi$ with the proviso that antipodal points on the sphere $k=\pi$
describe the same half-rotation, i.e., rotation by $\pi$ about a
given axis. For $k<\pi$ the representation is unique. The
magnitude $k$ equals the angle of rotation, whereas the versor
$\overline{n}:=\overline{k}/k$ represents the oriented rotation
axis in the right screw sense (for $k=\pi$ it does not matter
right or left ones; they coincide). In certain expressions it is
convenient to use the spherical coordinates $k$, $\vartheta$,
$\varphi$ in the $\overline{k}$-space, thus,
$k^{1}=k\sin\vartheta\cos\varphi$,
$k^{2}=k\sin\vartheta\sin\varphi$, $k^{3}=k\cos\vartheta$. For the
completeness, let us quote some important three-dimensional
formulas.

The "basic" matrices $E^{a}{}_{b}\in{\rm
SO}(3,\mathbb{R})^{\prime}$ are represented dually by the actually
basic system of $E_{a}$, where
$E^{a}{}_{b}=\epsilon^{a}{}_{b}{}^{c}E_{c}$,
$E_{a}=(1/2)\epsilon_{ab}{}^{c}E^{b}{}_{c}$,
$(E_{a})^{b}{}_{c}=-\epsilon_{a}{}^{b}{}_{c}$. The structure
constants are then given simply by "epsilons":
$[E_{a},E_{b}]=\epsilon_{ab}{}^{c}E_{c}$. For any rotation vector
$\overline{k}\in\mathbb{R}^{3}$ the corresponding matrices
$W\left(\overline{k}\right)\in{\rm SO}(3,\mathbb{R})$ act on
vectors $\overline{u}\in\mathbb{R}^{3}$ as follows:
\[
W\left(\overline{k}\right)\cdot\overline{u}=\cos
k\overline{u}+\frac{(1-\cos
k)}{k^{2}}\left(\overline{k}\cdot\overline{u}\right)\overline{k}+\frac{\sin
k}{k}\overline{k}\times\overline{u};
\]
obviously, the scalar and vector product are meant in the standard
$\mathbb{R}^{3}$-sense. The components of $\overline{k}$ are
canonical coordinates of the first kind on SO$(3,\mathbb{R})$,
\[
W\left(\overline{k}\right)= \exp\left(k^{a}E_{a}\right)=
\sum^{\infty}_{m=0}\frac{1}{m!}\left(k^{a}E_{a}\right)^{m}.
\]
One can show that
\[
W\left(\overline{k}\right)\cdot\overline{u}=\overline{u}+\overline{k}\times\overline{u}
+\frac{1}{2}\overline{k}\times\left(\overline{k}\times\overline{u}\right)+\cdots+
\frac{1}{n!}\overline{k}\times\left(\overline{k}\times\left(\overline{k}\times\cdots
\left(\overline{k}\times\overline{u}\right)\cdots\right)\right)+\cdots
\]
This infinite series is an alternative representation of the
exponential formula. The term with multiplicator $1/n!$ contains
the $n$-fold vector multiplication of $\overline{u}$ by
$\overline{k}$. Explicitly the matrix of
$W\left(\overline{k}\right)$ is given by
\[
W\left(\overline{k}\right)^{a}{}_{b}=\cos k\
\delta^{a}{}_{b}+(1-\cos k)\frac{k^{a}k_{b}}{k^{2}}+\sin k\
\epsilon^{a}{}_{bc}\frac{k^{c}}{k};
\]
obviously, the raising and lowering of indices is meant here in
the trivial (purely cosmetic) delta-sense.

One can show that the generators of right regular translations on
SO$(3,\mathbb{R})$ are given by the following expression:
\[
{\mathbf{\Upsilon}}_{a}=\frac{k}{2}{\rm
ctg}\frac{k}{2}\frac{\partial}{\partial
k^{a}}+\left(1-\frac{k}{2}{\rm
ctg}\frac{k}{2}\right)\frac{k_{a}k^{b}}{k^{2}}\frac{\partial}{\partial
k^{b}}-\frac{1}{2}\epsilon_{ab}{}^{c}k^{b}\frac{\partial}{\partial
k^{c}}.
\]
This is a common formula for ${\mathbf{\Upsilon}}_{a}(L)$,
${\mathbf{\Upsilon}}_{a}(R)$, and now for simplicity we again use
the analytical matrix representation, when $U$ and $V$ are
identified with $\mathbb{R}^{3}$ and the $L,R$-terms of the
two-polar decomposition are identified with elements of
SO$(3,\mathbb{R})$. To specify this formula to
${\mathbf{\Upsilon}}_{a}(L)$, ${\mathbf{\Upsilon}}_{a}(R)$ one
must replace the general symbol of the rotation vector
$\overline{k}$ on SO$(3,\mathbb{R})$ by the rotation vectors
$\overline{l}$, $\overline{r}$ parameterizing the $L,R$-terms:
$L\left(\overline{l}\right)=\exp\left(l^{a}E_{a}\right)$,
$R\left(\overline{r}\right)=\exp\left(r^{a}E_{a}\right)$.
Generators of the left regular translations on SO$(3,\mathbb{R})$
are as follows:
\[
{\mathbf{\Lambda}}_{a}=\frac{k}{2}{\rm
ctg}\frac{k}{2}\frac{\partial}{\partial
k^{a}}+\left(1-\frac{k}{2}{\rm
ctg}\frac{k}{2}\right)\frac{k_{a}k^{b}}{k^{2}}\frac{\partial}{\partial
k^{b}}+\frac{1}{2}\epsilon_{ab}{}^{c}k^{b}\frac{\partial}{\partial
k^{c}}.
\]
And this again specifies to ${\mathbf{\Lambda}}_{a}(L)$,
${\mathbf{\Lambda}}_{a}(R)$ when instead of $\overline{k}$ we
substitute respectively $\overline{l}$, $\overline{r}$, i.e.,
rotation vectors parameterizing the manifolds of $L,R$-factors in
the two-polar decomposition.

Let us observe that
${\mathbf{\Lambda}}_{a}-{\mathbf{\Upsilon}}_{a}={\bf D}_{a}=
\epsilon_{ab}{}^{c}k^{b}(\partial/\partial k^{c})$, and these
differential operators generate the group of inner automorphisms
of SO$(3,\mathbb{R})$: $W\left(\overline{k}\right)\mapsto
UW\left(\overline{k}\right)U^{-1}=W\left(U\overline{k}\right)$,
where $U$ runs over SO$(3,\mathbb{R})$. Roughly speaking, these
transformations result in rotations of the rotation vectors. And,
just as previously, substituting here $\overline{l}$ and
$\overline{r}$ in place of $\overline{k}$ we obtain the
corresponding transformations of the manifolds of
$L\left(\overline{l}\right)$- and
$R\left(\overline{r}\right)$-terms of the two-polar
decompositions. One can show that the generators of the left and
right regular translations on SO$(3,\mathbb{R})$ may be expressed
in terms of operators $\partial/\partial k$ and ${\bf D}_{a}$
acting, respectively, along the radius and tangently to spheres in
the representative spaces $\mathbb{R}^{3}$ of the rotation vector
$\overline{k}$, i.e.,
\[
{\mathbf{\Lambda}}_{a}=\frac{k_{a}}{k}\frac{\partial}{\partial
k}-\frac{1}{2}{\rm ctg}\frac{k}{2}\epsilon_{ab}{}^{c}k^{b}{\bf
D}_{c}+\frac{1}{2}{\bf D}_{a},\qquad
{\mathbf{\Upsilon}}_{a}=\frac{k_{a}}{k}\frac{\partial}{\partial
k}-\frac{1}{2}{\rm ctg}\frac{k}{2}\epsilon_{ab}{}^{c}k^{b}{\bf
D}_{c}-\frac{1}{2}{\bf D}_{a}.
\]
Obviously, $[{\bf D}_{a},{\bf D}_{b}]=-\epsilon_{ab}{}^{c}{\bf
D}_{c}$.

In many formulas we need orthogonal invariants like $\|{\bf
S}\|^{2}$, $\|{\bf V}\|^{2}$. They are based on the Casimir
invariants $C_{{\rm SO}(n,\mathbb{R})}(2)$ built of generators
${\mathbf{\Lambda}}_{a}$, ${\mathbf{\Upsilon}}_{a}$ of the left
and right regular translations on SO$(n,\mathbb{R})$. If $n=3$,
these Casimirs have the following form:
\begin{equation}\label{7.3}
{\mathbf{\Lambda}}^{2}={\mathbf{\Upsilon}}^{2}={\mathbf{\Lambda}}_{1}^{2}+{\mathbf{\Lambda}}_{2}^{2}+{\mathbf{\Lambda}}_{3}^{2}={\mathbf{\Upsilon}}_{1}^{2}+{\mathbf{\Upsilon}}_{2}^{2}+{\mathbf{\Upsilon}}_{3}^{2},
\end{equation}
and one can show that analytically
\[
{\bf C}_{{\rm
SO}(3,\mathbb{R})}(2)={\mathbf{\Lambda}}^{2}={\mathbf{\Upsilon}}^{2}=\left(\frac{\partial^{2}}{\partial
k^{2}}+{\rm ctg}\frac{k}{2}\frac{\partial}{\partial
k}\right)+\frac{1}{4\sin^{2}\frac{k}{2}}{\bf D}^{2},
\]
where ${\bf D}^{2}={\bf D}_{1}^{2}+{\bf D}_{2}^{2}+{\bf
D}_{3}^{2}$. Obviously, $\|{\bf S}\|^{2}=-\hbar^{2}{\bf C}_{{\rm
SO}(3,\mathbb{R})}\left(L\left(\overline{l}\right)\right)$ and
$\|{\bf V}\|^{2}=-\hbar^{2}{\bf C}_{{\rm
SO}(3,\mathbb{R})}\left(R\left(\overline{r}\right)\right)$, where
the last two terms multiplied by $-\hbar^{2}$ are obtained from
the previous ${\bf C}_{{\rm SO}(3,\mathbb{R})}$ by substituting
the $\overline{l}$- and $\overline{r}$-variables in place of
$\overline{k}$.

\noindent {\bf Remark:} Obviously, the equality (\ref{7.3}) of
${\mathbf{\Lambda}}^{2}$ and ${\mathbf{\Upsilon}}^{2}$ holds only
when ${\mathbf{\Lambda}}_{a}$ and ${\mathbf{\Upsilon}}_{a}$
involve the same kind of independent variables, e.g.,
$\overline{k}$ on the abstract SO$(3,\mathbb{R})$ as generators of
the left or right regular translations, $\overline{l}$ when both
operating on the left two-polar factor $L(\overline{l})$, or
$\overline{r}$ when both acting on the right two-polar factor. But
of course $\|{\bf S}\|^{2}$ and $\|{\bf V}\|^{2}$ are different
for any dimension $n$, although, of course, $\|{\bf
S}\|^{2}=\|{\bf \hat{r}}\|^{2}$ and $\|{\bf V}\|^{2}=\|{\bf
\hat{t}}\|^{2}$ always hold just on the basis of equations
(\ref{x29}).

\section{Expansion of wave functions}

When we use the two-polar decomposition $\varphi=LDR^{-1}$, then,
according to the Peter-Weyl theorem, the wave functions on
GL$^{+}(n,\mathbb{R})$ may be expanded in $L,R$-variables with
respect to matrix elements of irreducible representations of the
compact group SO$(n,\mathbb{R})$. Obviously, the expansion
coefficients depend on deformation invariants, i.e., on the
diagonal factor $D$ (equivalently, on the variables $Q^{a}$ or
$q^{a}=\ln Q^{a}$). In general, we have that
\begin{equation}\label{7.4}
\Psi(\varphi)=\Psi(L,D,R)=\sum_{\alpha,\beta\in\Omega}
\sum_{m,n=1}^{N(\alpha)}\sum_{k,l=1}^{N(\beta)}
\mathcal{D}^{\alpha}_{mn}(L)f^{\alpha\beta}_{{}^{nk}_{ml}}
(D)\mathcal{D}^{\beta}_{kl}(R^{-1}),
\end{equation}
where the meaning of symbols is as follows:
\begin{itemize}
\item $\Omega$ is the set of equivalence classes of unitary
irreducible representations of SO$(n,\mathbb{R})$.

\item $N(\alpha)$ is the dimension of the $\alpha$-th
representation class. It is finite because SO$(n,\mathbb{R})$ is
compact.

\item $\mathcal{D}^{\alpha}$ is the $\alpha$-th representation
matrix. For many classical groups $\mathcal{D}^{\alpha}$ are
explicitly known (at least in terms of some well-investigated
special functions).
\end{itemize}
Analytically $\mathcal{D}^{\alpha}(L)$,
$\mathcal{D}^{\beta}(R^{-1})$ are matrices depending on the group
coordinates $\omega_{L}{}^{a}{}_{b}$, $\omega_{R}{}^{a}{}_{b}$ of
$L$, $R$, e.g., rotation vectors $\overline{l}$, $\overline{r}$ if
$n=3$. The argument $D$ of $f$ is the system of $q$-variables
$q^{1},\ldots,q^{n}$. According to the mentioned multi-valuedness
of the two-polar decomposition, the reduced amplitudes
$f^{\alpha\beta}(q^{1},\ldots,q^{n})$ must obey some conditions,
because $\Psi$ must not distinguish triplets $(L,D,R)$
corresponding to the same configuration $\varphi=LDR^{-1}$.

Therefore, on the submanifold $M^{(n)}\subset{\rm
SO}(n,\mathbb{R})\times\mathbb{R}^{n}\times{\rm SO}(n,\mathbb{R})$
with non-degenerate systems of $(q^{1},\ldots,q^{n})$ (no
coincidences) we must have that
\[
f^{\alpha\beta}_{{}^{nk}_{ml}}(q^{\pi_{W}(1},\ldots,q^{n)})=
\sum_{r=1}^{N(\alpha)}\sum_{s=1}^{N(\beta)}
\mathcal{D}^{\alpha}_{nr}(W^{-1})f^{\alpha\beta}_{{}^{rs}_{ml}}
(q^{1},\ldots,q^{n})\mathcal{D}^{\beta}_{sk}(W)
\]
for any $W\in K^{+}$. The same holds on the subsets
$M^{(k;p_{1},\ldots,p_{k})}\subset{\rm
SO}(n,\mathbb{R})\times\mathbb{R}^{n}\times{\rm SO}(n,\mathbb{R})$
with degenerate systems $(q^{1},\ldots,q^{n})$ (coincidences of
some $q$'s). The difference is that in degenerate cases $W$ runs
over the continuous subgroups of SO$(n,\mathbb{R})$ generated by
$K^{+}$ and the subgroups $H^{(k;p_{1},\ldots,p_{k})}$ described
above. The special case of the total degeneracy is extreme and,
because of this, very simple one. Indeed, then in the two-polar
decomposition it is only $LR^{-1}$ that is meaningful whereas $L$,
$R$ separately are not well-defined. Therefore, if $D=cI_{n}$,
i.e., $q^{1}=\cdots=q^{n}=q$, then the reduced amplitude obeys
very severe restrictions, i.e., $f^{\alpha\beta}(cI_{n})=0$ if
$\alpha\neq\beta$, and
$f^{\alpha\alpha}_{{}^{rs}_{ml}}(cI_{n})=g_{ml}\delta_{rs}$. The
non-uniqueness is extreme here, namely, for any $Z\in{\rm
SO}(n,\mathbb{R})$ the triplets $(L,cI_{n},R)$, $(LZ,cI_{n},RZ)$
represent the same classical configuration, thus, the wave
functions do not distinguish them.

It is seen that if $q^{1},\ldots,q^{n}$ are interpreted as
coordinates of some fictitious material points on the real axis
$\mathbb{R}$, one is dealing with a very peculiar system of
identical para-statistical particles.

It is clear that in geodetic models or in models with doubly
isotropic potentials (ones depending only on deformation
invariants; dilatation-stabilizing potentials $V(q)$ provide the
simplest example), $m$ and $l$ in the Peter-Weyl expansion
(\ref{7.4}) are "good" quantum numbers. In other words, the spin
and vorticity operators ${\bf S}^{i}{}_{j}$, ${\bf V}^{A}{}_{B}$
do commute with the Hamilton operator ${\bf H}$. The same concerns
representation labels $\alpha,\beta\in\Omega$, i.e., finally, the
systems of eigenvalues for the Casimir operators of the groups
SO$(V,g)$, SO$(U,\eta)$ acting argument-wise on wave functions.
Let us remind that these Casimirs are given by
\begin{equation}\label{7.5}
{\bf C}_{{\rm SO}(V,g)}(p)\simeq {\bf S}^{i}{}_{k}{\bf
S}^{k}{}_{m}\cdots{\bf S}^{r}{}_{z}{\bf S}^{z}{}_{i}, \quad {\bf
C}_{{\rm SO}(U,\eta)}(p)\simeq {\bf V}^{A}{}_{K}{\bf
V}^{K}{}_{M}\cdots{\bf V}^{R}{}_{Z}{\bf V}^{Z}{}_{A},
\end{equation}
$p$ operator multipliers in every expression; $p\leq n$ and even.

In such situation it is convenient to keep $\alpha$, $\beta$, $m$,
$l$ fixed and use the following reduced amplitudes (with the same
as previously provisos concerning the one-valuedness of $\Psi$ as
a function of $\varphi$):
\begin{equation}\label{7.6}
\Psi(\varphi)=\Psi^{\alpha\beta}_{ml}(L,D,R)=
\sum_{n=1}^{N(\alpha)}\sum_{k=1}^{N(\beta)}
\mathcal{D}^{\alpha}_{mn}(L)f^{\alpha\beta}_{nk}(D)\mathcal{D}^{\beta}_{kl}(R^{-1}).
\end{equation}

In the physical case $n=3$, we have obviously the standard form of
SO$(3,\mathbb{R})$-Casimirs:
\[
{\bf C}_{{\rm SO}(V,g)}(2)={\bf S}^{2}_{1}+{\bf S}^{2}_{2}+{\bf
S}^{2}_{3}={\bf \hat{r}}^{2}_{1}+{\bf \hat{r}}^{2}_{2}+{\bf
\hat{r}}^{2}_{3}={\bf C}_{{\rm SO}(3,\mathbb{R})}(2),
\]
\[
{\bf C}_{{\rm SO}(U,\eta)}(2)={\bf V}^{2}_{1}+{\bf V}^{2}_{2}+{\bf
V}^{2}_{3}={\bf \hat{t}}^{2}_{1}+{\bf \hat{t}}^{2}_{2}+{\bf
\hat{t}}^{2}_{3}={\bf C}_{{\rm SO}(3,\mathbb{R})}(2).
\]
Our expansions for wave functions are then described in terms of
well-known expressions found by Wigner, and, of course, the family
of rotational Casimirs begins and terminates on $p=2$.

Obviously, for $n=3$, $\Omega$ is the set of non-negative integer,
$\alpha$, $\beta$ are traditionally denoted by symbols like
$s,j=0,1,2,\ldots$, etc., $N(s)=2s+1$, $N(j)=2j+1$, and the
indices $(m,n)$, $(k,l)$ are considered as jumping by $1$,
respectively, from $-s$ to $s$ and from $-j$ to $j$; here the
tradition is too strong to respect the formal logical conventions.
Thus, the expansion (\ref{7.4}) is written according to the
mentioned conventions:
\begin{equation}\label{7.7}
\Psi(\varphi)=\Psi(L,D,R)=\sum_{s,j=0}^{\infty}
\sum_{m,n=-s}^{s}\sum_{k,l=-j}^{j}
\mathcal{D}^{s}_{mn}(L)f^{sj}_{{}^{nk}_{ml}}(D)\mathcal{D}^{j}_{kl}(R^{-1}).
\end{equation}
Similarly, the reduced amplitudes (\ref{7.6}) are written as:
\begin{equation}\label{7.8}
\Psi(\varphi)=\Psi^{sj}_{ml}(L,D,R)=
\sum_{n=-s}^{s}\sum_{k=-j}^{j}
\mathcal{D}^{s}_{mn}(L)f^{sj}_{nk}(D)\mathcal{D}^{j}_{kl}(R^{-1}).
\end{equation}
Here $\mathcal{D}^{s}$ are celebrated Wigner matrices of
$(2s+1)$-dimensional irreducible representations of the
three-dimensional rotation group. They are well-known special
functions of mathematical physics and may be assumed to be
something in principle standard and well-know.

Obviously, the amplitudes $\Psi^{sj}_{ml}$ are eigenfunctions of
rotational Casimir invariants, i.e., essentially angular momentum
and vorticity:
\[
\|{\bf S}\|^{2}\Psi^{sj}_{ml}=\|{\bf
\hat{r}}\|^{2}\Psi^{sj}_{ml}=\hbar^{2}s(s+1)\Psi^{sj}_{ml}, \
\|{\bf V}\|^{2}\Psi^{sj}_{ml}=\|{\bf
\hat{t}}\|^{2}\Psi^{sj}_{ml}=\hbar^{2}j(j+1)\Psi^{sj}_{ml},
\]
where, let us remind, in three dimensions we have $\|{\bf
S}\|^{2}={\bf S}^{2}_{1}+{\bf S}^{2}_{2}+{\bf S}^{2}_{3}$, $\|{\bf
V}\|^{2}={\bf V}^{2}_{1}+{\bf V}^{2}_{2}+{\bf V}^{2}_{3}$, and
similarly for ${\bf \hat{r}}$, ${\bf \hat{t}}$. According to
tradition, one uses such a basis that $\Psi^{sj}_{ml}$ are also
eigenfunctions of the third components of rotational generators,
\[
{\bf S}_{3}\Psi^{sj}_{ml}=\hbar m\Psi^{sj}_{ml},\qquad {\bf
V}_{3}\Psi^{sj}_{ml}=\hbar l\Psi^{sj}_{ml}.
\]
And, obviously, when the values $n$, $k$ in the superposition
(\ref{7.8}) are kept fixed and we retain only the corresponding
single term, for the resulting $\Psi$ we have
\[
{\bf \hat{r}}_{3}\Psi^{sj}_{{}^{ml}_{nk}}=\hbar
n\Psi^{sj}_{{}^{ml}_{nk}},\qquad {\bf
\hat{t}}_{3}\Psi^{sj}_{{}^{ml}_{nk}}=\hbar
k\Psi^{sj}_{{}^{ml}_{nk}}.
\]

\section{Representation matrices}

In this way one is dealing with quantum states of well-definite
values of magnitudes and third components of the angular momentum
and vorticity. For the general $n$, the amplitudes
$\Psi^{\alpha\beta}_{ml}$ have, of course, the well-definite
values $(\hbar/i)^{p}C(\alpha,p)$, $(\hbar/i)^{p}C(\beta,p)$ of
the Casimirs (\ref{7.5}). And now it will be convenient to return
for a while (at least in a formal way) to the general case of
dimension $n$.

Let us again use the exponential formulas (\ref{kw301}) for the
elements of $W(\omega)\in{\rm SO}(V,g)$, $W(\omega)\in{\rm
SO}(U,\eta)$, and just their simply numerical counterparts in
SO$(n,\mathbb{R})$,
\[
W(\omega)=\exp\left(\frac{1}{2}\omega^{a}{}_{b}E^{b}{}_{a}\right),
\]
where the basic matrices $E^{b}{}_{a}$ are simply given by
$(E^{b}{}_{a})^{c}{}_{d}=\delta^{b}{}_{d}\delta^{c}{}_{a}-
\delta^{bc}\delta_{ad}$ (just simply the numerical counterpart of
(\ref{kw301}) showing that one works just in $\mathbb{R}^{n}$ and
SO$(n,\mathbb{R})^{\prime}$ not in $V$, $U$, SO$(V,g)$,
SO$(U,\eta)$ basis-identified with the previous ones). And from
now on let us again decide to work in purely analytical matrix
form using orthonormal coordinates in $V$, $U$ and identifying
them with $\mathbb{R}^{n}$. Representation matrices
$\mathcal{D}^{\alpha}$ are given by the following expresion:
\[
\mathcal{D}^{\alpha}(\omega)=\exp\left(\frac{1}{2}\omega^{a}{}_{b}
M^{\alpha}{}^{b}{}_{a}\right),
\]
where $N(\alpha)\times N(\alpha)$ anti-hermitian matrices
$M^{\alpha}{}^{b}{}_{a}$ form irreducible representations of the
Lie algebra SO$(n,\mathbb{R})^{\prime}$, thus, their commutation
rules are identical with those for $E^{b}{}_{a}$.

\noindent {\bf Remark:} For any $\alpha\in \Omega$ and for any
pair of indices $b$, $a$, $M^{\alpha}{}^{b}{}_{a}$ are just
matrices not $(b,a)$-matrix elements of some $M^{\alpha}$; let us
notice in this connection that $a,b=\overline{1,n}$, whereas any
$M^{\alpha}{}^{b}{}_{a}$ is an $N(\alpha)\times N(\alpha)$-matrix.
Obviously, when dealing with matrices $\mathcal{D}^{\alpha}(L)$,
$\mathcal{D}^{\beta}(R)$, we must specialize the redundant
"coordinates" $\omega^{a}{}_{b}$ to the ones parameterizing
respectively the $L$- and $R$-terms of the two-polar splitting,
writing, e.g.,
\[
\mathcal{D}^{\alpha}\left(L(l)\right)=\exp\left(\frac{1}{2}l^{a}{}_{b}
M^{\alpha}{}^{b}{}_{a}\right),\qquad
\mathcal{D}^{\beta}\left(R(r)\right)=\exp\left(\frac{1}{2}r^{a}{}_{b}
M^{\beta}{}^{b}{}_{a}\right).
\]
For example, in three dimensions, where the pseudovector
$\overline{k}$ may be used instead of the tensor
$\omega^{b}{}_{a}$, i.e.,
$\mathcal{D}^{s}\left(W\left(\overline{k}\right)\right)=\exp\left(k^{a}
M^{s}{}_{a}\right)$, we should write that
$\mathcal{D}^{s}\left(L\left(\overline{l}\right)\right)=
\exp\left(l^{a}M^{s}{}_{a}\right)$,
$\mathcal{D}^{j}\left(R\left(\overline{r}\right)\right)=
\exp\left(r^{a}M^{j}{}_{a}\right)$, where $M^{s}{}_{a}$ ($s$ being
non-negative integers and $a=1,2,3$) are basic
$(2s+1)\times(2s+1)$, thus, odd-dimensional, anti-hermitian
matrices representing in an irreducible way the Lie algebra
SO$(3,\mathbb{R})^{\prime}$. Therefore,
$[M^{s}{}_{a},M^{s}{}_{b}]=-\epsilon_{ab}{}^{c}M^{s}{}_{c}$, and
it is impossible to reduce simultaneously all $M^{s}{}_{a}$ to the
block form. The apparently impossible even dimension $(2s+1)$ of
$M^{s}{}_{a}$, thus, positive half-integer $s$ will be an
important point of our further analysis because
SO$(3,\mathbb{R})^{\prime}$ (just as any
SO$(n,\mathbb{R})^{\prime}$, $n\geq 3$) admits even-dimensional
representations corresponding to the half-integer angular
momentum, both for rigid and homogeneously deformable bodies.

\section{Algebraic form of equations}

Let us introduce Hermitian matrices
$S^{\alpha}{}^{a}{}_{b}=(\hbar/i)M^{\alpha}{}^{a}{}_{b}$, thus,
for $n=3$, $S^{j}{}_{a}=(\hbar/i)M^{j}{}_{a}$, and
\[
\frac{1}{i\hbar}[S^{j}{}_{a},S^{j}{}_{b}]=\epsilon_{ab}{}^{c}S^{j}{}_{c}.
\]
These are standard well-known matrices, possible to be determined
in purely algebraic terms, basing only on the commutation
relations \cite{Lan-Lif58}. And it was just a surprise that there
exist even-dimensional irreducible representations, experimentally
compatible with the half-integer internal angular momentum spin.
The $(2j+1)\times(2j+1)$ matrices $S^{j}$ provide the quantum
description of the angular momentum with the quantized magnitude
$\hbar^{2}j(j+1)$; $j$ being a non-negative integer, or also a
positive half-integer in the theory of fermionic objects.

The representation property of $\mathcal{D}^{\alpha}$, i.e.,
$\mathcal{D}^{\alpha}(R_{1}R_{2})=\mathcal{D}^{\alpha}(R_{1})\mathcal{D}^{\alpha}(R_{2})$,
together with the definition of generators (\ref{kw30}),
(\ref{7.2}) imply that certain obvious relationships which enable
one to replace some differential operations and equations by
algebraic ones. Namely, it is clear from the above formulas that
\[
\frac{\hbar}{i}{\mathbf{\Lambda}}^{i}{}_{j}(L)\mathcal{D}^{\alpha}(L)=
S^{\alpha i}{}_{j}\mathcal{D}^{\alpha}(L),\qquad
\frac{\hbar}{i}{\mathbf{\Lambda}}^{A}{}_{B}(R)\mathcal{D}^{\beta}(R)=
\mathcal{D}^{\beta}(R)S^{\beta A}{}_{B},
\]
\[
\frac{\hbar}{i}{\mathbf{\Upsilon}}^{a}{}_{b}(L)\mathcal{D}^{\alpha}(L)=
\mathcal{D}^{\alpha}(L)S^{\alpha a}{}_{b},\qquad
\frac{\hbar}{i}{\mathbf{\Upsilon}}^{a}{}_{b}(R)\mathcal{D}^{\beta}(R)=
S^{\beta a}{}_{b}\mathcal{D}^{\beta}(R);
\]
expressions on the right-hand side meant, obviously, in the sense
of the matrix multiplication.

In other words, ${\bf S}^{i}{}_{j}\Psi^{\alpha\beta}=S^{\alpha
i}{}_{j}\Psi^{\alpha\beta}$ and ${\bf
V}^{A}{}_{B}\Psi^{\alpha\beta}=\Psi^{\alpha\beta}S^{\beta
A}{}_{B}$, where $\Psi^{\alpha\beta}$ is an abbreviation for the
$N(\alpha)\times N(\beta)$ matrices
$\left[\Psi^{\alpha\beta}_{ml}\right]$ in (\ref{7.6}),
$m=\overline{1,N(\alpha)}$, $l=\overline{1,N(\beta)}$. Obviously,
everything is formally correct because $S^{\alpha i}{}_{j}$,
$S^{\beta A}{}_{B}$ are, respectively, $N(\alpha)\times
N(\alpha)$- and $N(\beta)\times N(\beta)$-matrices. Let us stress
once again that the indices $(i,j)$, $(A,B)$ label basic matrices
within their sets; they do not refer to matrix elements.

From now on it will be convenient to write also (\ref{7.6}),
(\ref{7.8}) in matrix terms,
\[
\Psi^{\alpha\beta}(L,D,R)=\mathcal{D}^{\alpha}(L)f^{\alpha\beta}(D)\mathcal{D}^{\beta}(R^{-1});
\]
obviously, the reduced amplitude $f^{\alpha\beta}(D)$ is an
$N(\alpha)\times N(\beta)$-matrix depending only on deformation
invariants $D_{aa}=Q^{a}=\exp(q^{a})$.

Similarly, ${\bf \hat{r}}^{a}{}_{b}$ and ${\bf \hat{t}}^{a}{}_{b}$
act on $\Psi^{\alpha\beta}$ as follows:
\[
{\bf \hat{r}}^{a}{}_{b}\Psi^{\alpha\beta}=
\mathcal{D}^{\alpha}(L)S^{\alpha
a}{}_{b}f^{\alpha\beta}(D)\mathcal{D}^{\beta}(R^{-1}),
\]
\[
{\bf
\hat{t}}^{a}{}_{b}\Psi^{\alpha\beta}=
\mathcal{D}^{\alpha}(L)f^{\alpha\beta}(D)S^{\beta
a}{}_{b}\mathcal{D}^{\beta}(R^{-1}).
\]
Therefore, this action reduces simply to the action on the reduced
amplitude $f^{\alpha\beta}$ only. It will be convenient to denote
it as follows:
$\overrightarrow{S^{\alpha}}{}^{a}{}_{b}f^{\alpha\beta}:=S^{\alpha
a}{}_{b}f^{\alpha\beta}$,
$\overleftarrow{S^{\beta}}{}^{a}{}_{b}f^{\alpha\beta}:=
f^{\alpha\beta}(D)S^{\beta a}{}_{b}$. By assumption, the
representations $\mathcal{D}^{\alpha}$ of SO$(n,\mathbb{R})$ are
irreducible, therefore, the matrices $C^{\alpha}(p)=S^{\alpha
a}{}_{b}S^{\alpha b}{}_{c}\cdots S^{\alpha u}{}_{w}S^{\alpha
w}{}_{a}$ (with $p$ factors) are proportional to the
$N(\alpha)\times N(\alpha)$ identity matrices,
\begin{equation}\label{x48}
C^{\alpha}(p)=\left(\frac{\hbar}{i}\right)^{p}C(\alpha,p)I_{N(\alpha)},
\end{equation}
where the numbers $C(\alpha,p)$ are eigenvalues of the
corresponding Casimir operators built of the generators of the
left and right regular translations on SO$(n,\mathbb{R})$, e.g.,
${\bf C}_{{\rm
SO}(n,\mathbb{R})}(p)=\mathbf{\Lambda}^{a}{}_{b}\mathbf{\Lambda}^{b}{}_{c}
\cdots \mathbf{\Lambda}^{u}{}_{w}\mathbf{\Lambda}^{w}{}_{a}$ (with
$p$ factors).

So, finally, let us summarize the corresponding formulas for the
physical case $n=3$,
\[
\|{\bf S}\|^{2}\Psi^{sj}=\|{\bf
\hat{r}}\|^{2}\Psi^{sj}=\hbar^{2}s(s+1)\Psi^{sj}, \qquad \|{\bf
V}\|^{2}\Psi^{sj}=\|{\bf
\hat{t}}\|^{2}\Psi^{sj}=\hbar^{2}j(j+1)\Psi^{sj},
\]
\[
{\bf S}_{a}\Psi^{sj}=S^{s}{}_{a}\Psi^{sj},\qquad {\bf
V}_{a}\Psi^{sj}=\Psi^{sj}S^{j}{}_{a},
\]
in particular, in the standard representation, ${\bf
S}_{3}\Psi^{sj}_{ml}=\hbar m\Psi^{sj}_{ml}$, ${\bf
V}_{3}\Psi^{sj}_{ml}=\hbar l\Psi^{sj}_{ml}$. And just as for the
general dimension value $n$, a little more complicated action of
${\bf \hat{r}}_{a}$, ${\bf \hat{t}}_{a}$ resulting in affecting
the reduced $f(D)$-amplitudes,
\[
{\bf \hat{r}}_{a}:\quad f^{sj}\mapsto
S^{s}{}_{a}f^{sj}=\overrightarrow{S^{s}}{}_{a}f^{sj},\qquad {\bf
\hat{t}}_{a}:\quad f^{sj}\mapsto
f^{sj}S^{j}{}_{a}=\overleftarrow{S^{j}}{}_{a}f^{sj}.
\]
In the standard representation we have ${\bf \hat{r}}_{3}:
\left[f^{sj}_{ml}\right]\mapsto \left[\hbar mf^{sj}_{ml}\right]$,
${\bf \hat{t}}_{3}: \left[f^{sj}_{ml}\right]\mapsto \left[\hbar
lf^{sj}_{ml}\right]$.

\section{Half-integer values of spin}

And now we are ready to return to the problem of covering spaces
and half-integer quantized angular momentum of rigid and
deformable bodies. The problem of half-integer spin appeared in
quantum mechanics due to experimental data concerning radiation
spectra of atoms and molecules. Later on some theoretical work
gave an evidence of the existence of even-dimensional
representations of the Lie algebra SO$(3,\mathbb{R})^{\prime}$.
Their exponentiation does not lead to representations of
SO$(3,\mathbb{R})$ but to representations of its universal
covering group SU$(2)$, roughly speaking to the double-valued
representations of SO$(3,\mathbb{R})$; in a sense to its
projective representations.

As mentioned, there are some arguments that, in contrast to the
current views, it need not be always the case that the wave
amplitudes must be one-valued functions on the configuration
space. In certain situations, when the homotopy group is finite,
it seems to be sufficient that they are correctly defined on the
universal covering manifold of the configuration space. A typical
example is rigid body mechanics \cite{ABB95,ABMB95,BBM92} and the
mechanics of affinely-rigid bodies.

Let us begin with the general $n$-dimensional case, $n\geq 3$. The
configuration space of the rigid body in $n$ dimensions may be
identified analytically with the special orthogonal group
SO$(n,\mathbb{R})$. For $n\geq 3$ the universal covering group
Spin$(n)$ is doubly-connected, and the corresponding canonical
projection $\tau:$ Spin$(n)\rightarrow$ SO$(n,\mathbb{R})$ is
$2:1$. The special case $n=2$ is completely different, because
then the homotopy group is $\mathbb{Z}$ just as in the covering of
the circle SO$(2,\mathbb{R})\simeq$ U$(1)$ by $\mathbb{R}$.
Therefore, in this case it does not seem possible to admit
multi-valued wave functions, i.e., ones defined on $\mathbb{R}$.
For $n=3$ the covering group Spin$(3)$ is isomorphic with the
group of special (determinant-one) unitary matrices SU$(2)$. For
any $u\in$ SU$(2)$ the matrices $\pm u\in$ SU$(2)$ project under
$\tau$ onto the same element of SO$(3,\mathbb{R})$. Therefore,
Ker$\ \tau=\tau^{-1}(I_{3})=\{I_{2},-I_{2}\}$, i.e., the kernel
consists of the unit $2\times 2$ matrix $I_{2}$ and $-I_{2}$. Lie
algebra SU$(2)^{\prime}$ consists of anti-hermitian traceless
complex matrices, i.e., such ones that
$\alpha^{+}=\overline{\alpha}^{T}=-\alpha$, Tr $\alpha=0$. The
most convenient choice of basis, commonly used in geometry and
physics, is the following system: $e_{a}=(1/2i)\sigma_{a}$,
$a=1,2,3$, where $\sigma_{a}$ are Pauli matrices:
\[
\sigma_{1}=\left[\begin{tabular}{cc}
  0 & 1 \\
  1 & 0
\end{tabular}\right],\qquad
\sigma_{2}=\left[\begin{tabular}{cc}
  0 & $-i$\\
  $i$ & 0
\end{tabular}\right],\qquad
\sigma_{3}=\left[\begin{tabular}{cc}
  1 & 0 \\
  0 & $-1$
\end{tabular}\right].
\]
The basis $\{e_{a}\}$ exactly corresponds to the basis $\{E_{a}\}$
of SO$(3,\mathbb{R})^{\prime}$, i.e.,
$[e_{a},e_{b}]=\epsilon_{ab}{}^{c}e_{c}$. Canonical coordinates of
the first kind are given by components of rotation vector
$\overline{k}\in\mathbb{R}^{3}$,
\begin{equation}\label{7.9}
u\left(\overline{k}\right)=\exp\left(k^{a}e_{a}\right)=\cos\frac{k}{2}I_{2}-
\frac{k^{a}}{k}\sin\frac{k}{2}i\sigma_{a};
\end{equation}
often one uses the relativistic convention $\sigma_{0}$ for
$I_{2}$. This parameterization exactly corresponds to the usual
rotation vector $\overline{k}\in\mathbb{R}^{3}$ in
SO$(3,\mathbb{R})^{\prime}$, thus,
$\tau\left(u\left(\overline{k}\right)\right)=R\left(\overline{k}\right)$
if $k\leq \pi$. The main difference is that $k$, the magnitude of
$\overline{k}$, runs over the doubled range $[0,2\pi]$.
Parameterization is singular in the sense that all points of the
limiting sphere $k=2\pi$ in $\mathbb{R}^{3}$ represent the same
point of SU$(2)$, namely, $-I=u(2\pi\cdot\overline{n})$ for any
versor $\overline{n}\in\mathbb{R}^{3}$,
$(\overline{n}\cdot\overline{n})=1$. All points of
$\mathbb{R}^{3}$ in the interior of the ball $k<2\pi$ represent
uniquely elements of SU$(2)$; in particular, unlike the situation
in SO$(3,\mathbb{R})$ there is no antipodal identification for
$k=\pi$, i.e., $u(\pi\cdot\overline{n})\neq
u(-\pi\cdot\overline{n})$. Obviously,
\[
\tau^{-1}\left(R\left(\overline{k}\right)\right)=
\left\{u\left(\overline{k}\right),-u\left(\overline{k}\right)\right\}=
\left\{u\left(\overline{k}\right),u\left(\left(1-\frac{2\pi}{k}\right)
\overline{k}\right)\right\}.
\]
The epimorphism $\tau:$ SU$(2)\rightarrow$ SO$(3,\mathbb{R})$ is
given by the assignment: SU$(2)\ni v\mapsto R\in$
SO$(3,\mathbb{R})$, where
$vu\left(\overline{k}\right)v^{-1}=u\left(R\overline{k}\right)$.
For any non-negative integer or positive half-integer
$s=0,1/2,1,3/2,\ldots$ ($s=j/2$, $j=0,1,2,3,\ldots$) the Lie
algebra SU$(2)^{\prime}$ does possess an irreducible
representation of dimension $(2s+1)$ (thus, all naturals admitted,
not only the even ones) in terms of anti-hermitian matrices
$M^{s}$, the basic ones $M^{s}{}_{a}$, $a=1,2,3$, chosen so as to
satisfy
$[M^{s}{}_{a},M^{s}{}_{b}]=-\epsilon_{ab}{}^{c}M^{s}{}_{c}$.
Obviously, the corresponding $s$-angular momentum matrices,
$S^{s}{}_{a}:=(\hbar/i)M^{s}{}_{a}$, are Hermitian and do commute:
$(1/i\hbar)[S^{s}{}_{a},S^{s}{}_{b}]=\epsilon_{ab}{}^{c}S^{s}{}_{c}$.
Exponentiation
\[
\mathcal{D}^{s}\left(u\left(\overline{k}\right)\right)=
\mathcal{D}^{s}\left(\exp\left(k^{a}e_{a}\right)\right):=
\exp\left(k^{a}M^{s}{}_{a}\right)=\exp\left(\frac{i}{\hbar}k^{a}S^{s}{}_{a}\right)
\]
leads to unitary irreducible $(2s+1)$-dimensional representations
of SU$(2)$. The matrices $M^{s}{}_{a}$, as mentioned, may be found
on the purely algebraic basis of commutation relations
\cite{Lan-Lif58}, thus, for $n=3$ they are explicitly known and
standard. And so are $\mathcal{D}^{s}$, or more precisely their
matrix elements as the special functions on SO$(3,\mathbb{R})$.
This, by the way, is not the only possible method. Another one is
solving of differential equations on the group manifold, or
symmetrized Kronecker products of the basic representation of
SU$(2)$ by itself.

For non-negative integers $s$, i.e., for the odd values of
$N(s)=(2s+1)$, $\mathcal{D}^{s}$ do not distinguish elements $\pm
u\in$ SU$(2)$ $\tau$-projecting onto the same elements $R\in$
SO$(3,\mathbb{R})$, so, as a matter of fact, they are
representations of SO$(3,\mathbb{R})$ (more precisely, they are
$\tau$-pull-backs of SO$(3,\mathbb{R})$-representations to
SU$(2)$), just the previously discussed $\mathcal{D}^{s}$,
\[
\mathcal{D}^{s}(u)=\mathcal{D}^{s}(-u).
\]
For the positive half-integers $s$, i.e., for the even values of
$N(s)=(2s+1)$, $\mathcal{D}^{s}$ differ in sign at $u,-u\in$
SU$(2)$,
\[
\mathcal{D}^{s}(u)=-\mathcal{D}^{s}(-u),
\]
thus, they are non-projectable to SO$(3,\mathbb{R})$. But for any
fixed $s$, the squared moduli of the matrix elements or those of
their linear combinations are pull-backs from SO$(3,\mathbb{R})$,
so the probabilistic interpretation of $\overline{\Psi}\Psi$ is
not violated. The same holds when we superpose matrix elements of
various $\mathcal{D}^{s}$, $\mathcal{D}^{j}$ but with the same
parity of $(2s+1)$, $(2j+1)$, i.e., with the same "half-nesses" of
$s$, $j$. But, in general, the probabilistic interpretation of
$\overline{\Psi}\Psi$ is violated when different "half-nesses" of
$s$, $j$ are superposed. This is a toy model of the superselection
between "fermionic" and "bosonic" situations. As we shall see, in
a much more drastic form the problem appears in quantum mechanics
of affinely-rigid bodies.

Having in view physical applications we do not consider the
general case with $n>3$, thus, our Spin$(n)$ will be
Spin$(3)\simeq$ SU$(2)$. The planar problems $n=2$ are of some
physical relevance and will be briefly reviewed. However, the
possibility of the half-integer spin does not appear then; at the
same time, some other problems difficult for $n=3$ become
drastically simplified, just trivialized, for $n=2$.

\section{Affine spinors and polar decompositions}

The configuration spaces of affinely-rigid body, i.e., roughly
speaking (if translational motion is neglected)
GL$^{+}(n,\mathbb{R})$, SL$(n,\mathbb{R})$, are also
doubly-connec\-ted, and the problem of physically admissible
two-valued wave functions also appears here. There is, however,
some difficulty, namely, the intriguing and interesting fact that
the universal covering groups $\overline{{\rm
GL}^{+}(n,\mathbb{R})}$, $\overline{{\rm SL}(n,\mathbb{R})}$ are
nonlinear, i.e., they do not possess faithful realizations in
terms of finite matrices. This, by the way, was a reason for
plenty of misunderstandings and vast time in field theory and
quantum mechanics \cite{HKVdH74,HLN77}. The fact was known long
ago to mathematicians, like, e.g., E.~Cartan, but was forgotten
and exotic for physicists. The nonlinearity of the mentioned
coverings implies, in particular, that affine spinors
(half-objects) must be either infinite-dimensional or ruled by
nonlinear realizations of $\overline{{\rm GL}^{+}(n,\mathbb{R})}$,
$\overline{{\rm SL}(n,\mathbb{R})}$ as abstract groups constructed
with the help of loops in GL$^{+}(n,\mathbb{R})$,
SL$(n,\mathbb{R})$.

However, in quantum mechanics of affinely-rigid bodies the
construction of multi-valued wave functions may be analytically
overcome with the use of polar and two-polar splittings. Let us
begin from the first one,
\[
\varphi=UA=BU=\left(UAU^{-1}\right)U,
\]
where $U\in$ SO$(n,\mathbb{R})$, and $A$, $B$ are symmetric and
positively definite (and in the case of SL$(n,\mathbb{R})$ their
determinants equal one). The splitting is unique and, because of
this, GL$^{+}(n,\mathbb{R})$ as a manifold (but not as a group)
may be identified with the Cartesian products
SO$(n,\mathbb{R})\times$ Sym$^{+}(n,\mathbb{R})$ or
Sym$^{+}(n,\mathbb{R})\times$ SO$(n,\mathbb{R})$. The manifold
Sym$^{+}(n,\mathbb{R})$ is diffeomorphic with
$\mathbb{R}^{n(n+1)/2}$ ($\mathbb{R}^{6}$ if $n=3$), therefore,
the covering manifold may be identified with Spin$(n)\times$
Sym$^{+}(n,\mathbb{R})$ or Sym$^{+}(n,\mathbb{R})\times$
Spin$(n)$. In the physical case $n=3$, these splittings become
${\rm SU}(2)\times{\rm Sym}^{+}(3,\mathbb{R})\simeq{\rm
SU}(2)\times\mathbb{R}^{6}$ or alternatively ${\rm
Sym}^{+}(3,\mathbb{R})\times{\rm
SU}(2)\simeq\mathbb{R}^{6}\times{\rm SU}(2)$. Topological
non-triviality is absorbed here by the factor SO$(3,\mathbb{R})$
(in general by SO$(n,\mathbb{R})$) and covered by SU$(2)$ (in
general by Spin$(n)$). Therefore, the admissible multi-valued wave
functions may be expanded as follows:
\[
\Psi(u,A)=\sum_{s}\sum^{s}_{m=-s}\sum^{s}_{k=-s}C^{s}{}_{mk}(A)\mathcal{D}^{s}{}_{mk}(u),
\]
where $s$ are non-negative integers or positive half-integers, and
the summation over $m$, $k$ is performed in steps by one,
$\mathcal{D}^{s}$ are matrices of irreducible unitary
representations of SU$(2)$, and (very important!) only
half-integer or integer values of $s$ may appear in a given
expansion if $\overline{\Psi}\Psi$ is to be one-valued on
GL$^{+}(3,\mathbb{R})$, or, more precisely, if it is to be a
pull-back from GL$^{+}(3,\mathbb{R})$ to $\overline{{\rm
GL}^{+}(3,\mathbb{R})}$. Therefore, in any admissible $\Psi$,
$C^{s}{}_{mk}=0$ either for all non-negative integer or for all
positive half-integer $s$. To be completely rigorous, we would
have to write
\[
\Psi(u,A)=\sum^{\infty}_{\sigma=1}\sum^{\sigma}_{\mu=0}
\sum^{\sigma}_{\kappa=0}C^{\frac{\sigma}{2}}{}_{\left(-\frac{\sigma}{2}+\mu\right),
\left(-\frac{\sigma}{2}+\kappa\right)}(A)\mathcal{D}^{\frac{\sigma}{2}}
{}_{\left(-\frac{\sigma}{2}+\mu\right),\left(-\frac{\sigma}{2}+\kappa\right)}(u)
\]
for half-integer spin ("fermionic") situations or, respectively,
\[
\Psi(u,A)=\sum^{\infty}_{s=0}\sum^{2s}_{\mu=0}
\sum^{2s}_{\kappa=0}C^{s}{}_{(-s+\mu),
(-s+\kappa)}(A)\mathcal{D}^{s}{}_{(-s+\mu),(-s+\kappa)}(u)
\]
for integer spin ("bosonic") situations. These formulas are valid
without any provisos, with summation over all indices meant in
steps by one. If $\overline{\Psi}\Psi$ is to be one-valued
probability distribution, then the superposing between indicated
subspaces of function series is forbidden (a kind of
superselection rule), and the admissible Hamiltonians must exclude
any transitions between them; otherwise they are not well-defined
on L$^{2}({\rm GL}^{+}(n,\mathbb{R}))$.

It was said that the two-polar decomposition is maximally
effective in problems on which we concentrate. Let us now describe
the covering manifold $\overline{{\rm GL}^{+}(n,\mathbb{R})}$ and
the corresponding two-valued wave functions on
GL$^{+}(n,\mathbb{R})$ in terms of the two-polar splitting. The
non-uniqueness of the two-polar splitting of
GL$^{+}(n,\mathbb{R})$ was described briefly at the beginning of
section 6 of Part I \cite{part1}. Certain modifications are
necessary when using this splitting for describing the covering
$\overline{{\rm GL}^{+}(n,\mathbb{R})}$.

The elements of GL$^{+}(n,\mathbb{R})$ were represented by the
triplets $(L,D,R)\in$ SO$(n,\mathbb{R})\times\mathbb{R}^{n}\times$
SO$(n,\mathbb{R})$ taken modulo certain identifications resulting
from the fact that it was just the product $\varphi=LDR^{-1}$ not
$(L,D,R)$ itself that was a true configuration. Now, when
describing $\overline{{\rm GL}^{+}(n,\mathbb{R})}$, we must start
from the triplets $(l,D,r)\in$ Spin$(n)\times\mathbb{R}^{n}\times$
Spin$(n)$, i.e., in the physical three-dimensional case
$(l,D,r)\in$ SU$(2)\times\mathbb{R}^{3}\times$ SU$(2)$. In this
last case, $l$ and $r$ will be analytically described by the
extended rotation vectors
$\overline{l},\overline{r}\in\mathbb{R}^{3}$ in the sense of
(\ref{7.9}) with $\overline{k}$ replaced respectively by
$\overline{l}$, $\overline{r}$. Similarly, $D$ is analytically
represented by the variables $q^{a}=\ln D_{aa}$, and the
dilatational degree of freedom by the centre
$q=\left(q^{1}+q^{2}+q^{3}\right)/3$. As above, $\tau:$
Spin$(n)\rightarrow$ SO$(n,\mathbb{R})$ denotes the canonical
projection ($2:1$ epimorphism).
$\overline{K^{+}}:=\tau^{-1}(K^{+})$ is a $2n\cdot n!$-element
subgroup of Spin$(n)$; the group $K^{+}\subset$ SO$(n,\mathbb{R})$
itself was defined in section 6 of Part I \cite{part1}. The
manifold $M^{(n)}$ introduced also there is covered by
$\overline{M^{(n)}}$, i.e., the subset of such triplets
$\left(l;q^{1},\ldots,q^{n};r\right)\in$
Spin$(n)\times\mathbb{R}^{n}\times$ Spin$(n)$ that all $q^{i}$'s
are pairwise distinct. The subgroup $\overline{K^{+}}$ induces on
$\overline{M^{(n)}}$ the transformation group $\overline{H^{(n)}}$
action of which is given by the following rule:
$\left(l;q^{1},\ldots,q^{n};r\right)\mapsto
\left(lu;q^{\pi_{\tau(u)}(1},\ldots,q^{n)};ru\right)$, where $u\in
\overline{K^{+}}\subset$ Spin$(n)$ (SU$(2)$ if $n=3$).

The corresponding generic part of $\overline{{\rm
GL}^{+}(n,\mathbb{R})}$ (non-degenerate deformation tensors) is
obtained as a quotient subset under the
$\overline{H^{(n)}}$-action, i.e., $Q^{(n)}\simeq
\overline{M^{(n)}}/\overline{H^{(n)}}$. Situation becomes more
complicated when some $q^{a}$'s coincide, i.e., when the spectra
of deformation tensors are degenerate. Let the symbols
$Q^{(k;p_{1},\ldots,p_{k})}=$ GL$^{+(k;p_{1},\ldots,p_{k})}$,
$p_{\sigma}$, $M^{(k;p_{1},\ldots,p_{k})}$, $k$, and
$H^{(k;p_{1},\ldots,p_{k})}$ have the same meaning as in the
beginning of section 6 of Part I \cite{part1} where the two-polar
splitting non-uniqueness was described. To describe the
half-integer angular momentum, we must take the manifold
$\overline{M^{(k;p_{1},\ldots,p_{k})}}$ consisting of triplets
$\left(l;q^{1},\ldots,q^{n};r\right)$, where $l,r\in$ Spin$(n)$
and the system $\left(q^{1},\ldots,q^{n}\right)$ is degenerate as
above. Let $\overline{H^{(k;p_{1},\ldots,p_{k})}}\subset$
Spin$(n)$ denote the subgroup
$\tau^{-1}\left(H^{(k;p_{1},\ldots,p_{k})}\right)$. The
corresponding manifolds of degenerate configurations are given by
the quotient subsets
$\overline{M^{(k;p_{1},\ldots,p_{k})}}/\overline{H^{(k;p_{1},\ldots,p_{k})}}$
in the sense of the following action:
$\left(l;q^{1},\ldots,q^{n};r\right)\mapsto
\left(lu;q^{\pi_{\tau(u)}(1},\ldots,q^{n)};ru\right)$; obviously,
$u$ runs over $\overline{H^{(k;p_{1},\ldots,p_{k})}}$.

The admissibly multi-valued wave functions on
GL$^{+}(n,\mathbb{R})$, i.e., the ones one-valued on
$\overline{{\rm GL}^{+}(n,\mathbb{R})}$, are represented by
complex amplitudes on Spin$(n)\times\mathbb{R}^{n}\times$
Spin$(n)$ which are invariant under the above actions of
$\overline{H^{(k;p_{1},\ldots,p_{k})}}$, i.e., are projectable
onto the resulting quotients
$\overline{M^{(k;p_{1},\ldots,p_{k})}}/\overline{H^{(k;p_{1},\ldots,p_{k})}}$.

\section{Three-dimensional physical case}

Let us now concentrate on the special case $n=3$, both the
practically important one and at the same time reducible in a
sense to the classical Wigner results \cite{Rose95,Wign31,Wign65}.

All the former expressions concerning function series,
eigenequations, etc. remain generally true with the following
changes: half-integer quantum numbers $s$, $j$, $m$, $k$, $l$,
$n$, etc. are admissible, and certain new complications appear
concerning the non-distinguishability of triplets
$\left(l;q^{1},q^{2},q^{3};r\right)$ by wave functions
representants. In particular, some correlation appears between
"half-nesses" of the quantum numbers $s$, $j$ (spin and vorticity)
in physically acceptable function series. Obviously, this is based
on the assumption (true or not?) that the wave functions $\Psi$
may be multi-valued, but their moduli $|\Psi|$ must be one-valued
in accordance with the statistical interpretation of
$\overline{\Psi}\Psi$.

It is known that $\mathcal{D}^{j}(u)=\pm \mathcal{D}^{j}(-u)$,
$u\in$ SU$(2)$, depending, respectively, on whether $j$ is integer
or half-integer. Therefore, the expansions (\ref{7.7}),
(\ref{7.8}) remain valid for half-integer spin and vorticity,
thus, within the framework SU$(2)\times\mathbb{R}^{3}\times$
SU$(2)$ provided that some care is taken what concerns the
superposition structure, more precisely, the $(s,j)$ correlation.
So, formally, we can rewrite (\ref{7.7}), (\ref{7.8}) as follows:
\begin{eqnarray}
\label{7.12}\Psi(u,D,v)&=&\sum_{s,j}\sum^{s}_{m,n=-s}\sum^{j}_{k,l=-j}
\mathcal{D}^{s}_{mn}(u)f^{sj}_{{}^{nk}_{ml}}(D)\mathcal{D}^{j}_{kl}(v^{-1}),
\\
\label{7.13}
\Psi^{sj}_{ml}(u,D,v)&=&\sum^{s}_{n=-s}\sum^{j}_{k=-j}
\mathcal{D}^{s}_{mn}(u)f^{sj}_{nk}(D)\mathcal{D}^{j}_{kl}(v^{-1}),
\end{eqnarray}
with the following descriptive comments. Summation over $(s,j)$ in
(\ref{7.12}) or the choice of particular $(s,j)$ in (\ref{7.13})
is extended over non-negative integers or positive half-integers,
but in such a way that either $(s,j)$ are simultaneously integers
or simultaneously half-integers. In other words,
$f^{sj}(q^{1},q^{2},q^{3})\equiv 0$ if the number $(j-s)$ is
half-integer, i.e., always the summation will be extended over
such pairs $(s,j)$ in (\ref{7.12}) or the values of $(s,j)$ will
be chosen in (\ref{7.13}) in such a way that $(j-s)$ will be an
integer number. The quantum numbers $(m,n)$, $(k,l)$ in
(\ref{7.12}) run over the ranges from $-s$ to $s$ and from $-j$ to
$j$ in integer jumps.

Just as for integer pairs $(s,j)$, we will use
$(2s+1)\times(2j+1)$ rectangular matrices
$\Psi^{sj}=\left[\Psi^{sj}_{ml}\right]$,
$f^{sj}=\left[f^{sj}_{nk}\right]$, where
\[
\Psi^{sj}\left(u;q^{1},q^{2},q^{3};v\right)=
\mathcal{D}^{s}(u)f^{sj}\left(q^{1},q^{2},q^{3}\right)\mathcal{D}^{j}(v^{-1}).
\]
As mentioned, $f^{sj}$ vanishes identically as a function of
$q^{a}$'s when $(j-s)$ is half-integer. The matrix elements of
$\Psi^{sj}$ with integer values of $(j-s)$ may be arbitrarily
superposed, and this correlation is a necessary condition if
$\Psi$, $\Psi^{sj}$ are to be well-defined on $\overline{{\rm
GL}(3,\mathbb{R})}$ not only on the auxiliary manifold ${\rm
SU}(2)\times\mathbb{R}^{3}\times{\rm SU}(2)$. If
$\overline{\Psi}\Psi$ is to be projectable onto GL$(3,\mathbb{R})$
(statistical interpretation), then we may superpose only terms
with half-integer $(s,j)$ or integer $(s,j)$ separately.

One is used to avoid in mathematical texts the descriptive
literature-like comments as above, however, sometimes the purely
formula-based presentation becomes more obscure. It is just the
case here, especially when we wish to retain the traditional
notation used in the theory of angular momentum. So, for example,
avoiding words following the formulas (\ref{7.12}), (\ref{7.13})
would be panished by the following rather obscure expressions:
\[
\Psi(u,D,v)=\Psi_{1}(u,D,v)+\Psi_{2}(u,D,v)=
\]
\[
\sum^{\infty}_{\sigma,\iota=1}\sum^{\sigma}_{\mu,\nu=0}
\sum^{\iota}_{\kappa,\lambda=0}
\mathcal{D}^{\frac{\sigma}{2}}_{\left(-\frac{\sigma}{2}+\mu\right),
\left(-\frac{\sigma}{2}+\nu\right)}(u)
f^{\frac{\sigma}{2},\frac{\iota}{2}}_{{}^{(-\sigma/2+\nu),(-\iota/2+\kappa)}
_{(-\sigma/2+\mu),(-\iota/2+\lambda)}}(D)\mathcal{D}^{\frac{\iota}{2}}
_{\left(-\frac{\iota}{2}+\kappa\right),\left(-\frac{\iota}{2}+\lambda\right)}(v^{-1})
\]
\[
+\sum^{\infty}_{s,j=0}\sum^{2s}_{\mu,\nu=0}
\sum^{2j}_{\kappa,\lambda=0}
\mathcal{D}^{s}_{(-s+\mu),(-s+\nu)}(u)
f^{sj}_{{}^{(-s+\nu),(-j+\kappa)}_{(-s+\mu),(-j+\lambda)}}(D)
\mathcal{D}^{j}_{(-j+\kappa),(-j+\lambda)}(v^{-1}).
\]
The first term $\Psi_{1}$ contains contributions with half-integer
spin and vorticity (simultaneously), the second one $\Psi_{2}$
involves only integer quantized values of both. And this fact
again means that $\Psi$ is well-defined on $\overline{{\rm
GL}(3,\mathbb{R})}$ not only on ${\rm
SU}(2)\times\mathbb{R}^{3}\times{\rm SU}(2)$. But if
$\overline{\Psi}\Psi$ is to be well-defined on GL$(3,\mathbb{R})$
itself, then only $\Psi_{1},\Psi_{2}$-terms are separately
admissible without being superposed.

\section{Reduction to Cartan subgroup}

Matrix elements of irreducible representations have important
well-investiga\-ted properties which enable one to algebraize a
good deal of differential equations problems and to perform an
effective reduction of the quantum dynamics. Roughly speaking,
this is reduction to the Cartan subgroup of GL$(n,\mathbb{R})$,
i.e., to its maximal Abelian subgroup. This is just the group of
diagonal matrices, i.e., degrees of freedom parameterized by
deformation invariants $q^{1},\ldots,q^{n}$. This reduction from
$n^{2}$ to $n$ degrees of freedom is possible for geodetic
problems, for dilatationally-stabilized problems (i.e.,
essentially for geodetic problems on SL$(n,\mathbb{R})$) and, more
generally, for doubly isotropic models when the potential energy
is non-trivial but depends only on the deformation invariants,
i.e., it has the form $V\left(q^{1},\ldots,q^{n}\right)$. Let us
remind that in this sense quantum mechanics of affine bodies is
"simpler" than the classical one where for $n>2$ there is no
simple way of reducing equations of motion to the Cartan subgroup.

It is convenient to start again with the general $n$, and later on
to restrict ourselves to the special cases $n=2,3$. Due to the
standard orthogonality properties of
$\mathcal{D}^{\alpha}{}_{mn}$, the scalar product of wave
functions $\Psi$ may be reduced to one for the amplitudes
$f^{\alpha\beta}$ depending only on deformation invariants, i.e.,
\[
<\Psi_{1}|\Psi_{2}>=\sum_{\alpha,\beta\in\Omega}\frac{1}{N(\alpha)N(\beta)}
\int\sum^{N(\alpha)}_{n,m=1}\sum^{N(\beta)}_{k,l=1}\overline{f_{1}}^{\alpha\beta}_{{}^{nk}_{ml}}
f_{2}{}^{\alpha\beta}_{{}^{nk}_{ml}}Pdq^{1}\cdots dq^{n},
\]
where, let us remind, the weight $P$ is given by the following
expression:
\[
P\left(q^{1},\ldots,q^{n}\right)=\prod_{i\neq j}\left|{\rm
sh}(q^{i}-q^{j})\right|.
\]
If we fix the labels $\alpha$, $\beta$, $m$, $l$ ("good" quantum
numbers for doubly-isotropic problems) and consider the simplified
$N(\alpha)\times N(\beta)$-matrix amplitudes,
\[
\Psi^{\alpha\beta}\left(L;q^{1},\ldots,q^{n};R\right)=\mathcal{D}^{\alpha}(L)f^{\alpha\beta}
\left(q^{1},\ldots,q^{n}\right)\mathcal{D}^{\beta}\left(R^{-1}\right),
\]
then the scalar product reduces to
\[
<\Psi^{\alpha\beta}_{1}|\Psi^{\alpha\beta}_{2}>=\frac{1}{N(\alpha)N(\beta)}
\int{\rm Tr}\left(f_{1}^{\alpha\beta +} f_{2}^{\alpha\beta}\right)
Pdq^{1}\cdots dq^{n},
\]
where, obviously, $f_{1}^{\alpha\beta +}$ denotes the Hermitian
conjugate of the matrix $f_{1}^{\alpha\beta}$.

Obviously, for the general expansion (\ref{7.4}) the corresponding
formula involves the summation over $\alpha$, $\beta$, and the
multiplication of reduced amplitudes and trace operation meant in
the sense of two-matrices with the entries labelled by two-indices
$f^{\alpha\beta}_{{}^{nk}_{ml}}$, i.e.,
\[
<\Psi_{1}|\Psi_{2}>=\sum_{\alpha,\beta\in\Omega}
\frac{1}{N(\alpha)N(\beta)} \int{\rm Tr}\left(f_{1}^{\alpha\beta
+} f_{2}^{\alpha\beta}\right) Pdq^{1}\cdots dq^{n}.
\]
For the sake of completeness, let us write explicitly
\[
{\rm Tr}\left(f_{1}^{\alpha\beta +}f_{2}^{\alpha\beta}\right)=
\sum^{N(\alpha)}_{n,m=1}\sum^{N(\beta)}_{k,l=1}\overline{f_{1}}^{\alpha\beta}_{{}^{nk}_{ml}}
{f_{2}}^{\alpha\beta}_{{}^{nk}_{ml}}.
\]
When we consider the class of problems with $\alpha$, $\beta$,
$m$, $l$ fixed once for all, then one can avoid the divisor
$N(\alpha)N(\beta)$, with the proviso of being careful with the
normalization of amplitudes so as not to violate the statistical
interpretation.

In certain problems it may be convenient to avoid the phase factor
$P$ in the above expressions for the scalar product. To achieve
this one should introduced rescaled amplitudes given by the
matrices $g^{\alpha\beta}:=\sqrt{P}f^{\alpha\beta}$. Then the
factor $P$ disappears from the above formulas, $f^{\alpha\beta}$
becomes replaced by $g^{\alpha\beta}$, and everything else remains
as previously.

\section{Metric tensors and arc elements}

Essentially everything said above remains valid when discussing
the half-integer angular momentum. Orthogonal groups
SO$(n,\mathbb{R})$ in the two-polar decomposition are then
replaced by their coverings Spin$(n)$, but it does not change
anything in local analytical expressions. Technically, the only
change is that the range of group parameters changes. And where
for different parameter values the corresponding elements of
SO$(n,\mathbb{R})$ were identical, in Spin$(n)$ they are
different. It was described above in some details for
SO$(3,\mathbb{R})$ and its covering Spin$(n)=$ SU$(2)$, where the
main analytical novelty was replacing the range $[0,\pi]$ for the
rotation vector magnitude $k$ with $[0,2\pi]$. All analytical
formulas remain formally the same, e.g., those for the generators
of left and right regular translations ${\mathbf{\Lambda}}_{a}$,
${\mathbf{\Upsilon}}_{a}$. The metric Killing tensors on
SO$(3,\mathbb{R})$ and SU$(2)$ normalized to be $\delta_{ij}$ in
$\overline{k}$-coordinates at the group identity (thus, differing
by the minus one-half factor in comparison with the general
Lie-algebraic definition), i.e., $\Gamma(a,b)=-(1/2){\rm Tr}(ab)$
and $\Gamma(a,b)=-2{\rm Tr}(ab)$, respectively, on
SO$(3,\mathbb{R})$ and SU$(2)$; in both cases they are
analytically given by the same formula:
\[
\Gamma_{ab}=\frac{4}{k^{2}}\sin^{2}\frac{k}{2}\delta_{ab}+
\left(1-\frac{4}{k^{2}}\sin^{2}\frac{k}{2}\right)\frac{k_{a}k_{b}}{k^{2}}.
\]
In other words, the corresponding arc element is as follows:
\[
ds^{2}=\Gamma_{ab}dk^{a}dk^{b}=dk^{2}+4\sin^{2}\frac{k}{2}
\left(d\vartheta^{2}+\sin^{2}\vartheta d\varphi^{2}\right).
\]
Obviously, this metric is conformally flat, for example, defining
new coordinates $\overline{r}=(a/k){\rm tg}(k/4)\overline{k}$,
$a>0$, we obtain that
\[
ds^{2}=\frac{16a^{2}}{a^{2}+r^{2}}\left(dr^{2}+
r^{2}\left[d\vartheta^{2}+\sin^{2}\vartheta
d\varphi^{2}\right]\right),
\]
where the second factor is just the arc element in Euclidean
$\mathbb{R}^{3}$ expressed in terms of spherical coordinates. This
is the conformal mapping of SU$(2)$ onto $\mathbb{R}^{3}$ if we
consider the total range $r\in [0,\infty]$. It is interesting that
$r\in [0,a]$ on SO$(3,\mathbb{R})$. This is also some kind of
arguments that SO$(3,\mathbb{R})$ is somehow "imperfect" in
comparison with its universal covering SU$(2)$.

The Haar measure $\mu$ in both cases is given by
\[
d\mu\left(\overline{k}\right)=\frac{4}{k^{2}}\sin^{2}\frac{k}{2}d_{3}\overline{k}=
4\sin^{2}\frac{k}{2}\sin\vartheta dk d\vartheta d\varphi
\]
if we wish its weight function to be equal one in
$\overline{k}$-coordinates at the unit element
$\left(\overline{k}=0\right)$. But if we wish, as we often do, to
normalize the total measure of the compact group to unity, then
both cases will differ by a constant factor.

\section{Quantizing affine models}

One can show after some calculations that the operator ${\bf
T}^{\rm aff-aff}_{\rm int}$ of kinetic energy invariant under both
spatial and material affine transformation is as follows:
\begin{equation}\label{7.14}
{\bf T}^{\rm aff-aff}_{\rm int}=-\frac{\hbar^{2}}{2A}{\bf
D}+\frac{\hbar^{2}B}{2A(A+nB)}\frac{\partial^{2}}{\partial q^{2}}
+ \frac{1}{32A}\sum_{a,b}\frac{({\bf M}^{a}{}_{b})^{2}}{{\rm
sh}^{2}\frac{q^{a}-q^{b}}{2}}-\frac{1}{32A}\sum_{a,b}\frac{({\bf
N}^{a}{}_{b})^{2}}{{\rm ch}^{2}\frac{q^{a}-q^{b}}{2}},\nonumber
\end{equation}
where $A$, $B$ are constants as previously in classical formulas,
${\bf M}^{a}{}_{b}=-{\bf \hat{r}}^{a}{}_{b}-{\bf
\hat{t}}^{a}{}_{b}$ and ${\bf N}^{a}{}_{b}={\bf
\hat{r}}^{a}{}_{b}-{\bf \hat{t}}^{a}{}_{b}$ (cf. (\ref{x29}),
(\ref{kw29})),
\[
{\bf D}=\frac{1}{P}\sum_{a}\frac{\partial}{\partial
q^{a}}P\frac{\partial}{\partial
q^{a}}=\sum_{a}\frac{\partial^{2}}{\partial (q^{a})^{2}}+
\sum_{a}\frac{\partial \ln P}{\partial
q^{a}}\frac{\partial}{\partial q^{a}}
\]
(every differentiation operator acts on everything on the right of
it), $P$ is the previously introduced weight factor.

It is seen that this is almost the previously used classical
formula with classical canonical quantities, e.g.,
$\hat{\rho}^{a}{}_{b}$, $\hat{\tau}^{a}{}_{b}$ replaced by the
corresponding operators ${\bf \hat{r}}^{a}{}_{b}$, ${\bf
\hat{t}}^{a}{}_{b}$. There is, however, some difference and
possibility of an easy mistake in the sector of
$(q^{a},p_{a})$-variables. Namely, the term involving
differentiation with respect to $q^{a}$ is not, as it might be
expected, the usual $\mathbb{R}^{n}$-Laplace operator in $q^{a}$
variables, although it contains such a term. Let us observe that
in the $\varphi=LDR^{-1}$-representation the $\partial/\partial
q^{a}$ operators act only on the $f^{\alpha\beta}$ amplitude,
whereas ${\bf \hat{r}}^{a}{}_{b}$, ${\bf \hat{t}}^{a}{}_{b}$ act
only, respectively, on the $L$- and $R$-variables. Therefore,
there is no problem of ordering of operators in ${\bf T}^{\rm
aff-aff}_{\rm int}$. One could get rid off the first derivatives
of $\Psi$ with respect to $q^{a}$ by the substitution which was
already used within a slightly different context, namely,
$\varphi=\sqrt{P}\Psi$. The action of the last three terms in
(\ref{7.14}) on $\varphi$ is exactly as that on $\Psi$ because
$\partial/\partial q^{a}$, ${\bf M}^{a}{}_{b}$, ${\bf
N}^{a}{}_{b}$ do not act on $(q^{a}-q^{b})$-quantities of which
$P$ is built; roughly speaking, the $\sqrt{P}$ is "transparent"
for these operators. It is no longer the case with the ${\bf
D}$-term, both in the good and in the bad senses. Namely, the
action of $-(\hbar^{2}/2A){\bf D}$ on $\Psi$ is represented by the
action of the following operator $-(\hbar^{2}/2A){\bf
\widetilde{D}}$ on $\varphi$:
\[
-\frac{\hbar^{2}}{2A}\widetilde{\bf
D}=-\frac{\hbar^{2}}{2A}\sum_{a}\frac{\partial^{2}}{\partial
(q^{a})^{2}}+{\bf \widetilde{V}},
\]
where ${\bf \widetilde{V}}$ is the following artificial potential
term:
\[
{\bf \widetilde{V}}=-\frac{\hbar}{2A}\frac{1}{P^{2}}+
\frac{\hbar^{2}}{4A}\frac{1}{P}\sum_{a}\left(\frac{\partial
P}{\partial q^{a}}\right)^{2}.
\]
In other words, ${\bf \widetilde{D}}\varphi=\sqrt{P}{\bf D}\Psi$.
There are no first derivatives of $\varphi$ with respect to
$q^{a}$, and the differential action is given by the usual
$\mathbb{R}^{n}$-Laplace operator, just as in mechanics of $n$
$q^{a}$-particles on $\mathbb{R}$. But this simplification is only
seeming one because, if $n>2$, it is completely destroyed by the
"potential" ${\bf \widetilde{V}}$. Obviously, in realistic
problems concerning deformable objects Hamiltonian should also
contain dilatation-stabilizing potential, i.e., ${\bf H}={\bf
T}^{\rm aff-aff}_{\rm int}+{\bf V}(q)$. And although such simple
SL$(n,\mathbb{R})$-geodetic models may successfully describe
elastic vibrations, some more general isotropic potentials
$V\left(q^{1},\ldots,q^{n}\right)$ are also acceptable and
compatible with the above description.

Quantizing metric-affine and affine-metric kinetic energies we
obtain, respectively, the following operators:
\[
{\bf T}^{\rm met-aff}_{\rm int}=-\frac{\hbar^{2}}{2\alpha}{\bf
D}-\frac{\hbar^{2}}{2\beta}\frac{\partial^{2}}{\partial q^{2}}+
\frac{1}{32\alpha}\sum_{a,b}\frac{({\bf M}^{a}{}_{b})^{2}}{{\rm
sh}^{2}\frac{q^{a}-q^{b}}{2}}-\frac{1}{32\alpha}\sum_{a,b}\frac{({\bf
N}^{a}{}_{b})^{2}}{{\rm
ch}^{2}\frac{q^{a}-q^{b}}{2}}+\frac{1}{2\mu}\|{\bf S}\|^{2},
\]
\[
{\bf T}^{\rm aff-met}_{\rm int}=-\frac{\hbar^{2}}{2\alpha}{\bf
D}-\frac{\hbar^{2}}{2\beta}\frac{\partial^{2}}{\partial q^{2}}+
\frac{1}{32\alpha}\sum_{a,b}\frac{({\bf M}^{a}{}_{b})^{2}}{{\rm
sh}^{2}\frac{q^{a}-q^{b}}{2}}-\frac{1}{32\alpha}\sum_{a,b}\frac{({\bf
N}^{a}{}_{b})^{2}}{{\rm
ch}^{2}\frac{q^{a}-q^{b}}{2}}+\frac{1}{2\mu}\|{\bf V}\|^{2},
\]
with the same meaning of operator symbols as above and the same
relationship between inertial constants $(\alpha,\beta,\mu)$ and
the primary ones $(I,A,B)$ as above.

\section{Potential case}

As mentioned, for Hamiltonians ${\bf H}={\bf T}+{\bf V}$ with
dilatation-stabilizing potentials $V(q)$, or more generally, with
doubly-isotropic potentials $V(q^{1},\ldots,q^{n})$, the action of
operators ${\bf M}^{a}{}_{b}$ and ${\bf N}^{a}{}_{b}$ become
algebraic and standard, and the stationary Schr\"odinger equation,
i.e., energy eigenproblem ${\bf H}\Psi=E\Psi$, splits into family
of eigenproblems for the amplitudes $f^{\alpha\beta}$; they are
partial differential equations involving $q^{a}$-variables only:
\[
{\bf
H}^{\alpha\beta}f^{\alpha\beta}=E^{\alpha\beta}f^{\alpha\beta},
\]
where $f^{\alpha\beta}$ for any $\alpha,\beta\in\Omega$ is an
$N(\alpha)\times N(\beta)$ matrix depending on
$q^{1},\ldots,q^{n}$. In a consequence of the double (spatial and
material) isotropy, this problem is $N(\alpha)\times
N(\beta)$-fold degenerate, i.e., for every component of
$f^{\alpha\beta}$ there exists an $N(\alpha)\times
N(\beta)$-dimensional subspace of solutions. Let us remind that in
the primary symbols $f^{\alpha\beta}_{{}^{nk}_{ml}}$ the indices
$m$, $l$ just label the degeneracy of solutions for every
$f^{\alpha\beta}_{nk}$. ${\bf H}^{\alpha\beta}$ is an
$N(\alpha)\times N(\beta)$-matrix of second-order differential
operators, ${\bf H}^{\alpha\beta}={\bf T}^{\alpha\beta}+{\bf V}$,
where ${\bf V}$ denotes a dilatation-stabilizing or general
doubly-isotropic potential, and ${\bf T}^{\alpha\beta}$ denotes
the kinetic energy operator. It is one of the previous ones
restricted to the corresponding ($\alpha,\beta$)-subspace.
Therefore, for the affine-affine, metric-affine, and affine-metric
models we have, respectively,
\begin{eqnarray}
{\bf T}^{\alpha\beta}f^{\alpha\beta}&=&-\frac{\hbar^{2}}{2A}{\bf
D}f^{\alpha\beta}+\frac{\hbar^{2}B}{2A(A+nB)}\frac{\partial^{2}}{\partial
q^{2}}f^{\alpha\beta}\label{A}
\\
&+&
\frac{1}{32A}\sum_{a,b}\frac{\left(\overleftarrow{S^{\beta}}{}^{a}{}_{b}-
\overrightarrow{S^{\alpha}}{}^{a}{}_{b}\right)^{2}}{{\rm
sh}^{2}\frac{q^{a}-q^{b}}{2}}f^{\alpha\beta}-
\frac{1}{32A}\sum_{a,b}\frac{\left(\overleftarrow{S^{\beta}}{}^{a}{}_{b}+
\overrightarrow{S^{\alpha}}{}^{a}{}_{b}\right)^{2}}{{\rm
ch}^{2}\frac{q^{a}-q^{b}}{2}}f^{\alpha\beta},\nonumber
\\
{\bf
T}^{\alpha\beta}f^{\alpha\beta}&=&-\frac{\hbar^{2}}{2\alpha}{\bf
D}f^{\alpha\beta}-\frac{\hbar^{2}}{2\beta}\frac{\partial^{2}}{\partial
q^{2}}f^{\alpha\beta}-\frac{\hbar^{2}}{2\mu}C(\alpha,2)
f^{\alpha\beta}\label{B}
\\
&+&\frac{1}{32\alpha}\sum_{a,b}
\frac{\left(\overleftarrow{S^{\beta}}{}^{a}{}_{b}-
\overrightarrow{S^{\alpha}}{}^{a}{}_{b}\right)^{2}}{{\rm
sh}^{2}\frac{q^{a}-q^{b}}{2}}f^{\alpha\beta}
-\frac{1}{32\alpha}\sum_{a,b}\frac{\left(\overleftarrow{S^{\beta}}{}^{a}{}_{b}+
\overrightarrow{S^{\alpha}}{}^{a}{}_{b}\right)^{2}}{{\rm
ch}^{2}\frac{q^{a}-q^{b}}{2}}f^{\alpha\beta},\nonumber
\\
{\bf
T}^{\alpha\beta}f^{\alpha\beta}&=&-\frac{\hbar^{2}}{2\alpha}{\bf
D}f^{\alpha\beta}-\frac{\hbar^{2}}{2\beta}\frac{\partial^{2}}{\partial
q^{2}}f^{\alpha\beta}-\frac{\hbar^{2}}{2\mu}C(\beta,2)
f^{\alpha\beta}\label{C}
\\
&+&\frac{1}{32\alpha}\sum_{a,b}
\frac{\left(\overleftarrow{S^{\beta}}{}^{a}{}_{b}-
\overrightarrow{S^{\alpha}}{}^{a}{}_{b}\right)^{2}}{{\rm
sh}^{2}\frac{q^{a}-q^{b}}{2}}f^{\alpha\beta}
-\frac{1}{32\alpha}\sum_{a,b}\frac{\left(\overleftarrow{S^{\beta}}{}^{a}{}_{b}+
\overrightarrow{S^{\alpha}}{}^{a}{}_{b}\right)^{2}}{{\rm
ch}^{2}\frac{q^{a}-q^{b}}{2}}f^{\alpha\beta},\nonumber
\end{eqnarray}
where the meaning of Casimir eigenvalues $C(\alpha,2)$,
$C(\beta,2)$ like in (\ref{x48}). The constants $\alpha$, $\beta$,
$\mu$  are exactly as previously; do not confuse them with labels
$\alpha$, $\beta$ at $f^{\alpha\beta}$. In the physical case
$n=3$, $\alpha=s=0,1/2,1,\ldots\in \mathbb{N}/2\cup\{0\}$ and
similarly $\beta=j=0,1/2,1,\ldots\in \mathbb{N}/2\cup\{0\}$
assuming that the half-integer values of angular momentum and
vorticity are admitted. Otherwise we would have
$s,j\in\mathbb{N}\cup\{0\}$. Obviously, in this case
$C(s,2)=-s(s+1)$, $C(j,2)=-j(j+1)$, and the additional constants
in the last two formulas are simply $(\hbar^{2}/2\mu)s(s+1)$,
$(\hbar^{2}/2\mu)j(j+1)$, expressions close to the heart of any
physicist. Let us stress that, even if half-integers are admitted,
there is a restriction that $(j-s)$ must be integer, i.e., $j$ and
$s$ have the same "half-ness". In any case, it must be so if wave
functions are to be well-defined on $\overline{{\rm
GL}(3,\mathbb{R})}$ not only on the "artificial" configuration
space ${\rm SU}(2)\times\mathbb{R}^{3}\times{\rm SU}(2)$. If they
are to be statistically interpretable in ${\rm GL}(3,\mathbb{R})$
itself, then only the terms with half-integer $(s,j)$ or integer
$(s,j)$ may be separately superposed, no mutual superposition
admissible (although some blasphemic doubts may be raised against
this superselection, i.e., against statistical interpretation in
${\rm GL}(3,\mathbb{R})$).

In three-dimensional case the above-mentioned additional terms
\[
(\hbar^{2}/2\mu)s(s+1),\quad (\hbar^{2}/2\mu)j(j+1) 
\]
seem to be
physically interesting and, at least qualitatively, compatible
with some experimental data. It is so as if the doubly affine
background (affine invariance in space and in the body) was
responsible for some fundamental part of the spectra, which later
on, the more the $\mu$ is smaller, splits due to some internal
rotations. The term $(\hbar^{2}/2\mu)s(s+1)$ is physically
intuitive and classically corresponds to the situation when in the
system some regime of rigid rotations was established after time
of transition processes. But, perhaps, $(\hbar^{2}/2\mu)j(j+1)$
appearing in the affine-metrical model is even more interesting.
Being a formal analogue of certain aspects of angular momentum, it
is not angular momentum and may be perhaps semiclassically related
to the isotopic spin or similar internal quantities ruled by
SU$(2)$ and appearing in nuclear and elementary particle physics.

\noindent {\bf Remark:} Just as previously, the terms with the
first-order derivatives of $f^{\alpha\beta}$ with respect to
$q^{a}$ may be avoided by the substitution
\[
g^{\alpha\beta}:=\sqrt{P}f^{\alpha\beta},
\]
which was also used for simplifying the scalar product. But then
again the artificial potential ${\bf V}$ appears in all reduced
Schr\"odinger equations.

By the way, one can have both things, i.e.,
$(\hbar^{2}/2\mu)s(s+1)$ and $(\hbar^{2}/2\mu)j(j+1)$ terms. For
this purpose we would have to use the kinetic energy consisting of
four terms:
\[
T_{\rm
int}=\frac{I_{1}}{2}g_{ik}g^{jl}\Omega^{i}{}_{j}\Omega^{k}{}_{l}+
\frac{I_{2}}{2}\eta_{KL}\eta^{MN}\hat{\Omega}^{K}{}_{M}\hat{\Omega}^{L}{}_{N}+
\frac{A}{2}\hat{\Omega}^{K}{}_{L}\hat{\Omega}^{L}{}_{K}+
\frac{B}{2}\hat{\Omega}^{K}{}_{K}\hat{\Omega}^{L}{}_{L},
\]
where the last two terms might be as well written as
$(A/2)\Omega^{i}{}_{j}\Omega^{j}{}_{i}+
(B/2)\Omega^{i}{}_{i}\Omega^{j}{}_{j}$. In matrix language, using
Cartesian coordinates $g_{ik}=_{\ast}\delta_{ik}$,
$\eta_{AB}=_{\ast}\delta_{AB}$, we would simply write that
\begin{eqnarray}\label{7.15}
T_{\rm int}&=&\frac{I_{1}}{2}{\rm Tr}(\Omega^{T}\Omega)+
\frac{I_{2}}{2}{\rm Tr}(\hat{\Omega}^{T}\hat{\Omega})+
\frac{A}{2}{\rm Tr}(\hat{\Omega}^{2})+ \frac{B}{2}({\rm
Tr}\hat{\Omega})^{2}\\
&=&\frac{I_{1}}{2}{\rm Tr}(\Omega^{T}\Omega)+ \frac{I_{2}}{2}{\rm
Tr}(\hat{\Omega}^{T}\hat{\Omega})+ \frac{A}{2}{\rm
Tr}(\Omega^{2})+ \frac{B}{2}({\rm Tr}\Omega)^{2}.\nonumber
\end{eqnarray}
But now some reproach might be raised that, doing as above, we
forget our primary motivation concerning the dynamical
GL$(n,\mathbb{R})$-invariance and return to models which are only
orthogonally invariant (geometrically speaking, O$(V,g)$- and
O$(U,\eta)$-invariant), and it is again only pure kinematics that
is ruled by affine group. This would be true, and we indeed do not
insist on the above model. Let us notice, however, that this
model, having still high dynamical symmetry, may also work as a
purely geodetic model encoding a kind of elastic bounded
vibrations without any extra introduced potential. Moreover, due
to the lack of dilatational invariance, it is not excluded (we are
not yet sure; this is a conjecture) that even
dilatation-stabilizing potentials would not be necessary.

\section{Doubly-isotropic d'Alembert models}

The above remarks about models (\ref{7.15}) again put our
attention on the doubly isotropic "d'Alembert" models of classical
kinetic energy (\cite{part1}2.1), i.e., (\cite{part1}4.21) with
the factorization
$\mathcal{A}^{K}{}_{i}{}^{L}{}_{j}=Ig_{ij}\eta^{KL}$. The
corresponding kinetic part of the classical kinetic Hamiltonian
$\mathcal{T}^{\rm d.A}_{\rm int}$ was given by (\cite{part1}6.68)
with the same meaning of $M^{a}{}_{b}$, $N^{a}{}_{b}$ as above,
$Q^{a}=D_{aa}$ are diagonal elements of $D$, and $P_{a}$ are
canonical momenta conjugate to $Q^{a}$. This time, as a measure
particularly convenient for quantization, the usual Lebesgue
measure $l$ on L$(n)$ should be used,
$dl(\varphi)=d\varphi^{1}{}_{1}\cdots\varphi^{n}{}_{n}$. In terms
of the two-polar splitting,
$dl(L,D,R)=P_{l}(Q)d\mu(L)d\mu(R)dQ^{1}\cdots dQ^{n}$, where
$\mu$, as previously, is the Haar measure on SO$(n,\mathbb{R})$,
and the weight factor $P_{l}$ is now given by the following
expression:
\[
P_{l}=\prod_{a\neq
b}\left|(Q^{a})^{2}-(Q^{b})^{2}\right|=\prod_{a\neq
b}\left|(Q^{a}+Q^{b})(Q^{a}-Q^{b})\right|.
\]
Everything concerning quantization looks in a similar way like
previously for affinely-invariant models. For example, expansion
of wave functions $\Psi$ with respect to
$\mathcal{D}^{\alpha}(L)$, $\mathcal{D}^{\beta}(R)$ with
$f^{\alpha\beta}(D)$-reduced amplitudes is exactly the same. The
difference appears in details concerning the integration
procedure, just the weight factor $P_{l}$ is substituted instead
of $P$. Also, in spite of formal similarities, the particular form
of the kinetic energy operator is different,
\[
{\bf T}^{\rm d.A}_{\rm int}=-\frac{\hbar^{2}}{2I}{\bf D}_{l}+
\frac{1}{8I}\sum_{a,b}\frac{({\bf
M}^{a}{}_{b})^{2}}{(Q^{a}-Q^{b})^{2}}+\frac{1}{8I}\sum_{a,b}\frac{({\bf
N}^{a}{}_{b})^{2}}{(Q^{a}+Q^{b})^{2}},
\]
where now
\[
{\bf D}_{l}=\frac{1}{P_{l}}\sum_{a}\frac{\partial}{\partial
Q^{a}}P_{l}\frac{\partial}{\partial
Q^{a}}=\sum_{a}\frac{\partial^{2}}{\partial (Q^{a})^{2}}+
\sum_{a}\frac{\partial \ln P_{l}}{\partial
Q^{a}}\frac{\partial}{\partial Q^{a}}.
\]
Just as previously, the weight factor $P_{l}$ in the scalar
product and first-order differentiations $\partial/\partial Q^{a}$
may be avoided by rescaling $\varphi=\sqrt{P_{l}}\Psi$, but in the
resulting differential operator acting on $\varphi$ also some
rather unpleasant potential term appears, i.e.,
\[
{\bf \widetilde{V}}_{l}=-\frac{\hbar}{2I}\frac{1}{P^{2}_{l}}+
\frac{\hbar^{2}}{4I}\frac{1}{P_{l}}\sum_{a}\left(\frac{\partial
P_{l}}{\partial Q^{i}}\right)^{2}.
\]
It is obvious that without an appropriate potential term ${\bf V}$
the geodetic Hamiltonian ${\bf T}^{\rm d.A}$ cannot work in theory
of deformable objects because just as on the classical level it
describes only purely scattering, non-bounded motions. Indeed, the
above operator
\[
{\bf T}^{\rm d.A}=-\frac{\hbar^{2}}{2I}\Delta^{n^{2}}=
-\frac{\hbar^{2}}{2I}\sum_{i,A}\frac{\partial^{2}}{\partial
(\varphi^{i}{}_{A})^{2}}
\]
is simply proportional to the usual Laplace operator in
$\mathbb{R}^{n^{2}}$ written in non-typical coordinates.

Therefore, the only realistic applications of the above ${\bf T}$
are those as a term of some doubly isotropic Hamiltonian ${\bf
H}={\bf T}^{\rm d.A}+{\bf V}\left(Q^{1},\ldots,Q^{n}\right)$. Just
as previously, due to the double isotropy of the model, the
resulting stationary Schr\"odinger equation ${\bf H}\Psi=E\Psi$
splits into the family of equations for partial amplitudes
$f^{\alpha\beta}$ depending only on $q^{a}$-variables, ${\bf
H}^{\alpha\beta}f^{\alpha\beta}=E^{\alpha\beta}f^{\alpha\beta}$,
where
\begin{eqnarray}
{\bf H}^{\alpha\beta}f^{\alpha\beta}&=&-\frac{\hbar^{2}}{2I}{\bf
D}_{l}f^{\alpha\beta}+\frac{1}{8I}\sum_{a,b}
\frac{\left(\overleftarrow{S^{\beta}}{}^{a}{}_{b}-
\overrightarrow{S^{\alpha}}{}^{a}{}_{b}\right)^{2}}{(Q^{a}-Q^{b})^{2}}
f^{\alpha\beta}\label{184}\\
&+&\frac{1}{8I}\sum_{a,b}\frac{\left(\overleftarrow{S^{\beta}}{}^{a}{}_{b}+
\overrightarrow{S^{\alpha}}{}^{a}{}_{b}\right)^{2}}{(Q^{a}+Q^{b})^{2}}f^{\alpha\beta}+
V\left(Q^{1},\ldots,Q^{n}\right)f^{\alpha\beta}.\nonumber
\end{eqnarray}

For d'Alembert models, the problem of coverings and multi-valued
wave functions looks exactly like in affine theories. Simply
SO$(n,\mathbb{R})$-groups in the two-polar decomposition must be
replaced by the coverings Spin$(n)$. In particular, for $n=3$ when
$\alpha,\beta=s,j=0,1/2,1,\ldots,\ $everything said above remains
true, and $S^{s}{}^{a}{}_{b}$, $S^{j}{}^{a}{}_b{}$ are replaced by
the standard Wigner matrices of angular momentum, $S^{s}{}_{a}$,
$S^{j}{}_{a}$.

\section{Usual Wigner matrices of angular momentum}

In three dimensions those terms of the affine-affine reduced
operator ${\bf T}^{\alpha\beta}$ (\ref{A}) which contain the
factor $1/32A$ may be written in the following form involving the
usual Wigner matrices $S^{j}{}_{a}$:
\begin{eqnarray}\label{7.16}
\sum_{a=1}^{3}\left[\frac{(S^{s}{}_{a})^{2}f^{sj}-
2S^{s}{}_{a}f^{sj}S^{j}{}_{a}+f^{sj}(S^{j}{}_{a})^{2}}{16A{\rm
sh}^{2}\frac{q^{b}-q^{c}}{2}}\right.\nonumber\\
\left.-\frac{(S^{s}{}_{a})^{2}f^{sj}+
2S^{s}{}_{a}f^{sj}S^{j}{}_{a}+f^{sj}(S^{j}{}_{a})^{2}}{16A{\rm
ch}^{2}\frac{q^{b}-q^{c}}{2}}\right],
\end{eqnarray}
where in any $a$-th term of both summations we have obviously
$b\neq a$, $c\neq a$, $b\neq c$ (it is clear that it does not
matter what is the sequence of $b$, $c$).

The same holds for the metric-affine and affine-metric models
(\ref{B}), (\ref{C}), with the proviso that the inertial factor
$A$ is replaced by $\alpha$. As mentioned, the last
constant-multiplicator terms are respectively
$(\hbar^{2}/2\mu)s(s+1)f^{sj}$ and $(\hbar^{2}/2\mu)j(j+1)f^{sj}$.
Similarly, in reduced d'Alembert expressions (\ref{184}) the terms
with the $1/8I$-factor become for $n=3$:
\begin{eqnarray}\label{7.17}
\sum_{a=1}^{3}\left[\frac{(S^{s}{}_{a})^{2}f^{sj}-
2S^{s}{}_{a}f^{sj}S^{j}{}_{a}+f^{sj}(S^{j}{}_{a})^{2}}{4I(Q^{b}-Q^{c})^{2}}\right.\nonumber\\
\left.+\frac{(S^{s}{}_{a})^{2}f^{sj}+
2S^{s}{}_{a}f^{sj}S^{j}{}_{a}+f^{sj}(S^{j}{}_{a})^{2}}{4I(Q^{b}+Q^{c})^{2}}\right]
\end{eqnarray}
with the same as previously convention concerning indices $a$,
$b$, $c$.

For affinely-invariant geodetic models the bounded state
L$^{2}$-solutions appear for particular relationships between $s$
and $j$ ($\alpha$ and $\beta)$ in $n$ dimensions). For the
d'Alembert models of kinetic energy this is impossible, an
appropriate potential $V\left(Q^{1},\ldots,Q^{n}\right)$ must be
always used.

Both the affine and d'Alembert expressions (\ref{7.16}),
(\ref{7.17}) become particularly simple for the lowest possible
values of rotational quantum numbers $s$, $j$, and then there
exists some hope for rigorous or at least numerical solutions.
Thus, for $s=j=0$ the corresponding expressions vanish at all, and
the resulting Schr\"odinger equations for $f^{00}$ are purely
scalar. For $s=j=1/2$ we obtain the spinor-spinor state, which is
also relatively simple because then
$S^{1/2}{}_{a}=(\hbar/2)\sigma_{a}$,
$\left(S^{1/2}{}_{a}\right)^{2}=(\hbar^{2}/4)I_{2}$, where,
obviously, $\sigma_{a}$ are Pauli matrices, and $I_{2}$ is the
unit $2\times 2$ matrix.

\section{Two-dimensional case on the classical level}

In some physical problems also the two-dimensional case $n=2$ may
be physically interesting \cite{Mart04}. And in any case it is
mathematically exceptionally simple. This is, so to speak,
"pathological" simplicity following from the commutativity of
SO$(2,\mathbb{R})$. Although this exceptional simplicity is rather
"exotic" from the point of view of the general $n$, it may suggest
some guiding hints for analysis of this general situation.

The main two-dimensional peculiarity is that
\[
\hat{\rho}=\rho=S,\qquad \hat{\tau}=\tau=-V.
\]
This is exactly due to the commutativity of SO$(2,\mathbb{R})$.
Because of this, the convenient quantities $\hat{\rho}$,
$\hat{\tau}$ are constants of motion for geodetic models and
models with doubly-invariant potentials. It was not the case for
$n>2$, where only $S$, $V$ are constants of motion (for invariant
geodetic models and, more generally, for doubly-isotropic models).
But it is just the use of $\hat{\rho}$ and $\hat{\tau}$, or
equivalently $M$ and $N$, that simplifies the problem and enables
one to perform a partial separation of variables, especially
effective on the quantum level. If $n=2$, the two things coincide,
and the problem may be effectively reduced to the Cartan subgroup
of diagonal matrices (deformation invariants) even on the
classical level.

Let us begin with the classical description. In the two-polar
decomposition $\varphi=LDR^{-1}$ we shall use the following
parameterization:
\[
L=\left[\begin{tabular}{cc}
  $\cos\alpha$ & $-\sin\alpha$ \\
  $\sin\alpha$ & $\cos\alpha$
\end{tabular}\right],\
R=\left[\begin{tabular}{cc}
  $\cos\beta$ & $-\sin\beta$ \\
  $\sin\beta$ & $\cos\beta$
\end{tabular}\right],\
D=\left[\begin{tabular}{cc}
  $\exp q^{1}$ & $0$ \\
  $0$ & $\exp q^{2}$
\end{tabular}\right].
\]
The splitting GL$^{+}(2,\mathbb{R})=\mathbb{R}^{+}{\rm
SL}(2,\mathbb{R})$ is well-suited to coordinates
\[
q=\left(q^{1}+q^{2}\right)/2,\quad x=q^{2}-q^{1}, 
\]
and their
conjugate canonical momenta, respectively, 
\[
p=p_{1}+p_{2},\quad p_{x}=\left(p_{2}-p_{1}\right)/2. 
\]
Before using these convenient
coordinates, let us express classical kinetic energies in terms of
primary variables. First of all, let us notice the obvious fact
that the angular velocities of $L$- and $R$-rotators are given,
respectively, by
\[
\chi=\frac{dL}{dt}L^{-1}=L^{-1}\frac{dL}{dt}=\hat{\chi}=
\frac{d\alpha}{dt}\left[\begin{tabular}{cc}
 $0$ & $-1$ \\
 $1$ & $0$
\end{tabular}\right],
\]
\[
\vartheta=\frac{dR}{dt}R^{-1}=R^{-1}\frac{dR}{dt}=\hat{\vartheta}=
\frac{d\beta}{dt}\left[\begin{tabular}{cc}
 $0$ & $-1$ \\
 $1$ & $0$
\end{tabular}\right].
\]
The corresponding spin and vorticity quantities are given (in
canonical representation) by the following expressions:
\[
S=\rho=\hat{\rho}=p_{\alpha}\left[\begin{tabular}{cc}
 $0$ & $1$ \\
 $-1$ & $0$
\end{tabular}\right],\qquad V=-\tau=-\hat{\tau}=
p_{\beta}\left[\begin{tabular}{cc}
 $0$ & $1$ \\
 $-1$ & $0$
\end{tabular}\right],
\]
where $p_{\alpha}$, $p_{\beta}$ are, respectively, canonical
momenta conjugate to $\alpha$, $\beta$. The corresponding duality
pairings are as follows:
\[
p_{\alpha}\frac{d\alpha}{dt}=\frac{1}{2}{\rm
Tr}(S\chi)=\frac{1}{2}{\rm Tr}(\rho\chi)=\frac{1}{2}{\rm
Tr}(\hat{\rho}\hat{\chi}),
\] 
\[
p_{\beta}\frac{d\beta}{dt}=\frac{1}{2}{\rm
Tr}(V\vartheta)=-\frac{1}{2}{\rm
Tr}(\tau\vartheta)=-\frac{1}{2}{\rm
Tr}(\hat{\tau}\hat{\vartheta}),
\]
where $d\alpha/dt$, $d\beta/dt$ are arbitrary virtual velocities
of the variables $\alpha$, $\beta$.

The corresponding classical quantities $M=-\hat{\rho}-\hat{\tau}$,
$N=\hat{\rho}-\hat{\tau}$ are, respectively, given by the
following expressions:
\[
M=\textsf{m}\left[\begin{tabular}{cc}
 $0$ & $1$ \\
 $-1$ & $0$
\end{tabular}\right],\qquad
N=\textsf{n}\left[\begin{tabular}{cc}
 $0$ & $1$ \\
 $-1$ & $0$
\end{tabular}\right],
\]
where $\textsf{m}:=p_{\beta}-p_{\alpha}$,
$\textsf{n}:=p_{\beta}+p_{\alpha}$ may be interpreted as canonical
momenta conjugate to the corresponding "mixtures" of angles
$\beta$, $\alpha$: $\gamma:=(\beta-\alpha)/2$,
$\delta:=(\beta+\alpha)/2$, i.e., $\alpha=\delta-\gamma$,
$\beta=\delta+\gamma$. In fact, one can easily show that
$\textsf{m}\dot{\gamma}+\textsf{n}\dot{\delta}=
p_{\alpha}\dot{\alpha}+p_{\beta}\dot{\beta}$ for arbitrary virtual
velocities occurring in these formulas, thus,
$\textsf{m}=p_{\gamma}=p_{\beta}-p_{\alpha}$,
$\textsf{n}=p_{\delta}=p_{\beta}+p_{\alpha}$, and conversely,
$p_{\alpha}=(\textsf{n}-\textsf{m})/2$,
$p_{\beta}=(\textsf{n}+\textsf{m})/2$. The previously used
magnitudes of $S$, $V$ become:
\[
\|S\|=|p_{\alpha}|=\frac{1}{2}|\textsf{n}-\textsf{m}|,\qquad
\|V\|=|p_{\beta}|=\frac{1}{2}|\textsf{n}+\textsf{m}|.
\]
For the classical affine-affine kinetic energy (\cite{part1}6.69)
in Hamiltonian representation we obtain the following expression:
\[
\mathcal{T}^{\rm aff-aff}_{\rm
int}=\frac{1}{2A}\left(p^{2}_{1}+p^{2}_{2}\right)-\frac{B}{2A(A+2B)}p^{2}+
\frac{1}{16A}\frac{\textsf{m}^{2}}{{\rm
sh}^{2}\frac{q^{2}-q^{1}}{2}}-\frac{1}{16A}\frac{\textsf{n}^{2}}{{\rm
ch}^{2}\frac{q^{2}-q^{1}}{2}};
\]
meaning of symbols $A$, $B$ is like previously, and $n=2$ is
substituted to constant factors.

Similarly, for the metrical-affine and affine-metrical models we
obtain, respectively,
\[
\mathcal{T}^{\rm met-aff}_{\rm
int}=\frac{1}{2\alpha}(p^{2}_{1}+p^{2}_{2})+\frac{1}{2\beta}p^{2}+
\frac{1}{16\alpha}\frac{\textsf{m}^{2}}{{\rm
sh}^{2}\frac{q^{2}-q^{1}}{2}}-\frac{1}{16\alpha}\frac{\textsf{n}^{2}}{{\rm
ch}^{2}\frac{q^{2}-q^{1}}{2}}+\frac{1}{8\mu}(\textsf{n}-\textsf{m})^{2},
\]
\[
\mathcal{T}^{\rm aff-met}_{\rm
int}=\frac{1}{2\alpha}(p^{2}_{1}+p^{2}_{2})+\frac{1}{2\beta}p^{2}+
\frac{1}{16\alpha}\frac{\textsf{m}^{2}}{{\rm
sh}^{2}\frac{q^{2}-q^{1}}{2}}-\frac{1}{16\alpha}\frac{\textsf{n}^{2}}{{\rm
ch}^{2}\frac{q^{2}-q^{1}}{2}}+\frac{1}{8\mu}(\textsf{n}+\textsf{m})^{2},
\]
where meaning of constants $\alpha$, $\beta$, $\mu$ is like
previously, but with $n=2$ substituted, thus, $\alpha=I+A$,
$\beta=-(I+A)(I+A+2B)/B$, $\mu=\left(I^{2}-A^{2}\right)/I$. As
$\textsf{m}$ and $\textsf{n}$, or equivalently $p_{\alpha}$ and
$p_{\beta}$, are now constants of motion, it is seen that for
geodetic problems and for problems with doubly-isotropic
potentials $V\left(q^{1},q^{2}\right)$, e.g., with
dilatation-stabilizing ones $V(q)$, everything reduces to the
two-dimensional dynamics in variables $q^{1}$, $q^{2}$ ruled by
the effective Hamiltonian obtained by the formal substitution of
fixed values $p_{\alpha}$, $p_{\beta}$ (or $\textsf{m}$,
$\textsf{n}$) to the above expressions. Moreover, for
SL$(2,\mathbb{R})$-geodetic problems, or for
GL$(2,\mathbb{R})$-problems with separated variables potentials
$V\left(q,x\right)=V_{\rm dil}(q)+V_{\rm sh}(x)$, everything
reduces trivially to independent one-dimensional motions. In the
above geodetic models it is only the relationship between constant
values of $\textsf{m}$, $\textsf{n}$ that decides whether the
motion is oscillatory or unbounded. The first case happens,
obviously, when $|\textsf{n}|>|\textsf{m}|$; then at large
"distances" $\left|q^{2}-q^{1}\right|$ the attractive ${\rm
ch}^{-2}$-term prevails. On the contrary, if
$|\textsf{n}|<|\textsf{m}|$, one deals with the repulsive case,
i.e., with the decaying motion of invariants $q^{1}$, $q^{2}$.
This is the simplest example of the fact mentioned above that
affinely-invariant geodetic models admit an open family of bounded
(vibrating) and an open family of non-bounded (decaying) motions.
Obviously, for general $n>2$ the situation is more complicated
because then $M^{a}{}_{b}$, $N^{a}{}_{b}$ fail to be constants of
motion and perform oscillations somehow coupled with those of
$q^{a}$. Using new variables $q$, $x$, $p$, $p_{x}$, we can
rewrite the above models of $\mathcal{T}$ in the following forms:
\begin{eqnarray}
\mathcal{T}^{\rm aff-aff}_{\rm
int}&=&\frac{p^{2}}{4(A+2B)}+\frac{p^{2}_{x}}{A}+
\frac{(p_{\alpha}-p_{\beta})^{2}}{16A{\rm
sh}^{2}\frac{x}{2}}-\frac{(p_{\alpha}+p_{\beta})^{2}}{16A{\rm
ch}^{2}\frac{x}{2}},\nonumber\\
\mathcal{T}^{\rm met-aff}_{\rm
int}&=&\frac{p^{2}}{4(I+A+2B)}+\frac{p^{2}_{x}}{I+A}
+\frac{Ip^{2}_{\alpha}}{I^{2}-A^{2}}\nonumber\\
&+&\frac{(p_{\alpha}-p_{\beta})^{2}}{16(I+A){\rm
sh}^{2}\frac{x}{2}}-\frac{(p_{\alpha}+p_{\beta})^{2}}{16(I+A){\rm
ch}^{2}\frac{x}{2}},
\nonumber\\
\mathcal{T}^{\rm aff-met}_{\rm
int}&=&\frac{p^{2}}{4(I+A+2B)}+\frac{p^{2}_{x}}{I+A}
+\frac{Ip^{2}_{\beta}}{I^{2}-A^{2}}\nonumber\\
&+&\frac{(p_{\alpha}-p_{\beta})^{2}}{16(I+A){\rm
sh}^{2}\frac{x}{2}}-\frac{(p_{\alpha}+p_{\beta})^{2}}{16(I+A){\rm
ch}^{2}\frac{x}{2}}.\nonumber
\end{eqnarray}
In the special case $n=2$, it is easily seen that on the level of
variables $q$, $x$ all these geodetic models have identical
dynamics. The difference appears  only on the level of angular
variables $\alpha$, $\beta$. And, just as for the general $n$, the
same is true if we introduce to Hamiltonians some doubly-isotropic
potentials $V(q,x)$. In particular, this is true for
dilatation-stabilizing potentials $V(q)$, i.e., in a sense, for
geodetic invariant models on SL$(2,\mathbb{R})$ (incompressible
bodies).

\section{Quantization of two-dimensional models}

Let us now turn to quantization. The Haar measure $\lambda$ on
GL$(2,\mathbb{R})$ is given by the following expression:
$d\lambda\left(\alpha;q^{1},q^{2};\beta\right)=\left|{\rm
sh}\left(q^{1}-q^{2}\right)\right|d\alpha d\beta dq^{1}dq^{2}$,
i.e., $d\lambda\left(\alpha;q,x;\beta\right)=\left|{\rm
sh}x\right|d\alpha d\beta dqdx$, $P=\left|{\rm sh}x\right|$. The
Peter-Weyl expansion with respect to the $L,R$-factors of the
two-polar splitting is just the usual double Fourier series:
\[
\Psi\left(\alpha;q,x;\beta\right)=\sum_{m,n\in\mathbb{Z}}f^{mn}(q,x)e^{im\alpha}e^{in\beta}.
\]
The reduced kinetic Hamiltonian corresponding to $\mathcal{T}^{\rm
aff-aff}_{\rm int}$ is as follows:
\[
{\bf T}^{mn}f^{mn}=-\frac{\hbar^{2}}{A}{\bf
D}_{x}f^{mn}-\frac{\hbar^{2}}{4(A+2B)}
\frac{\partial^{2}f^{mn}}{\partial q^{2}}
\]
\[
+\frac{\hbar^{2}(n-m)^{2}}{16A{\rm
sh}^{2}\frac{x}{2}}f^{mn}-\frac{\hbar^{2}(n+m)^{2}}{16A{\rm
ch}^{2}\frac{x}{2}}f^{mn},
\]
where
\[
{\bf D}_{x}f^{mn}=\frac{1}{|{\rm sh}x|}\frac{\partial}{\partial
x}\left(|{\rm sh}x|\frac{\partial f^{mn}}{\partial x}\right).
\]
For the metric-affine and affine-metric models $\mathcal{T}^{\rm
met-aff}_{\rm int}$, $\mathcal{T}^{\rm aff-met}_{\rm int}$ we
obtain, respectively, the following expressions:
\begin{eqnarray}
{\bf T}^{mn}f^{mn}&=&-\frac{\hbar^{2}}{I+A}{\bf D}_{x}f^{mn}-
\frac{\hbar^{2}}{4(I+A+2B)}\frac{\partial^{2}f^{mn}}{\partial
q^{2}}\nonumber\\
&+&\frac{\hbar^{2}(n-m)^{2}}{16(I+A){\rm
sh}^{2}\frac{x}{2}}f^{mn}-\frac{\hbar^{2}(n+m)^{2}}{16(I+A){\rm
ch}^{2}\frac{x}{2}}f^{mn}+\frac{I\hbar^{2}m^{2}}{I^{2}-A^{2}}f^{mn},\nonumber\\
{\bf T}^{mn}f^{mn}&=&-\frac{\hbar^{2}}{I+A}{\bf D}_{x}f^{mn}-
\frac{\hbar^{2}}{4(I+A+2B)}\frac{\partial^{2}f^{mn}}{\partial
q^{2}}\nonumber\\
&+&\frac{\hbar^{2}(n-m)^{2}}{16(I+A){\rm
sh}^{2}\frac{x}{2}}f^{mn}-\frac{\hbar^{2}(n+m)^{2}}{16(I+A){\rm
ch}^{2}\frac{x}{2}}f^{mn}+\frac{I\hbar^{2}n^{2}}{I^{2}-A^{2}}f^{mn},\nonumber
\end{eqnarray}
It is seen that in all these expressions the complete separation
between dilatational and incompressible motion is very effectively
described in analytical terms just due to the use of coordinates
$q$, $x$. Obviously, for geodetic Hamiltonians on
GL$(2,\mathbb{R})$ the energy spectrum is continuous (and
classical trajectories are unbounded; in a sense equivalent facts)
because dilatational motion is free. As in the general case, this
fact is physically avoided by introducing to the Hamiltonian some
dilatation-stabilizing potential $V_{\rm dil}(q)$. On the quantum
level the simplest possible model is the potential well.

This is, in a sense, reduction to the geodetic quantum problem on
SL$(2,\mathbb{R})$. Obviously, the problem with $V_{\rm dil}(q)$
remains explicitly separable. It remains so also for a more
general class of doubly isotropic potentials, e.g., for ones
explicitly splitting, $V(q,x)=V_{\rm dil}(q)+V_{\rm sh}(x)$, but
perhaps also for more general ones. Solutions of the corresponding
stationary Schr\"odinger equations may be sought in the following
form: $f^{mn}(q,x)=\varphi^{mn}(q)\chi^{mn}(x)$; the problem
reduces then to one-dimensional Schr\"odinger equations for
$\varphi^{mn}$ and $\chi^{mn}$. And now, in the special
two-dimensional case, it is explicitly seen that there exists a
discrete spectrum (bounded situations) for $\chi$-functions, i.e.,
for the isochoric SL$(2,\mathbb{R})$-problem, even in the purely
geodetic case without any potential $V_{x}(x)$. And this is true
in spite of the non-compactness of the
SL$(2,\mathbb{R})$-configuration space. Everything depends on the
mutual relationship between "rotational" quantum numbers $m$, $n$.
If $|n+m|>|n-m|$, the attractive ${\rm ch}^{-2}$-term prevails at
large "distances" $|x|\rightarrow\infty$ and the spectrum is
discrete. In the opposite case, if $|n+m|<|n-m|$, it is
continuous.

For the affine-affine geodetic model on SL$(2,\mathbb{R})$, the
total spectrum (total in the sense of solutions for all possible
$m,n\in\mathbb{Z}$) is not bounded from below; this might seem
undesirable. For the metric-affine and affine-metric geodetic
problems on SL$(2,\mathbb{R})$, the spectrum may be bounded from
below (and so is the corresponding kinetic energy). Everything
depends on the mutual relationship between inertial constants $I$,
$A$, $B$, which play the role of some controlling parameters.

\section{Usual two-dimensional d'Alembert models}

For comparison, let us quote a few corresponding formulas for the
"usual" d'Alembert model in two dimensions. We restrict ourselves
to the doubly-isotropic model. The classical kinetic Hamiltonian
may be expressed as follows:
\[
\mathcal{T}^{\rm d.A}_{\rm
int}=\frac{1}{2I}\left(P^{2}_{1}+P^{2}_{2}\right)
+\frac{1}{4I}\frac{\textsf{m}^{2}}{\left(Q^{1}-Q^{2}\right)^{2}}+
\frac{1}{4I}\frac{\textsf{n}^{2}}{\left(Q^{1}+Q^{2}\right)^{2}},
\]
with the same meaning of symbols as previously. Let us stress that
$Q^{a}$ are diagonal elements of $D$, and now the variables
$q^{a}=\ln Q^{a}$ would be completely useless. The quantity
$P_{l}$ is given simply by the following expression:
\[
P_{l}=\left|\left(Q^{1}\right)^{2}-\left(Q^{2}\right)^{2}\right|=
\left|\left(Q^{1}+Q^{2}\right)\left(Q^{1}-Q^{2}\right)\right|,
\]
and the usual Lebesgue measure on
L$(2,\mathbb{R})\simeq\mathbb{R}^{4}$ is expressed as follows:
\[
dl\left(\alpha;Q^{1},Q^{2};\beta\right)=P_{l}\left(Q^{1},Q^{2}\right)d\alpha
d\beta dQ^{1}dQ^{2}.
\]
As mentioned, geodetic models are non-physical (and, by the way,
the above coordinates would be completely artificial for them).
There is, however, a class of physically reasonable doubly
isotropic potentials $V\left(Q^{1},Q^{2}\right)$ for which the
corresponding Hamiltonians $H=\mathcal{T}+V$ describe integrable
systems admitting solutions in terms of separation of variables.
This fact is obvious when, instead of $Q^{1}$, $Q^{2}$, the
$(\pi/4)$-rotated coordinates $Q^{+}$, $Q^{-}$ on the plane of
deformation invariants are used, $Q^{\pm}:=\left(Q^{1}\pm
Q^{2}\right)/\sqrt{2}$. The polar and elliptic coordinates on the
$(Q^{+},Q^{-})$-plane are also convenient, i.e., $Q^{+}=r \cos
\varphi$, $Q^{-}=r \sin \varphi$ and $Q^{+}={\rm ch}\rho \cos
\lambda$, $Q^{-}={\rm sh}\rho \sin \lambda$.

There exist physically reasonable potentials $V$ for which the
corresponding Hamiltonian problems are separable (thus, obviously,
integrable) in coordinates $(Q^{+},Q^{-})$, $(r,\varphi)$, or
$(\rho,\lambda)$. There are also interesting superintegrable
(degenerate) models separable simultaneously in two or even three
of the above coordinate systems.

On the quantized level the reduced Schr\"odinger equation has the
following form: ${\bf H}^{mn}f^{mn}=E^{mn}f^{mn}$, where
\[
{\bf H}^{mn}f^{mn}={\bf T}^{mn}f^{mn}+{\bf
V}\left(Q^{1},Q^{2}\right)f^{mn}
\]
\[
=-\frac{\hbar^{2}}{2I}{\bf
D}_{l}f^{mn}+\frac{\hbar^{2}m^{2}}{4I(Q^{1}-Q^{2})^{2}}f^{mn}
+\frac{\hbar^{2}n^{2}}{4I(Q^{1}+Q^{2})^{2}}f^{mn}+{\bf
V}\left(Q^{1},Q^{2}\right)f^{mn}.
\]
Obviously,
\[
{\bf D}_{l}f=\frac{1}{P_{l}}\sum_{a=1}^{2}\frac{\partial}{\partial
Q^{a}}\left(P_{l}\frac{\partial f}{\partial Q^{a}}\right).
\]
Everything said above about separability of the classical problems
remains true on the quantized level. Again the coordinate systems
$(Q^{+},Q^{-})$, $(r,\varphi)$, $(\rho,\lambda)$ are crucial.

\section{Hamiltonian systems on U$(n)$}

To finish these quantization remarks let us mention briefly about
Hamiltonian systems on U$(n)$, i.e., in a sense, affine systems
with "compactified deformation invariants" (\cite{part1}6.69). The
resulting kinetic energy operator has the following form:
\[
{\bf T}=-\frac{\hbar^{2}}{2A}{\bf
D}_{U}+\frac{\hbar^{2}B}{2A(A+nB)}\frac{\partial^{2}}{\partial
q^{2}}+\frac{1}{32A}\sum_{a,b}\frac{({\bf
M}^{a}{}_{b})^{2}}{\sin^{2}\frac{q^{a}-q^{b}}{2}}
+\frac{1}{32A}\sum_{a,b}\frac{({\bf
N}^{a}{}_{b})^{2}}{\cos^{2}\frac{q^{a}-q^{b}}{2}},
\]
where
\[
{\bf D}_{U}=\frac{1}{P_{U}}\sum_{a}\frac{\partial}{\partial
q^{a}}P_{U}\frac{\partial}{\partial
q^{a}}=\sum_{a}\frac{\partial^{2}}{\partial (q^{a})^{2}}+
\sum_{a}\frac{\partial\ln P_{U}}{\partial
q^{a}}\frac{\partial}{\partial q^{a}},
\]
\[
P_{U}=\prod_{a\neq
b}\left|\sin(q^{a}-q^{b})\right|.
\]
The Haar measure is given by the expression
\[
d\lambda_{U}(L,D,R)=P_{U}d\mu(L)d\mu(R)dq^{1}\cdots dq^{n},
\]
where $\mu$, as previously, denotes the Haar measure on
SO$(n,\mathbb{R})$.

Obviously, U$(n)$ is compact, thus, all classical trajectories for
geodetic models are bounded and the corresponding quantum spectrum
is discrete. Nevertheless, more general models with
doubly-isotropic potentials, i.e., ${\bf H}={\bf T}+{\bf
V}\left(q^{1},\ldots,q^{n}\right)$, may be also of physical
interest.

The problem splits again, just as in the GL$(n,\mathbb{R})$-case,
into the family of reduced problems resulting from the Fourier
analysis on SO$(n,\mathbb{R})$ performed both in the $L$- and
$R$-variables: ${\bf
H}^{\alpha\beta}f^{\alpha\beta}=E^{\alpha\beta}f^{\alpha\beta}$,
where
\begin{eqnarray}
{\bf H}^{\alpha\beta}f^{\alpha\beta}&=&-\frac{\hbar^{2}}{2I}{\bf
D}_{U}f^{\alpha\beta}-\frac{\hbar^{2}}{2\beta}
\frac{\partial^{2}f^{\alpha\beta}}{\partial q^{2}}+
\frac{1}{32A}\sum_{a,b}\frac{\left(\overleftarrow{S^{\beta}}{}^{a}{}_{b}
-\overrightarrow{S^{\alpha}}{}^{a}{}_{b}\right)^{2}}
{\sin^{2}\frac{q^{a}-q^{b}}{2}}f^{\alpha\beta}\nonumber\\
&+&\frac{1}{32A}\sum_{a,b}\frac{\left(\overleftarrow{S^{\beta}}{}^{a}{}_{b}
+\overrightarrow{S^{\alpha}}{}^{a}{}_{b}\right)^{2}}
{\cos^{2}\frac{q^{a}-q^{b}}{2}}f^{\alpha\beta}+ {\bf
V}\left(q^{1},\ldots,q^{n}\right)f^{\alpha\beta}.\nonumber
\end{eqnarray}

Just as in the GL$(n,\mathbb{R})$-models, particularly simple are
physical dimensions $n=2,3$. The former one has also certain very
peculiar features and admits simple calculations based on
integrable models and separability techniques. Namely, ${\bf
H}^{mn}$ acts as follows:
\begin{eqnarray}
{\bf H}^{mn}f^{mn}&=&-\frac{\hbar^{2}}{A}{\bf
D}_{x}f^{mn}+\frac{\hbar^{2}(n-m)^{2}}{16A\sin^{2}\frac{x}{2}}f^{mn}
+\frac{\hbar^{2}(n+m)^{2}}{16A\cos^{2}\frac{x}{2}}f^{mn}+{\bf
V}_{x}(x)f^{mn}
\nonumber\\
&-&\frac{\hbar^{2}}{4(A+2B)}\frac{\partial^{2}f^{mn}}{\partial
q^{2}}+{\bf V}_{q}(q)f^{mn},\nonumber
\end{eqnarray}
where the Haar measure has the expression
$d\lambda_{U}\left(\alpha;q,x;\beta\right)=|\sin x|d\alpha d\beta
dqdx$ and
\[
{\bf D}_{x}f=\frac{1}{|\sin x|}\frac{\partial}{\partial x}\left(
|\sin x|\frac{\partial f}{\partial x}\right).
\]
The problem also separates, in particular, for geodetic problems,
$V=0$, or for potentials of the above-mentioned form
$V(q,x)=V_{\rm dil}(q)+V_{\rm sh}(x)$.

\section*{Acknowledgements}

This paper is the second in the series which is thought on as a
draft of some future monograph. It contains results obtained
during our work within the framework of the research project
8T07A04720 of the Committee of Scientific Research (KBN)
"Mechanical Systems with Internal Degrees of Freedom in Manifolds
with Nontrivial Geometry". Also results obtained within the KBN
Supervisor Programme 5T07A04824 "Group-Theoretical Models of
Internal Degrees of Freedom" are included here. Authors are
greatly indebted to KBN for the financial support.

Included are also some results obtained or inspired during the
stay of one of us (Jan Jerzy S\l awianowski) in Pisa in 2001, at
the Istituto Nazionale di Alta Matematica "Francesco Severi",
Universit\`{a} di Pisa. It is a pleasure to express the deep
gratitude to Professor Gianfranco Capriz for his hospitality,
discussions, and inspiration. They contributed in an essential way
to the results presented here. The same concerns discussions with
Professor Carmine Trimarco, Dipartimento di Matematica Applicata
"U. Dini", Universit\`{a} di Pisa. Also our "three-body
interactions" ("three-soul interactions"), i.e., common
discussions in our small group (Gianfranco Capriz, Carmine
Trimarco, Jan Jerzy S\l awianowski) contributed remarkably to
results presented in this treatise. One of us (Jan Jerzy S\l
awianowski) is deeply indebted to Istituto Nazionale di Alta
Matematica "Francesco Severi", Gruppo Nazionale per la Fizica
Matematica Firenze, for the fellowship which made this
collaboration possible.

Discussions with Professor Paolo Maria Mariano during his stay at
the Institute of Fundamental Technological Research of Polish
Academy of Sciences in Warsaw also influenced this paper and are
cordially acknowledged.


\begin{thebibliography}{99}

\bibitem{ABB95}
D.~Arsenovi\'{c}, A.~O.~Barut  and M.~Bo\v{z}i\'{c}: {\em Il Nuovo
Cimento\/} {\bf 110B}, 2, 177--188 (1995).

\bibitem{ABMB95}
D.~Arsenovi\'{c}, A.~O.~Barut, Z.~Mari\'{c} and M.~Bo\v{z}i\'{c}:
{\em Il Nuovo Cimento\/} {\bf 110B}, 2, 163--175 (1995).

\bibitem{BBM92}
A.~O.~Barut, M.~Bo\v{z}i\'{c} and Z.~Mari\'{c}: {\em Annals of
Physics\/} {\bf 214}, 1, 53--83 (1992).

\bibitem{Bar-Racz77}
A.~O.~Barut and R.~R\c{a}czka: {\em Theory of Group
Representations and Applications\/}, PWN --- Polish Scientific
Publishers, Warsaw 1977.

\bibitem{God03}
P.~Godlewski: {\em Int.\ J.\ Theor.\ Phys.\/} {\bf 42}, 12,
2863--2875 (2003).

\bibitem{HKVdH74}
F.~W.~Hehl, G.~D.~Kerlick and P.~Van der Heyde: {\em Phys.\ Rev.\
D\/} {\bf 10}, 1066 (1974).

\bibitem{HLN77}
F.~W.~Hehl, E.~A.~Lord and Y.~Ne'eman: {\em Phys.\ Lett.\/} {\bf
71B}, 432 (1977).

\bibitem{Lan-Lif58}
L.~D.~Landau and E.~M.~Lifshitz: {\em Quantum Mechanics\/},
Pergamon Press, London 1958.

\bibitem{Loom53}
L.~H.~Loomis: {\em An Introduction to Abstract Harmonic
Analysis\/}, D. Van Nostrand Company, Inc., Princeton-New
Jersey-Toronto-London-New York 1953.

\bibitem{Mac63}
G.~W.~Mackey: {\em The Mathematical Foundations of Quantum
Mechanics\/}, Benjamin, New York 1963.

\bibitem{Mart03}
A.~Martens: {\em Rep.\ on Math.\ Phys.\/} {\bf 51}, 2/3, 287--295
(2003).

\bibitem{Mart04}
A.~Martens: {\em J.\ of Nonlin.\ Math.\ Phys\/} {\bf 11},
Supplement, 151--156 (2004).

\bibitem{Maur68}
K.~Maurin: {\em General Eigenfunction Expansions and Unitary
Representations of Topological Groups\/}, PWN --- Polish
Scientific Publishers, Warsaw 1968.

\bibitem{Rose95}
M.~E.~Rose: {\em Elementary Theory of Angular Momentum\/}, Dover
Publications 1995.

\bibitem{AKS89}
A.~K.~S\l awianowska: {\em Arch.\ of Mech.\/} {\bf 41}, 5,
619--640 (1989).

\bibitem{JJS02_2}
J.~J.~S\l awianowski: {\em Quantum and Classical Models Based on
GL$(n,\mathbb{R})$-Symmetry\/}, in: Proc. of the Second
International Symposium on Quantum Theory and Symmetries,
Krak\'ow, Poland, July 18-21, 2001, E. Kapu\'scik and A. Horzela
(eds), World Scientific, New Jersey-London-Singapore-Hong Kong
2002, 582--588.

\bibitem{JJS04}
J.~J.~S\l awianowski: {\em Classical and Quantum Collective
Dynamics of Deformable Objects. Symmetry and Integrability
Problems\/}, in: Proc. of the Fifth International Conference on
Geometry, Integrability and Quantization, June 5-12, 2003, Varna,
Bulgaria, Iva\"{\i}lo M. Mladenov and Allen C. Hirshfeld (eds),
SOFTEX, Sofia 2004, 81--108.

\bibitem{JJS04_2}
J.~J.~S\l awianowski: {\em J.\ of Nonlin.\ Math.\ Phys\/} {\bf
11}, Supplement, 130--137 (2004).

\bibitem{JJS-VK04_2}
J.~J.~S\l awianowski and V.~Kovalchuk: {\em J.\ of Nonlin.\ Math.\
Phys\/} {\bf 11}, Supplement, 157--166 (2004).

\bibitem{part1} J.~J.~S\l awianowski, V.~Kovalchuk, A.~S\l awianowska,
B.~Go\l ubowska, A.~Martens, E.~E.~Ro\.zko, and Z.~J.~Zawistowski:
{\em Rep.\ on Math.\ Phys.\/} (part I of this series submitted for
publication).

\bibitem{JJS-AKS93}
J.~J.~S\l awianowski and A.~K.~S\l awianowska: {\em Arch.\ of
Mech.\/} {\bf 45}, 3, 305--331 (1993).

\bibitem{Wign31}
E.~P.~Wigner: {\em Gruppentheorie und Ihre Anwendung auf die
Quantenmechanik der Atomspektren\/}, F. Viewag und Sohn,
Braunschweig 1931 (eng. translation by J. J. Griffin, Academic
Press, New York 1959).

\bibitem{Wign65}
E.~P.~Wigner, in: {\em Quantum Theory of Angular Momentum\/}, L.
C. Biedenharn and H. van Dam (eds), Academic Press, New York 1965.

\bibitem{Zhel78}
D.~P.~Zhelobenko: {\em Compact Lie Groups and Their
Representations\/}, Translations of Mathematical Monographs,
vol.~{\bf 40}, AMS 1978.

\bibitem{Zhel83}
D.~P.~Zhelobenko: {\em Representations of Lie Groups\/}, Nauka,
Moscow 1983 (in Russian).

\end{thebibliography}
\end{document}